\documentclass{aa}  

\usepackage[switch]{lineno}

\usepackage{bbm}
\usepackage{physics}
\usepackage{units}
\usepackage{cancel}
\usepackage{xcolor} 
\usepackage{float}
\usepackage{graphicx}
\usepackage{multirow}
\usepackage{txfonts}
\usepackage[export]{adjustbox}
\usepackage{url}
\colorlet{orange}{green!10!orange}
\usepackage{hyperref}
\hypersetup{
colorlinks=false, 
linktoc=all,    
linkcolor=black,  
}
\begin{document}

   \title{Evolution of Chemistry in the envelope of HOt corinoS (ECHOS)}

   \titlerunning{Sulphur chemistry in the Class 0 objects HH\,212 and NGC\,1333 IRAS\,4A}

   \subtitle{III. Sulphur chemistry in the Class 0 objects HH\,212 and NGC\,1333 IRAS\,4A}

   \author{P. Fernández-Ruiz
          \inst{1,2}\fnmsep\thanks{Corresponding author; \texttt{\href{mailto:pfernandez@cab.inta-csic.es}{pfernandez@cab.inta-csic.es}}}\and A. Fuente\inst{1}\and G. Esplugues\inst{3}\and D. Navarro-Almaida\inst{1}\and P. Rivière-Marichalar\inst{3}\and T. Alonso-Albi\inst{3}\and M. Rodríguez-Baras\inst{4}\and A. Asensio Ramos\inst{5,6}\and C. Westendorp Plaza\inst{5,6}\and L. Moral-Almansa\inst{1,2}
          }

   \institute{Centro de Astrobiología (CAB), CSIC-INTA, Carretera de Ajalvir Km. 4, Torrejón de Ardoz, 28850 Madrid, Spain
              \and Departamento de Física de la Tierra y Astrofísica, Facultad de Ciencias Físicas, Universidad Complutense de Madrid, 28040 Madrid,
Spain
         \and
              Observatorio Astronómico Nacional (OAN), Alfonso XII 3, 28014 Madrid Spain
               \and European Space Agency (ESA), European Space Astronomy Centre (ESAC), Camino Bajo del Castillo s/n, 28692 Villanueva de la Cañada, Madrid, Spain
        \and Instituto de Astrofísica de Canarias (IAC), Avda Vía Láctea s/n, E-38200 La Laguna, Tenerife, Spain
        \and Departamento de Astrofísica, Universidad de La Laguna, E-38205 La Laguna, Tenerife, Spain 
              }

   \abstract
   {The physics and chemistry of molecular gas change with the dynamical evolution of the newly formed protostars. The details of this evolution are not yet fully understood.}
   {Our goal is to identify chemical diagnostics to determine the physical conditions in protostellar envelopes and to help establish the development of matter during the formation of a low-mass star. In particular, we aim to investigate a possible variation of sulphur depletion during the star formation process at the scale of the cold envelope.}
   {We used observations obtained with the Yebes-40m and IRAM-30m telescopes to estimate column densities of CS, C$_2$S, C$_3$S, OCS, HCS$^+$, SO, H$_2$CS, SO$_2$, NS, NS$^+$, H$_2$S, HSCN, HNCS, CH$_3$SH, C$^{18}$O, C$^{17}$O, CH$_3$OH, H$_2$CO, and their isotopologues in the Class 0 objects HH\,212 and NGC\,1333 IRAS\,4A, using both {local thermodynamic equilibrium} (LTE) and non-LTE calculations. We used a neural emulator of the chemical code Nautilus to investigate the physical (density, gas temperature, cosmic-ray ionisation rate, and chemical time) and chemical (sulphur elemental abundance) conditions driving the evolution of the chemical composition of the cold envelope.}
   {A comparison of the abundances of these species with those of the Class 0 object B\,335 reveals a differentiated chemistry. While sulphur-bearing species containing carbon chains (C$_2$S, C$_3$S) are between three and seven times more abundant in B\,335 than in the other two objects, suggesting an earlier chemistry in this protostar, sulphur oxides and nitrogen-bearing species are approximately three times more abundant in NGC\,1333 IRAS\,4A. {Comparing with} similar studies in pre-stellar and protostellar cores, we derive an increase of about two orders of magnitude in the SO/CS and SO$_2$/C$_2$S ratios and a potential decrease in the HCS$^+$/CS ratio of a factor of ten during the transition from the pre-stellar phase to the Class 0 phase. {Our chemical modelling shows that while the chemistry of HH\,212 and NGC\,1333 IRAS\,4A is} well reproduced assuming a gas temperature of $\sim$25 K, we find significant differences in their average density and cosmic-ray ionisation rate. We estimate a sulphur depletion of a factor of $\sim$100 in their protostellar envelopes, which is similar to that measured in starless cores.  }   
   {Sulphur compounds are good evolutionary tracers of the pre- to protostellar phase transition, with oxygen-bearing species being more abundant than carbon-bearing species in evolved sources. These chemical differences are the consequence of envelope heating by the forming star and the environmental conditions. We do not detect any significant change in sulphur depletion in the cold envelope during the Class 0 phase.  }


   \maketitle

\section{Introduction}

{Class 0 protostars form within the filaments of molecular clouds,} which are characterised by {the presence of} dense ($n$\,>\,10$^4$~cm$^{-3}$) clumps, with sizes of $\sim$0.1-0.01~pc and temperatures around $\sim$10~K. Some of these cores, called prestellar cores, slowly accrete matter towards the centre, leading to the formation of a central protostar, which, together with the infalling envelope, forms the so-called Class~0 protostar. During the progressive collapse of a prestellar core, the density rapidly rises to 10$^8$-10$^9$~cm$^{-3}$ within a radius of $\sim$100~au, leading to a temperature increase of the inner envelope up to several hundred kelvin. This enables the sublimation of ice mantles from dust grain surfaces, significantly enriching {their chemistry. These regions that arise during the formation of Solar-type protostars are known as hot corinos, and they are relatively rich in interstellar complex organic molecules \citep[e.g.][]{Cazaux2003,herbstvandishoeck2009,jorgensen2020,vangelder2020}.} These complex organic molecules (COMs) have been observed in high-mass and low-mass star-forming regions, both in young protostars  (\citealt{caselliCeccarelli2012}; \citealt{jorgensen2016}) and in protostellar outflows (\citealt{arce2008}; \citealt{oberg2010}). By contrast, in the outer envelope, molecules are mostly depleted \citep{Ceccarelli2004}, reflecting pre-collapse conditions. As a protostellar system develops from the embedded Class~0/I phase to the more evolved Class~II phase, gas molecules may eventually be destroyed or incorporated into ice mantles on dust grains or planetesimals in discs (\citealt{visser2009}; \citealt{Drozdovskaya2016}). Studying Class~0 objects is therefore pivotal for understanding the physical and chemical transition that takes place between the birth of a Solar-type protostar and the formation of protoplanetary discs, which are believed to arise during the Class~0/I phase owing to the conservation of angular momentum (see e.g. \citealt{bianchi2019}; \citealt{jorgensen2020}).

During low-mass star formation, chemistry changes with the dynamical evolution of newly formed cores and can be traced through molecule deuteration, since deuterium fractionation changes with their evolution (e.g. \citealt{Aikawa2012}; \citealt{esplugues2022}). In addition to deuteration, the evolution of the early stages of star formation in dense clouds can also be traced from sulphur (S) chemistry through the so-called sulphur depletion problem, since at low ($\sim$10~K) temperatures and high densities (>\,$10^4$~$\mathrm{cm^{-3}}$), S molecules such as CS condense onto dust grain surfaces (e.g. {\citealt{Crapsi2005}}), and the depletion level increases with time (e.g. \citealt{BerginLanger1997}; \citealt{Aikawa2003}; \citealt{esplugues2022}). This leads to a sulphur depletion factor of up to 1000 compared to the estimated cosmic abundance (\citealt{Ruffle1999}; \citealt{Wakelam2004}; \citealt{esplugues2014}; \citealt{fuente2023}).
By contrast, in other regions of the interstellar medium (ISM), such as the diffuse ISM and photon-dominated regions, the observed gaseous sulphur accounts for the total solar abundance (e.g. \citealt{Howk2006}), with S/H$\sim$$1.5\times10^{-5}$ (\citealt{OSratiosol}).

Chemical models predict that in the dense ISM, atomic S sticks onto grains and is mostly hydrogenated, forming H$_2$S, especially at low (<\,20~K) temperatures (\citealt{millarherbst1990}; \citealt{vidal2017}; \citealt{Navarro-Almaida2020}), making H$_2$S the main sulphur reservoir in the ice. However, H$_2$S has never been detected in interstellar ices{, and computational studies on its binding energies show that the newly generated H$_2$S may {be desorbing} from the ice surface after its formation \citep{Bariosco2024,Bariosco2026}.} 
In fact, OCS and SO$_2$ are the only S-bearing molecules unambiguously detected in ice mantles to date (\citealt{palumbo1995}; \citealt{boogert1997}; \citealt{macclure2023}; \citealt{rocha2024}). Depleted sulphur could also be locked into (semi)refractory materials, such as species containing Na, K, Ca, Mg, and Al. Recent detections of sulphur refractory species in the ISM include MgS and NaS \citep{rey-montejo2024}, and CaS \citep{tasa-chaveli2025}; however, their column densities are about three to four orders of magnitude lower than those for SO$_2$ and CH$_3$SH (\citealt{tasa-chaveli2025}). This rules out sulphur refractory species as the main sulphur reservoirs in the ISM. 
Models and laboratory experiments also suggest that S could be locked into pure S-allotropes or hydrogen sulphides \citep{Cazaux2022}. In this regard, HS$_2$ is the only molecule containing more than one sulphur atom {that has been} reported in the ISM (\citealt{fuente2025}; \citealt{esplugues2025}). Nevertheless, its abundance is not sufficient to explain the missing S in dense regions of the ISM.
This missing sulphur, known as the sulphur depletion problem, continues to challenge our understanding of chemistry in star-forming regions.

The role of sulphur as a powerful tracer of dense core evolution has recently been reinforced by the study of S ratios and their variation throughout the prestellar to protostellar transition (\citealt{esplugues2023evolution}) within the Evolution of Chemistry in the envelope of Hot Corinos (ECHOS) project. This is an international long-term project (\citealt{esplugues2023evolution,esplugues2024}) whose main goal is to provide a complete chemical inventory of a large sample of hot corinos, including Class~0 and Class~I objects, located in several sky regions (Ophiuchus, Taurus, Perseus, and Orion), which are characterised by different star formation {activity}. The sample was selected to study how both the physical environment and the evolutionary stage of the associated protostar influence molecular complexity in these regions. The ECHOS project is based on homogeneous and systematic spectral surveys covering the 30-300~GHz range, carried out with the Yebes-40m and {Institut de radioastronomie millimétrique} (IRAM) 30m single-dish telescopes. {A multi-transition study with radiative transfer models allows for the derivation of accurate abundances from the cooler outer envelope to the warm hot corino.} The source sample considered in ECHOS contains several hot corinos at different stages (Class~0 and Class~I), such as NGC\,1333~IRAS\,4A (Perseus), HH\,212 (Orion), SVS13-A (Perseus), and B\,335 (a dark globule).

{Within the ECHOS project, we {have} analysed the abundances of several sulphur-bearing molecules in protostellar cold envelopes \citep{esplugues2023evolution,esplugues2024}. Our goal {is} to identify chemical diagnostics that trace the evolution of star formation and to understand the origin of their variations and their impact on sulphur chemistry, which {can be} shaped by the initial physical conditions and environmental factors.  Sulphur depletion is expected to decrease from the starless core phase in the vicinity of hot corinos {as} sulphur-bearing molecules are injected into the gas phase. In that regard, our study also aims to assess whether ice sublimation in the warm and hot inner regions of the protostellar gas influences the chemistry of the cold outer envelope. The first hot corino extensively analysed in the ECHOS project was B\,335 (Class~0).} \cite{esplugues2023evolution} derived abundances for 20 detected sulphur species, highlighting B\,335 as especially rich in S-bearing carbon chains compared to other Class~0 sources, with low SO {abundance} and no detection of SO$^+$. The characteristic chemistry found in this very early evolutionary stage (<\,$10^5$~yr) source, compared to other young protostars, suggested a possible accretion of surrounding material from the diffuse cloud onto the envelope of B\,335. In this paper, we continue the chemical characterisation of hot corinos within the ECHOS project by focusing on two new targets, HH\,212 and NGC\,1333~IRAS\,4A, located in Orion and Perseus, respectively. The analysis of these sources allows {a} comparison with the results for B\,335 (an isolated source) and enables a study of environmental effects on the chemistry of hot corinos, given the different star formation activity observed in Orion, Perseus, and the isolated dense globule B\,335. 

\section{Sample}
This work focuses on {the} Class~0 protostars from the ECHOS project. We present, for the first time, ECHOS observations of HH\,212 and NGC\,1333~IRAS\,4A. 
In the following, we briefly describe each target.

{HH\,212} is a young Class 0 protostellar system deeply embedded in a compact molecular cloud core in the L1630 cloud of Orion, at a distance of $\sim$400 pc (\citealt{Kounkel2017}). {HH\,212} is one of the very few hot corinos detected thus far in the Orion cloud, making it an interesting case to study a hot corino prototype in a massive star-forming region. The central protostar has a mass of $\sim$0.25 $M_\odot$ (\citealt{lee2017c}) and drives a powerful spinning bipolar jet (\citealt{Zinnecker1998}; \citealt{lee2017a}). Observations with the Atacama Large Millimeter Array (ALMA) towards the centre show a spatially resolved, nearly edge-on dusty disc with a radius of $\sim$60 au (\citealt{lee2017b}). \cite{lee2017c} detected a warm disc atmosphere containing a few organic molecules, and more recent ALMA observations reveal additional COMs characteristics of a hot corino, showing that the hot corino in HH\,212 corresponds to the warm disc atmosphere, {within 30-40 au of the protostar  (\citealt{lee2019})}. 

{NGC\,1333~IRAS\,4A} (hereafter IRAS\,4A) is a proto-binary system located in the NGC\,1333 reflection nebula in the Perseus cloud {($\sim$290 pc; \citealt{Zucker2018})} and consists of components IRAS\,4A1 and 4A2. This makes it an interesting source for probing the chemistry of star formation in the Perseus cloud, as well as the role of the environment in an active region. The two Class~0 protostars are separated by 1.8 arcsec (538 au; see e.g. \citealt{Looney2000}; \citealt{lopez-sepulcre2017}; \citealt{tobin2018}), and another nearby component, IRAS\,4B, is separated by an angular distance of 31 arcsec from IRAS\,4A (\citealt{marvel2008}). The IRAS\,4A system is surrounded by a massive and dense envelope of 5.6~$M_\odot$ {\citep{Kristensen2012} and has a bolometric luminosity of 14~$L_\odot$ (\citealt{Karska2013}, after scaling to a distance of $\sim$290~pc). Moreover, the dust temperature is estimated to be between 50 and 200 K (\citealt{lopez-sepulcre2017}). Many COMs such as CH$_3$OH, CH$_3$CN, HCOOH, CH$_3$OCHO, CH$_3$OCH$_3$, C$_2$H$_5$CN, and CH$_2$OHCHO have been detected towards IRAS\,4A with the IRAM-30m, Plateau de Bure interferometer (PdBI) and ALMA telescopes (\citealt{Bottinelli2004}; \citealt{Santangelo2015}; \citealt{lopez-sepulcre2017}; \citealt{Quitian-Lara2024}). {In particular, studies find that the hot corino emission size in IRAS\,4A2 ranges from 20 to 40 au, depending on the specific COM \citep{Frediani2025}}. In addition, the IRAS\,4A system also hosts a chemically rich bipolar molecular outflow that has been detected in many molecular tracers, such as CO, CS, H$_2$O, HCO$^+$, and SiO (\citealt{Lefloch1998a, Lefloch1998b}; \citealt{Santangelo2014}){, as well as COMs CH$_3$OH, CH$_3$CHO, NH$_2$CHO, and CH$_3$OCH$_3$ \citep{DeSimone2020}.} 

\section{Observations}

The observations of HH\,212 and IRAS\,4A were carried out with the Yebes-40m and IRAM-30m telescopes within the ECHOS and {Astrochemical Surveys At IRAM} \citep[ASAI;][]{Lefloch2018} projects.
Previous observations of IRAS\,4A with IRAM-30m were carried out within the IRAM Large Program ASAI, covering the 73-117~GHz, 130-179~GHz, and 200-276~GHz bands. To complement these data, ECHOS observed IRAS\,4A in the 31.0-50.3~GHz band with the Yebes-40m telescope. For HH\,212, ECHOS covered the 31.0-50.3~GHz band with the Yebes-40m telescope, as well as {the 73-103~GHz, 110-117~GHz, and 130-179~GHz frequency ranges }with the IRAM-30m telescope. These telescopes pointed to the position with coordinates $\alpha$$_{J2000.0}$ = 05$^{\mathrm{h}}$43$^{\mathrm{m}}$51.41$^{\mathrm{s}}$, $\delta$$_{J2000.0}$ = -01$^{\mathrm{o}}$02$\arcmin$53.1" for HH\,212, and $\alpha$$_{J2000.0}$ = 03$^{\mathrm{h}}$29$^{\mathrm{m}}$10.42$^{\mathrm{s}}$, $\delta$$_{J2000.0}$ = +31$^{\mathrm{o}}$13$\arcmin$32.2" for IRAS\,4A, {which corresponds to the position of component IRAS\,4A2.} 

\subsection{Yebes-40m telescope}

The observations with the Yebes-40m radiotelescope, located in Yebes (Guadalajara, Spain), were obtained using a receiver that consists of two cold high-electron-mobility transistor amplifiers that cover the 31.0-50.3 GHz Q-band with horizontal and vertical polarizations \citep{Tercero2021}. The backends consist of 2$\times$8$\times$2.5~GHz fast Fourier transform spectrometers with a spectral resolution of 38.15 kHz, providing full coverage of the Q-band in both polarizations. The observations were performed in position-switching mode with (-400$''$, 0$''$) as the reference position. The main beam efficiency varies from 0.6 at 32 GHz to 0.43 at 50 GHz. Pointing corrections were derived from nearby quasars and SiO masers, and errors remained within 2-3$''$.

\subsection{IRAM-30m telescope}

The IRAM Large Program ASAI targets were observed over six semesters from September 2012 to March 2015 using {Eight MIxer Receiver} (EMIR) E090, E150, and E230 connected to the Fast Fourier Transform Spectrometer (FFTS) in the {200~kHz} spectral resolution mode. They covered the 73-117 GHz, 130-179 GHz, and 200-276 GHz spectral ranges in wobbler-switching mode. For IRAS\,4A, complementary observations were obtained in January 2016. The observation strategy is fully described in the article that presents the ASAI project (\citealt{Lefloch2018}).

The ECHOS observations with the IRAM-30m radio telescope, located in Pico Veleta (Granada, Spain), cover the {73-103~GHz, 110-117~GHz, and 130-175~GHz spectral ranges for HH\,212.} These observations were carried out in a single session in July 2021 in wobbler-switching mode ($\pm$120\,$\arcsec$), using the broad-band EMIR (E090 and E150) receivers and the FFTS with a spectral resolution of 200 kHz. 
The intensity scale in antenna temperature ($T$$^{\star}_{\mathrm{A}}$, which is corrected for atmospheric absorption and for antenna ohmic and spillover losses) was calibrated using two absorbers at different temperatures and the atmospheric transmission model \citep[ATM;][]{Cernicharo1985, Pardo2001}. The calibration uncertainties were assumed to be 10$\%$. 

To convert to the main-beam brightness temperature ($T$$_{\mathrm{MB}}$), we used the expression
\begin{equation}
T_{\mathrm{MB}}=\left(F_{\mathrm{eff}}/B_{\mathrm{eff}}\right)\times{T^{\star}_{\mathrm{A}}}=\left(T^{\star}_{\mathrm{A}}/\eta_{\mathrm{MB}}\right),
\end{equation}
where $F$$_{\mathrm{eff}}$ is the telescope forward efficiency and $B$$_{\mathrm{eff}}$ is the main beam efficiency\footnote{\href{https://rt40m.oan.es/rt40m_en.php}{https://rt40m.oan.es/rt40m\_en.php}}$^,$\footnote{\href{http://www.iram.es/IRAMES/mainWiki/Iram30mEfficiencies}{http://www.iram.es/IRAMES/mainWiki/Iram30mEfficiencies}}. 
Table \ref{table:tablaeficiencias} shows a summary of the telescope beam sizes and efficiencies as a function of frequency. {For the IRAM-30m telescope, we used Ruze's equation to obtain precise beam efficiencies.} 

For both telescopes, we reduced and processed the data using the {Continuum and Line Analysis Single-dish Software} (CLASS) and {Grenoble Graphic} (GREG) packages provided within the {Grenoble Image and Line Data Analysis Software} (GILDAS)\footnote{\href{http://www.iram.fr/IRAMFR/GILDAS}{http://www.iram.fr/IRAMFR/GILDAS}}, developed by the IRAM institute.}

\section{Data analysis and results}

We searched for all possible transitions of 26 sulphur-bearing species with upper energy levels below 100 K. Along with the detected S-bearing molecules, we included C$^{18}$O, C$^{17}$O, CH$_3$OH, and H$_2$CO in our chemical analyses to compare their abundances. We took line parameters from the Cologne Database for Molecular Spectroscopy (CDMS\footnote{\href{https://cdms.astro.uni-koeln.de/classic/}{https://cdms.astro.uni-koeln.de/classic/}}; \citealt{Muller2001,Muller2005,cdms}) and the Jet Propulsion Laboratory (JPL) Molecular Spectroscopy catalog\footnote{\href{https://spec.jpl.nasa.gov/}{https://spec.jpl.nasa.gov/}} \citep{jpl}. Table \ref{detectadas} shows the searched species, along with the number of detected emission lines above 3$\sigma$. {Figs.~S.1-S.15 and S.16-S.41 of the supporting material show the detected spectral lines for HH\,212 and IRAS\,4A, respectively.} We did not include detected lines that overlapped other species or lines of the same molecule, as occurs for several CH$_3$OH, CH$_3$SH, H$_2$CS, and NS transitions. All studied species except C$_3$$^{34}$S, OC$^{33}$S, HC$^{34}$S$^+$, and HNCS have been detected in IRAS\,4A. {However, only 14 species have been detected in HH\,212, in addition to a tentative detection of CH$_3$SH (see Fig. \ref{CH3SH_tentativo}). Upper energy levels in HH\,212 only reach $\sim$33~K, whereas in IRAS\,4A, we still detect transitions up to our maximum searched value of 100~K. }

\begin{table}[t]
\centering
\caption{Number of detected lines for each searched molecule in HH\,212 and IRAS\,4A, together with the maximum and minimum upper energy levels in K ($E_\text{u}^\text{max}$, $E_\text{u}^\text{min}$).}
\begin{tabular}{l|ccc|ccc}
\hline\hline         
\noalign{\smallskip}
     \multicolumn{1}{l}{}   & \multicolumn{3}{c}{HH\,212} & \multicolumn{3}{c}{IRAS\,4A} \\
\noalign{\smallskip}
\hline
\noalign{\smallskip}
Species & Lines &$E_\text{u}^\text{min}$&$E_\text{u}^\text{max}$        & Lines &$E_\text{u}^\text{min}$&$E_\text{u}^\text{max}$   \\
\noalign{\smallskip}
\hline
\noalign{\smallskip}
CS                      &     3    &  2.4  &  14.1   &  4          & 2.4   & 35.3    \\
$\mathrm{C{^{34}S}}$    &     3    &  2.3  &  11.8   &  4          & 2.3   & 27.8    \\
$\mathrm{C^{33}S}$      &     -    &  -    &  -      &  4$^{(a)}$  & 2.3   & 35.0    \\
$\mathrm{^{13}CS}$      &     2    &  2.2  &  6.7    &  4          & 2.2   & 33.3    \\
$\mathrm{C_2S}$         &     6    &  3.2  &  19.9   &  24         & 3.2   & 65.3    \\
$\mathrm{C_2{^{34}S}}$  &     -    &  -    &  -      &  3          & 3.2   & 36.3    \\
$\mathrm{C_3S}$         &     4    &  5.8  &  33.3   &  9          & 5.8   & 47.4    \\
$\mathrm{C_3{^{34}S}}$  &     -    &  -    &  -      &  -          & -     & -       \\
$\mathrm{OCS}$          &     -    &  -    &  -      &  12         & 3.5   & 99.8    \\
$\mathrm{OC^{34}S}$     &     -    &  -    &  -      &  6          & 20.5  & 97.4    \\
$\mathrm{OC^{33}S}$     &     -    &  -    &  -      &  -          & -     & -       \\
$\mathrm{HCS^+}$        &     1    &  2.0  &  2.0    &  6          & 2.0   & 43.0    \\
$\mathrm{HC^{34}S^+}$   &     -    &  -    &  -      &  -          & -     & -       \\
$\mathrm{SO}$           &     3    &  9.2  &  28.7   &  17         & 9.2   & 56.5    \\
$\mathrm{^{34}SO}$      &     1    &  9.1  &  9.1    &  13         & 9.1   & 55.7    \\
$\mathrm{S^{18}O}$      &     -    &  -    &  -      &  3          & 8.7   & 22.9    \\
$\mathrm{H_2CS}$        &     2    &  1.6  &  22.9   &  20         & 1.6   & 98.8    \\
$\mathrm{SO_2}$         &     1    &  15.7 &  15.7   &  28         & 7.7   & 82.8    \\
$\mathrm{^{34}SO_2}$    &     -    &  -    &  -      &  3          & 7.7   & 15.6    \\
$\mathrm{NS}$           &     -    &  -    &  -      &  8          & 16.6  & 38.8    \\
$\mathrm{NS^+}$         &     -    &  -    &  -      &  2          & 7.2   & 14.4    \\
$\mathrm{H_2S}$         &     1    &  27.9 &  27.9   &  2          & 27.9  & 84.0    \\
$\mathrm{H_2{^{34}S}}$  &     -    &  -    &  -      &  2          & 27.8  & 83.8    \\
$\mathrm{HSCN}$         &     -    &  -    &  -      &  1          & 24.8  & 24.8    \\
$\mathrm{HNCS}$         &     -    &  -    &  -      &  -          & -     & -       \\
CH$_3$SH                &1$^{(b)}$ & 13.5  & 13.5    &  8          & 16.7  & 66.7    \\
\noalign{\smallskip}
\hline
\noalign{\smallskip}
C$^{18}$O       &   -$^{(c)}$ & -     & -     &  2         & 5.3   & 15.8  \\
C$^{17}$O       &   1         & 5.4   & 5.4   &  1 & 16.2  & 16.2  \\

CH$_3$OH        &   9         & 2.3   & 27.1  &  58        & 2.3  & 98.5   \\

H$_2$CO         &   3         & 10.5  & 22.6  &  12        &  3.5 & 72.4   \\
\noalign{\smallskip}
\hline
\noalign{\smallskip}
\end{tabular}
\tablefoot{$^{(a)}$Number of rotational lines without considering their hyperfine components. $^{(b)}$Tentative detection. $^{(c)}$None of the lines fell within our spectral range.}
\label{detectadas}
\end{table}

We fit the detected lines with Gaussian profiles using the CLASS software from the GILDAS package, thus deriving their radial velocity $\left(v_\text{LSR}\right)$, linewidth $\left(\Delta v \right)$, and intensity $\left( T_\text{MB} \right)$.  Tables S.1 and S.2 of the supporting material show these parameters, for HH\,212 and IRAS\,4A, respectively, {with uncertainties that include the 10\% calibration error as well as the {root mean square} (rms) at the original spectral resolution around each line.} {The C$^{33}$S lines present hyperfine (HF) splitting that can be resolved in some cases, where we fitted individual Gaussians to each component, as shown in Table S.2.} {Taking into account the rms of each detected line, the mean is around 7~mK for both the Yebes-40m observations and the IRAM-30m IRAS\,4A observations. This value is slightly higher for the IRAM-30m HH\,212 observations, at $\sim$9~mK.}

\begin{figure}[t]
    \centering
    \includegraphics[width=\linewidth]{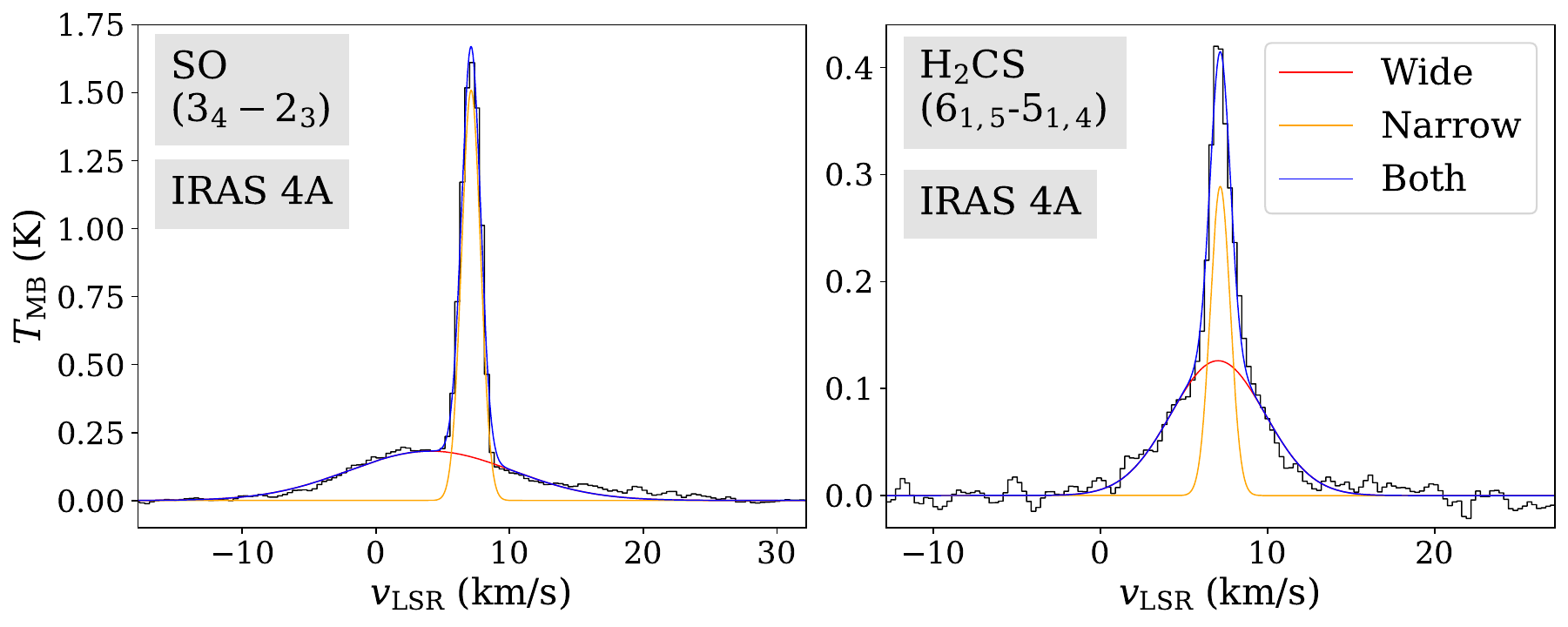}
    \caption{Gaussian fits {of} two selected IRAS\,4A lines that {present} narrow and wide components, expressed in terms of main-beam brightness temperature $(T_\text{MB})$ and {velocity of} local standard of rest $(v_\text{LSR})$.}
    \label{espectros_wide}
\end{figure}

Some transitions show complex profiles with intense high-velocity wings. In these cases, we added a secondary wide Gaussian component to properly reproduce the line shape, as shown in Fig. \ref{espectros_wide}.
We detect this wide component in many line transitions towards IRAS\,4A, including CS, C$^{34}$S, $^{13}$CS, OCS, HCS$^+$, SO, H$_2$CS, SO$_2$, H$_2$S, CH$_3$OH, and H$_2$CO. In HH\,212, we only identify this component in the SO~(4$_3$-3$_2$) line. This feature detected towards the Class 0 sources of our sample is likely due to {the} outflows associated with these young protostars{. It reaches very broad linewidths in IRAS\,4A, especially in SO and SO$_2$, with a maximum width of 19\,$\mathrm{km\,s^{-1}}$, while the linewidth of the wide component in HH\,212 is only 2.8\,$\mathrm{km\,s^{-1}}$. Nonetheless, the specific origin of this wide component is difficult to determine without interferometric observations, as higher-frequency transitions may include emission from more compact regions.}

Figs. \ref{EuHH212} and \ref{EuIRAS4A} show the physical and kinematical structure of the protostellar envelopes of HH\,212 and  IRAS\,4A from the linewidths of the different molecular lines as a function of the upper energy level of the transition. We only considered molecular lines fitted with a single narrow component since these are the ones coming from the envelope. Linewidths in HH\,212 are within the errors associated with the channel resolution, and we obtain an average linewidth value of  $\sim$1~$\mathrm{km\,s^{-1}}$. We observe a more complex kinematical structure for IRAS\,4A, as expected given the large size of the inner hot corino. For most molecules, we do not detect any clear trend, although we find a large dispersion around an average value of $\sim$1.5~$\mathrm{km\,s^{-1}}$.
However, we find differences for C$_3$S and C$^{18}$O. In the case of C$_3$S, the linewidth is larger for high-energy transitions, consistent with the interpretation that intense C$_3$S emission {may originate from an inner region of the envelope}. We find the opposite trend for C$^{18}$O, suggesting that {the} emission in this molecule is dominated by the outer layer of the protostellar envelope. 

\subsection{Rotational diagrams and LTE column densities}
\label{rotdiag}

For molecules with two or more rotational lines detected in each object, we performed a rotational diagram analysis \citep{goldsmith1999population}, assuming {local thermodynamic equilibrium} (LTE) and an optically thin regime. With this technique, we computed the rotational temperature ($T_\text{rot}$) and column density ($N$) of each molecule using the following relation:
\begin{equation}
    \ln \left( \dfrac{N_\text{u}}{g_\text{u}}\right)=\ln \left(\dfrac{N}{Z}\right)-\dfrac{E_\text{u}}{kT_\text{rot}},
    \label{gold1}
\end{equation}
where
\begin{equation}
    \dfrac{N_\text{u}}{g_\text{u}}=\dfrac{8k\pi \nu_\text{ul}^2}{hc^3A_\text{ul}g_\text{u}} \cdot f_\text{bf}^{-1}\cdot \int T_\text{MB}\,\mathrm{d}v.
    \label{gold2}
\end{equation}
Here, $N_\text{u}$ is the column density of the upper level, $g_\text{u}$ is the statistical weight of the upper state of each level, $Z$ is the partition function at temperature $T_\text{rot}$, $E_\text{u}$ is the energy of the upper level, $\nu_\text{ul}$ is the frequency of the transition, $A_\text{ul}$ is the Einstein coefficient of the transition, $f_\text{bf}$ is the beam filling factor, and $\int T_\text{MB}\,\mathrm{d}v$ is the velocity-integrated line intensity.
The beam-filling factor for a round source is $f_\text{bf}=\theta_\text{s}^2/(\theta_\text{s}^2+\theta_\text{b}^2)$, where $\theta_\text{s}$ is the diameter of the Gaussian source and $\theta_\text{b}$ is the {half-power beam width} (HPBW) of the telescope, both in arcseconds. This factor accounts for cases where the source does not completely fill the telescope beam, and equals one when $\theta_\text{s}\gg\theta_\text{b}$. We assumed this value to be one, unless stated otherwise, {as} the studied molecules are expected to have extended emission.

We analysed rotational diagrams for the wide and narrow components of each species separately. {For $\mathrm{C^{33}S}$, which shows some differentiated HF components, we summed the integrated intensities of the components for each rotational transition.} Moreover, we find that the H$_2$CO~($3_{2,2}$-$2_{2,1}$) and H$_2$CO~($3_{2,1}$-$2_{2,0}$) transitions in IRAS\,4A did not follow the trend of the other ten transition lines in the rotational diagram for either the wide or narrow components, so we excluded them from the fit. As these lines show broad wings, they may be contaminated with the outflow or be optically thick. 

\begin{figure}[t]
    \centering
    \includegraphics[width=\linewidth]{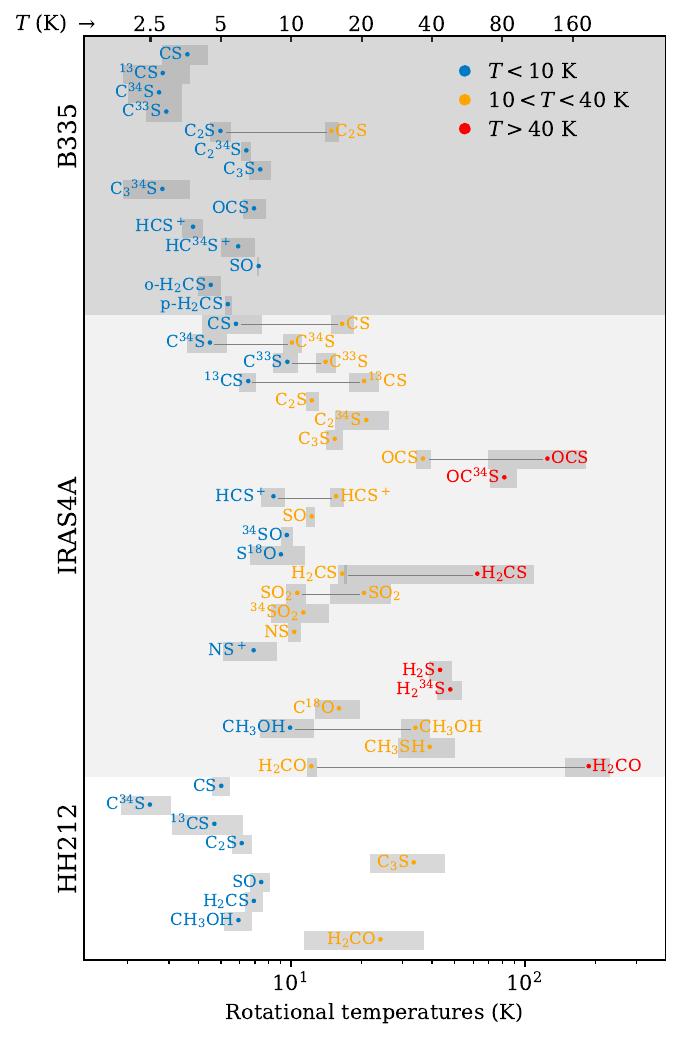}
    \caption{Rotational temperatures obtained from the narrow component of detected species in HH\,212, IRAS\,4A (this work), and B\,335 \citep{esplugues2023evolution}. Horizontal lines indicate species with two rotational temperatures {and shaded areas represent the uncertainties.}}
    \label{fig:temperaturas}
\end{figure}

Figs. \ref{diagramasHH212} and \ref{diagramasIRAS4A1} show rotational diagrams and fittings of narrow components, and Table \ref{ncol} lists their results. The uncertainties are derived from {the} least-squares linear {fitting}, including {the} 10\% flux-calibration error. We find that several detected species in IRAS\,4A require two $T_\text{rot}$ to fit their data: CS, C$^{34}$S, C$^{33}$S, $^{13}$CS, OCS, HCS$^+$, H$_2$CS, SO$_2$, CH$_3$OH, and H$_2$CO, whereas this is not required for any molecule in HH\,212. In those cases, a higher rotational temperature is {needed} for higher upper energy level transitions. 
However, {it is important to note that} for OCS, H$_2$CS, and H$_2$CO we have only a limited number of high $E_\text{u}$ data points, and the uncertainties in the warmer rotational temperature are large. These molecules trace gas that {arises} from different layers of the envelope: a colder gas from the outer envelope and a warmer gas that originates in an inner layer or in the walls of the bipolar outflow. 
Figure~\ref{fig:temperaturas} shows the rotational temperatures obtained for molecules with two or more detected lines in HH\,212 and IRAS\,4A, as well as the previously derived temperatures for the Class 0 source B\,335 \citep{esplugues2023evolution}. We find key differences between the three objects. Most species in B\,335 and HH\,212 have rotational temperatures below 10~K, with only three exceptions: C$_3$S and H$_2$CO in HH\,212, and C$_2$S in B\,335. This is not the case for IRAS\,4A molecules, which have temperatures mostly in the 10-40~K range. Interestingly, OC$^{34}$S, H$_2$S, and H$_2$$^{34}$S in IRAS\,4A reach temperatures above 40 K with a single rotational temperature in their fitting, suggesting that they exclusively trace the inner part of the envelope. This is particularly important for OC$^{34}$S, which has a very high rotational temperature of 82\,K and may be used as a tracer of the hot corino. Overall, IRAS\,4A shows a broader range of rotational temperatures than the other two objects, with low temperatures likely arising from the outer envelope, intermediate values originating from the outflow walls or an inner region of the envelope, and high temperatures that could {be stemming} from the hot corino.

Fig. \ref{diagramasIRAS4Awide} shows rotational diagrams of the wide components.
Most of them {present} greater dispersion than their narrow counterparts due to the difficulty in separating the wide component of each line.
We were unable to obtain {a} reliable fit in the rotational diagrams of the wide component of C$^{34}$S, $^{13}$CS, and OCS in IRAS\,4A when assuming a beam filling factor of {one}. 
For these molecules, the fit improves when assuming that the emission is point-like with respect to the angular resolution of our observations. 
Since we do not have information on the source size, we assumed $\theta_\text{s}$ to be the smallest $\theta_\text{b}$ in each diagram {($\sim$11~arcsec)}. {A much smaller source size, such as 0.5~arcsec, reduces the derived $T_\text{rot}$ by only 30\%.} Table~\ref{ncolwide} shows the results of the rotational diagram fits for the wide component. The rotational temperatures of the wide components in IRAS\,4A are mainly between 20-30~K, which is consistent with the interpretation that narrow component emission at those temperatures may arise from the outflow walls. For each species in Table \ref{ncolwide}, we also estimated the mass of the molecules associated with the wide component ($m_\text{wide}$) relative to the mass of the molecules associated with the narrow component ($m_\text{narr}$), where $m$ is proportional to the column density times $\theta_\text{s}^2$. 
This ratio may be an indicator of the relative abundances in the outflow with respect to the envelope gas. {We assumed the emission size to be equally extended for both the wide and narrow components, except for species for which we {included} a beam-filling factor in the wide component (C$^{34}$S, $^{13}$CS, and OCS in IRAS\,4A). In those cases, we estimate the diameter of the narrow-component emission as the maximum $\theta_\text{b}$ in the frequency range of the detected transitions ($\sim$30~arcsec). } 
With this in mind, most species in the two objects show a lower or similar number of molecules in the outflow compared to the envelope. However, this is not the case for CS, SO$_2$, H$_2$S, and CH$_3$OH in IRAS\,4A, which show approximately twice as many molecules in the outflow and are therefore good tracers of the outflow in this object.

To compute the LTE column densities of species with one single detected transition, we assumed an appropriate rotational temperature for each molecule. For most species, we adopted a temperature of 10~K, assuming that each detected line is associated with the cold envelope. For isotopologues whose stable molecule has a derived rotational temperature, we used that value in the calculation. If we estimate a conservative 25\% uncertainty in the assumed rotational temperatures and a 10\% error due to calibration, we derive a maximum $\sim$50\% error in the column densities. We also derived upper column density limits for the undetected species, {expecting} the lowest upper energy level transition of each molecule {to have} an integrated intensity below $3\sigma$ and an estimated linewidth of 1~$\mathrm{km\,s^{-1}}$. Tables \ref{ncol} and \ref{ncolwide} list column densities, upper limits, and assumed rotational temperatures. We find that the column densities of the narrow component of sulphur-bearing species are dominated by CS, SO, and H$_2$S in HH\,212 and by OCS in IRAS\,4A. In contrast, we find the lowest values for C$_3$S in HH\,212, and NS$^+$ and HSCN in IRAS\,4A.

\subsection{Non-LTE column densities and H$_2$ densities\label{h2densities}}

This section focuses only on the detected lines associated with the lower rotational temperature of narrow components, in order to exclusively target the gas stemming from the outer envelope of the objects. In addition, we do not include OC$^{34}$S, H$_2$S and H$_2$$^{34}$S in IRAS\,4A, since their rotational temperatures are well over 40~K and likely come from an inner region of the protostar instead.

We used the non-LTE radiative transfer code RADEX\footnote{\href{https://home.strw.leidenuniv.nl/~moldata/radex.html}{https://home.strw.leidenuniv.nl/~moldata/radex.html}} \citep{radex} to derive the column densities of species with available collisional rate coefficients, as well as the H$_2$ volume density associated with each molecule's region. The RADEX code provides the intensities of molecular lines for a species in a uniform medium for a given set of parameters. These parameters include the column density ($N_\chi$), H$_2$ density $(n_\mathrm{H_2})$, kinetic temperature $(T_\text{k})$ and linewidth. To run the code on a specific molecule, we used a molecular data file that includes its collisional rate coefficients. Files were either downloaded or constructed {from data in} the Leiden Atomic and Molecular Database (LAMBDA\footnote{\href{https://home.strw.leidenuniv.nl/~moldata/}{https://home.strw.leidenuniv.nl/~moldata/}}; \citealt{lambda}), BASECOL\footnote{\href{https://basecol.vamdc.eu/}{https://basecol.vamdc.eu/}} \citep{basecol}, the {Centre d'Analyse Scientifique de Spectres Instrumentaux et Synthétiques} (CASSIS) Collision Database\footnote{\href{https://cassis.irap.omp.eu/?page=catalogs-collision}{https://cassis.irap.omp.eu/?page=catalogs-collision}} and the Excitation of Molecules and Atoms for Astrophysics (EMAA\footnote{\href{https://emaa.osug.fr/}{https://emaa.osug.fr/}}). Table \ref{table:colisiones} provides the references for the collisional rate coefficients used in each case, noting that H$_2$CS, H$_2$S, and H$_2$CO are separated in their para and ortho forms, and CH$_3$OH in its A and E types.
As a general rule, we preferred recently published coefficients that were not scaled from analogue molecules. For the isotopologues whose coefficients were not available, we constructed their molecular files using their spectroscopic data and the same rate coefficients as the main isotopologue. Since no collisional coefficients were available for HSCN, HNCS, and CH$_3$SH, we adopted the LTE column densities from the previous calculations. We also assumed the LTE approximation for C$_3$S and C$_3$$^{34}$S, since the available collisional rate coefficients for C$_3$S only reach $J$=11 \citep{C3Scoeff}.

We computed a grid of H$_2$ density and column density values in RADEX for each detected line of a given species, {assuming a beam filling factor of one, as in the LTE analysis of the narrow components.}
{The grid was sampled with 50 H$_2$ density and 50 column density values scaled logarithmically over the ranges listed in Table S.3 of the supporting material for each species.}
Thus, for each transition, the code computes a synthetic line intensity at each point of the grid. We obtained the best fitting pair of $N_\chi$ and $n_\mathrm{H_2}$ by evaluating the $\chi^2$ estimator at every intersection in the grid using all detected lines as follows:
\begin{equation}
    \chi^2_{N_\chi,n_\mathrm{H_2}}=\dfrac{1}{\sqrt{n}}\sum_i^n \dfrac{\left(T_i^\text{syn}-T_i^\text{obs}\right)^2}{T_i^\text{obs}},
\end{equation}
where $n$ is the number of detected lines, $T_i^\text{syn}$ is the intensity given by RADEX in each of the $n$ transitions, and $T_i^\text{obs}$ is each of the line intensities of our observational data. The minimum $\chi^2$ in the grid provides the H$_2$ density and non-LTE column density for each molecule. To estimate the uncertainties, we calculated a $\Delta \chi^2$ of the $\chi^2_\text{min}$ using error propagation:
\begin{equation}
    \Delta \chi ^2 = \sqrt{\dfrac{1}{n}\sum_i^n \left(\dfrac{\partial \chi_\text{min}^2}{\partial T_i^\text{obs}}\Delta T_i^\text{obs}\right)^2+\dfrac{1}{n}\sum_i^n\left(\dfrac{\partial \chi^2_\text{min}}{\partial T^\text{syn}_i}\Delta T^\text{syn}_i\right)^2},
\end{equation}
where
\begin{align}
    &\dfrac{\partial \chi^2_\text{min}}{\partial T_i^\text{obs}} = \dfrac{\left(T_i^\text{obs}\right)^2-\left(T_i^\text{syn}\right)^2}{\left(T_i^\text{obs}\right)^2},\\
    &\dfrac{\partial \chi^2_\text{min}}{\partial T^\text{syn}_i}= \dfrac{2\left(T_i^\text{syn}-T_i^\text{obs}\right)}{T_i^\text{obs}}.
\end{align}
The uncertainty in each observed line intensity, $\Delta T_i^\text{obs}$, is assumed to be {of a conservative} 20\%, while $\Delta T^\text{syn}_i$ is the maximum variation in the synthetic intensity of each line associated with the size of the cell in the grid. Therefore, grid points satisfying $\chi^2_\text{min}+\Delta \chi^2$ provide the errors in the H$_2$ densities and column densities for each molecule.

\begin{figure*}[t]
    \centering
    \includegraphics[width=\textwidth]{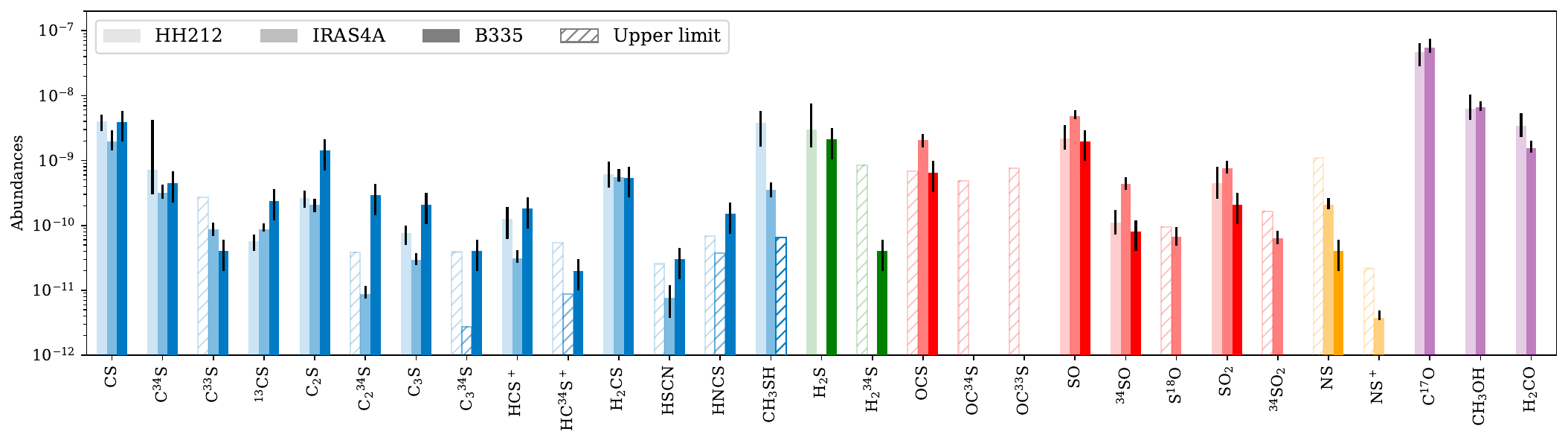}
    \caption{Abundances associated with the cold component of  HH\,212, IRAS\,4A, and B\,335 \citep{esplugues2023evolution}. The S-bearing species are colour-coded as blue for the carbonated species, green for the hydrogenated species, red for the oxygenated species, and yellow for the nitrogenated species. Non S-bearing species are coloured purple.}
    \label{fig:abundancias}
\end{figure*}

As the kinetic temperature is unknown for each species and represents the largest source of error in the H$_2$ densities, we adopted a set of three kinetic temperatures: 10, 20, and 30~K. We ran the code for each $T_\text{k}$, provided it was above the rotational temperature of the species plus its uncertainty. For molecules with a rotational temperature above 30\,K, we calculated the results for a singular kinetic temperature higher than their $T_\text{rot}$. 
We combined para and ortho synthetic lines in the $\chi^2$ calculation under the assumption that the ortho-to-para column density ratio is three for H$_2$CS, H$_2$S, and H$_2$CO.

Similar to the LTE column densities, we also derived an $N_\chi$ estimation for species with only one detected transition by assuming an H$_2$ density value. For HH\,212 and IRAS\,4A, we adopted the H$_2$ density from SO at each kinetic temperature, as this species shows several bright lines, which minimise observational errors, and has well-known collisional rate coefficients \citep{SOcoeff}. Moreover, the spatial distribution of SO is not expected to differ greatly from the other species considered in this analysis{, as they {present} similar rotational temperatures} (see Fig.~\ref{fig:temperaturas}). We derived the $N_\chi$ uncertainties by accounting for the H$_2$ density error obtained from SO and the 20\% error in the line intensity.

 Tables \ref{RADEXHH212} and \ref{RADEXIRAS4A} list all densities and non-LTE column densities for each molecule and object. We observe that the H$_2$ densities decrease with kinetic temperature in most molecules, and this variation can be up to a factor of approximately four between the 10~K and 30~K densities. There is also a large variation in the densities obtained from the different molecules. They are mostly in the $10^4$-$10^5$ cm$^{-3}$ range, with two exceptions in IRAS\,4A: $^{34}$SO$_2$ and NS, at the $10^6$ cm$^{-3}$ level, which might indicate that these molecules arise from an inner region. Overall, we also {notice} that the derived densities are always lower in HH\,212 than in IRAS\,4A for all considered temperatures. The mean density value at 20~K for HH\,212 is $3.6\times10^4$ cm$^{-3}$, while for IRAS\,4A, excluding the previously mentioned outliers, this value is $2.4\times10^5$~cm$^{-3}$.
Nevertheless, it should be noted that the uncertainties on the densities derived with this conservative method are large.

In contrast, column density uncertainties are much smaller and are always determined, which allows for more robust results. Since non-LTE column densities vary by less than a factor of 1.5 for $T_\text{k}$=10-30 K (see Tables \ref{RADEXHH212} and \ref{RADEXIRAS4A}), we assumed a representative
$N_\chi$ for each species in Table \ref{ncol}. We selected the column density value associated with $T_\text{k}$=20 K. If $N_\chi$ is not available for that $T_\text{k}$, we used the column density for the lowest $T_\text{k}$. 
We also provide a non-LTE column density upper limit for the undetected species using a kinetic temperature of 20\,K or the lowest possible $T_\text{k}$, taking the rotational temperatures of their isotopologues into account. Most species in HH\,212 and IRAS\,4A show a good agreement between their LTE and non-LTE column densities, with differences within a factor $\le$1.6. However, this factor is slightly higher for  C$^{34}$S and H$_2$S in HH\,212, reaching approximately three. This is not surprising for H$_2$S, since only one line was available and we assumed the rotational temperature, kinetic temperature, and H$_2$ density.

\subsection{Molecular abundances in the cold envelope}\label{sect:abundances}

\label{abun}

Reliable estimates of the H$_2$ column densities are required to derive molecular abundances. One possibility is to derive them from dust continuum maps, but this requires assuming a dust temperature and a dust spectral index, which can introduce important uncertainties in these protostars, where heating from the central star and grain growth are proceeding. In addition, the calculations depend on the spatial resolution of the observations. Instead, we used the C$^{18}$O column density derived in this work to estimate the H$_2$ column density, {assuming that the emission from all the molecules has a similar spatial extent.} This estimate is based on the same spectral survey and thus {with} a similar angular resolution, and the C$^{18}$O/C$^{17}$O ratio computed {in Sect.~\ref{secisotopic} (Fig.~\ref{fig:isotopic})} shows that the C$^{18}$O column density estimate is not affected by opacity effects. Furthermore, the C$^{18}$O abundance is relatively stable on molecular cloud scales, with an average value of $X(\mathrm{C^{18}O})$ equal to $1.7\times 10^{-7}$ \citep{trevino2019dynamics}. Since C$^{18}$O was not detected in HH\,212, we derived its column density from its $^{17}$O isotopologue by assuming a $^{18}$O/$^{17}$O ratio of $\sim$3.6 \citep{ratioO18}.

We used the non-LTE column densities, when available, to calculate the abundances relative to C$^{18}$O for the observed species in HH\,212 and IRAS\,4A. 
Fig. \ref{fig:abundancias} {and Table~\ref{abundances}} show the resulting abundances, where we also include the abundances reported for B\,335 by \cite{esplugues2023evolution}. Our results show that the abundances of S-bearing species can be grouped according to distinct chemical behaviour. The first group consists of molecules containing carbon, hydrogen, and sulphur, and is shown in blue in Fig. \ref{fig:abundancias}. The abundances of CS and H$_2$CS are similar within the uncertainties across the three protostellar envelopes. However, we detect a clear overabundance for the carbon-chain sulphur compounds C$_2$S and C$_3$S in B\,335.  \cite{esplugues2023evolution} attribute the unusually high abundance of these compounds  to a high cosmic ray ionisation rate ($\zeta_\mathrm{H_2}$ = 1.3 $\times$ 10$^{-16}$ cm$^{-3}$) and an early chemical age ($<$ 0.1 Myr) in this particular globule. \cite{Cabedo2023} also report evidence of a high cosmic ray ionisation rate in this source, based on high spatial resolution observations of CO, HCO$^+$, and DCO$^+$ isotopologues. Within the group of carbon species, the abundance of HCS$^+$ is particularly low in IRAS\,4A compared to the other two objects, which suggests a lower cosmic ray ionisation rate. In this group, we also include the complex organic molecule CH$_3$SH, which has been tentatively detected towards HH\,212 and shows a significantly higher abundance than  in IRAS\,4A. This indicates that the previously discussed low CH$_3$OH/CH$_3$SH ratio in HH\,212 is due to an overabundance of CH$_3$SH. 

\begin{table}[t]
\centering
\caption{Abundances associated with the cold component of HH\,212, IRAS\,4A, and B\,335 \citep{esplugues2023evolution}. }
\begin{tabular}{llll}
\hline\hline         
\noalign{\smallskip}
        & \multicolumn{1}{c}{HH\,212} & \multicolumn{1}{c}{IRAS\,4A} & \multicolumn{1}{c}{B\,335}\\
\noalign{\smallskip}
\hline
\noalign{\smallskip}
Species & $X\ (\times\,10^{-9})$ &$X\ (\times\,10^{-9})$&$X\ (\times\,10^{-9})$   \\
\noalign{\smallskip}
\hline
\noalign{\smallskip}
CS & $ 17 $$^{(a)}$ &
 $ 7.6 $$^{(a)}$ &
 $ 11 $$^{(a)}$ \\
C$^{34}$S & $ 0.7 $ &
 $ 0.31 $ &
 $ 0.45 $ \\
C$^{33}$S & $< 0.27 $ &
 $ 0.088 $ &
 $ 0.040 $ \\
$^{13}$CS & $ 0.056 $ &
 $ 0.088 $ &
 $ 0.24 $ \\
C$_2$S & $ 0.27 $ &
 $ 0.22 $$^{(a)}$ &
 $ 1.4 $ \\
C$_2$$^{34}$S & $< 0.039 $ &
 $ 0.0088 $ &
 $ 0.29 $ \\
C$_3$S & $ 0.076 $$^{(b)}$ &
 $ 0.030 $$^{(b)}$ &
 $ 0.21 $ \\
C$_3$$^{34}$S & $< 0.039 $$^{(b)}$ &
 $< 0.0027 $$^{(b)}$ &
 $ 0.040 $ \\
OCS & $< 0.68 $ &
 $ 2.1 $ &
 $ 0.65 $ \\
OC$^{34}$S & $< 0.49 $ &
- &
- \\
OC$^{33}$S & $< 0.77 $ &
- &
- \\
HCS$^+$ & $ 0.13 $ &
 $ 0.031 $ &
 $ 0.50 $$^{(a)}$ \\
HC$^{34}$S$^+$ & $< 0.054 $ &
 $< 0.0088 $ &
 $ 0.020 $ \\
SO & $ 2.7 $$^{(a)}$ &
 $ 11 $$^{(a)}$ &
 $ 2.0 $ \\
$^{34}$SO & $ 0.11 $ &
 $ 0.44 $ &
 $ 0.080 $ \\
S$^{18}$O & $< 0.094 $ &
 $ 0.066 $ &
- \\
H$_2$CS & $ 0.62 $ &
 $ 0.55 $ &
 $ 0.54 $ \\
SO$_2$ & $ 0.45 $ &
 $ 1.6 $$^{(a)}$ &
 $ 0.21 $ \\
$^{34}$SO$_2$ & $< 0.17 $ &
 $ 0.063 $ &
- \\
NS & $< 1.1 $ &
 $ 0.21 $ &
 $ 0.040 $ \\
NS$^+$ & $< 0.022 $ &
 $ 0.0038 $ &
- \\
H$_2$S & $ 3.0 $ &
- &
 $ 2.1 $ \\
H$_2$$^{34}$S & $< 0.85 $ &
- &
 $ 0.040 $ \\
HSCN & $< 0.025 $$^{(b)}$ &
 $ 0.0077 $$^{(b)}$ &
 $ 0.030 $ \\
HNCS & $< 0.068 $$^{(b)}$ &
 $< 0.037 $$^{(b)}$ &
 $ 0.15 $ \\
 CH$_3$SH & $ 3.8 $$^{(b)}$ &
 $ 0.35 $$^{(b)}$ &
 $< 0.065 $ \\
  \noalign{\smallskip}
\hline
\noalign{\smallskip}
C$^{18}$O &- &
 $ 170 $ &
- \\
C$^{17}$O & $ 47 $ &
 $ 58 $ &
- \\
CH$_3$OH & $ 6.2 $ &
 $ 6.6 $ &
- \\

H$_2$CO & $ 3.4 $ &
 $ 1.6 $ &
- \\
 \noalign{\smallskip}
\hline
\noalign{\smallskip}
Total $X$ of & \multirow{2}{*}{$2.9\times 10^{-8}$} & \multirow{2}{*}{$2.4\times 10^{-8}$} & \multirow{2}{*}{$2.0\times 10^{-8}$}\\
S-bearing species\\
\noalign{\smallskip}
\hline
\noalign{\smallskip}
\end{tabular}
\tablefoot{{HH\,212 and IRAS\,4A abundances are derived from the non-LTE column densities of the cold component in Table~\ref{ncol}. We adopted a kinetic temperature of 20~K to derive the column densities used, except for H$_2$CO in HH\,212, for which a $T_\text{k}$ of 40~K is used, C$_2$$^{34}$S in IRAS\,4A ($30$~K), and OCS in IRAS\,4A ($40$~K).} $^{(a)}$Obtained from the corresponding $^{34}$S isotopologue assuming $^{32}$S/$^{34}$S = 24.4 \citep{ratioS}. $^{(b)}${Derived from LTE column densities of the cold component in Table~\ref{ncol} for the lack of available collisional rate coefficients.}}
\label{abundances}
\end{table}
Oxygen-bearing compounds (OCS, SO, and SO$_2$) show higher abundances in IRAS\,4A than in HH\,212 and B\,335. Interestingly, we do not detect OCS in HH 212, although it is one of the most abundant species in IRAS\,4A and B\,335.
On the other hand, SO and SO$_2$ can form both in the gas phase and on grain surfaces. In particular, SO$_2$ is thought to form in the gas phase over long timescales ($\sim$1 Myr) under the physical conditions prevailing in starless cores \citep{B1b, Vidal2018}. The compound SO$_2$ is known to be a major component of interstellar ice \citep{macclure2023} and is released into the gas phase in warm regions \citep{esplugues2014,Artur2023}.
The nitrogen bearing species NS and NS$^+$ show the same behaviour.
Table \ref{abundances} shows the abundances for HH\,212 and IRAS\,4A, deriving those of the $^{32}$S species from their $^{34}$S isotopologues, when available, to account for opacity effects. Similarly, the abundances for B\,335 \citep{esplugues2023evolution} are included in Table \ref{abundances}, with CS and HCS$^+$ also {considered} to be optically thick. We summed the abundances of all detected sulphur compounds to identify any evolution in sulphur depletion during protostellar evolution. The sulphur budget in the detected molecules ranges from 2.0 to $2.9 \times10^{-8}$, with B\,335 and HH\,212 being the least and most sulphur-rich targets, respectively. This value is similar to that found in starless cores such as TMC-1 (CP), but is several orders of magnitude lower than those found in photon-dissociation regions, hot cores, and hot corinos \citep{Goicoechea2006,Riviere-Marichalar2019, esplugues2014, fuente2024, fuente2025, Miranzo-Pastor2025}. This suggests that sulphur depletion remains high in the cold envelopes of young protostars, although the detailed chemical composition of the sulphur budget may evolve with environmental conditions.

\section{Chemical modelling}\label{sect:modeling}

\begin{figure*}[t]
    \sidecaption
    \includegraphics[width=0.3\linewidth,valign=c]{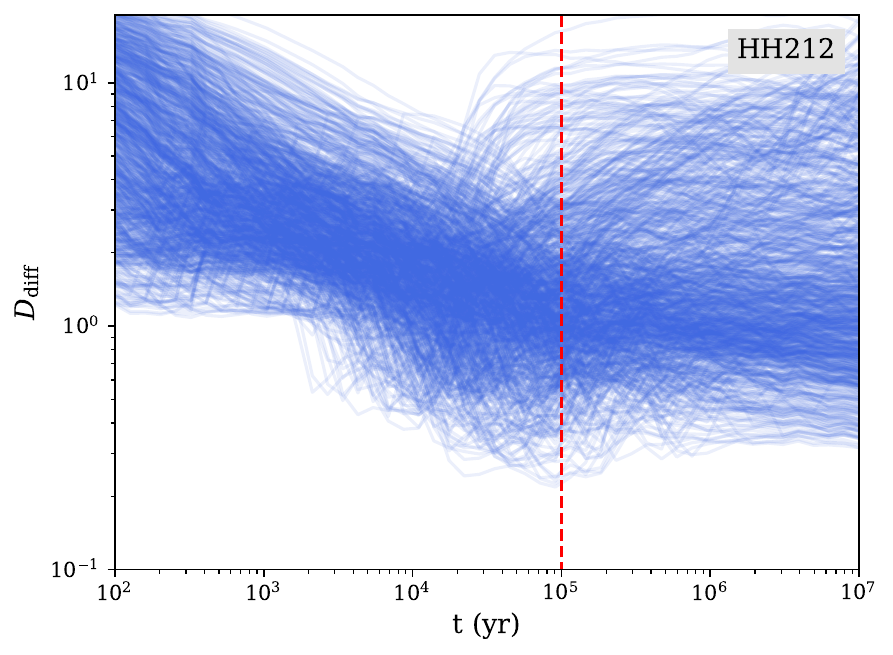}
    \includegraphics[width=0.3\linewidth,valign=c]{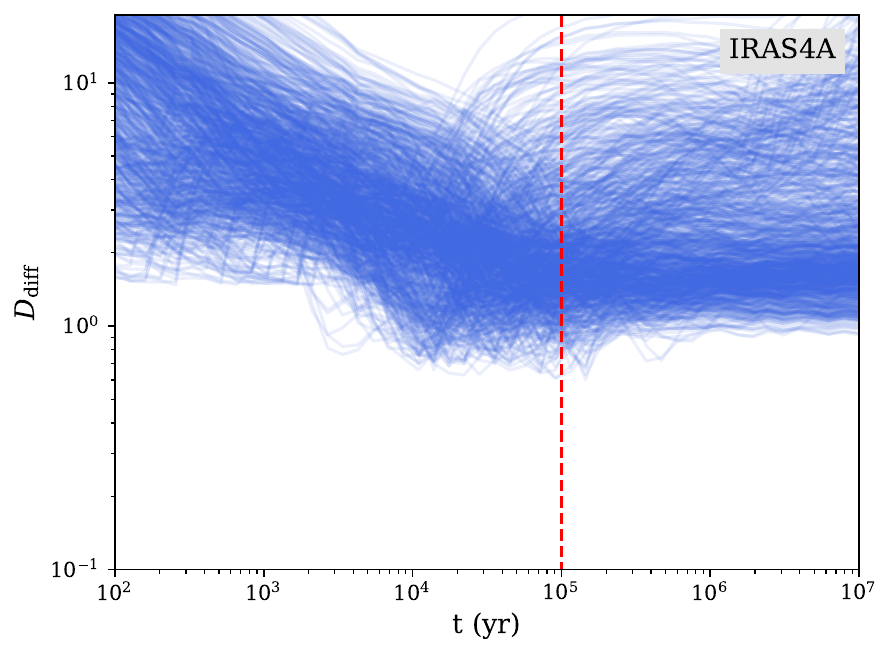}
    \caption{Distribution of $D_{\text{diff}}(t)$ for HH\,212 and IRAS\,4A from a set of $10^3$ models considering the physical conditions in Table \ref{parameters}. The dashed red line indicates $t=10^5$~yr.}
    \label{fig:modelos_tiempo}
\end{figure*}

To further investigate the physical and chemical properties of these Class 0 objects, we compared the observational data (Table~\ref{abundances}) with a set of chemical models. We computed these chemical models using a fast neural emulator of the astrochemical code Nautilus 1.1 based on conditional neural fields \citep{AsensioRamos2024}. Nautilus 1.1 \citep{Ruaud2016} is a three-phase model that includes gas, grain surface, and grain-mantle phases, together with their interactions. 
{It provides abundances of chemical species for a given set of parameters by solving the rate equations for the gas phase, the surface of the icy mantle, and the bulk of the ice mantle on interstellar dust grains. Both accretion and desorption {can} occur between the gas and the grain surface, with desorption proceeding through thermal and non-thermal mechanisms.} The neural emulator {has the advantage of accurately predicting} abundances with uncertainties well below 0.2 dex, with a computing time of the order of $10^4$ shorter than Nautilus. This enables a rapid in-depth exploration of the parameter space that is not possible with complex chemical models. 

{Despite the accuracy provided by the neural emulator, we must also consider the limitations of chemical modelling, which are mainly associated with the adopted chemical network (the number and type of included reactions and chemical parameters, such as binding and diffusion energies) and the assumed physical model (such as time-dependent or fixed physical parameters and the presence of turbulence). Assuming constant physical properties (such as temperature or density) over time is a significant simplification that might affect the {calculation of} molecular abundances, since chemical evolution is tightly linked to environmental conditions, which change with time (e.g. \citealt{NavarroAlmaida2024}; \citealt{Jensen2026}). However, introducing time dependence for the different physical parameters would significantly increase both the complexity of the model and the degeneracy of the results, making their interpretation challenging. Chemical networks are also an important source of uncertainty because, among the thousands of chemical reactions used to model interstellar regions, most have not been studied experimentally or theoretically, especially under conditions relevant to the ISM (\citealt{Wakelam2010}). {In this regard,} we used the Nautilus version that was recently updated with the new astrochemical network kida.uva.2024, together with the ice chemical network and the Fortran code to compute the time-dependent compositions of the gas, ice surface, and ice mantles under physical conditions in the ISM (\citealt{Wakelam2024}). Despite these updates, the chemical network may still be incomplete for some species. This is particularly relevant for surface reactions involving sulphur species, such as S-allotropes and hydrogen sulphides, whose chemistry is still very limited due to the lack of laboratory experiments. Nevertheless, in this work we use the most up-to-date astrochemical network which best reproduces interstellar conditions, thus minimising the modelling caveats.  }

For both objects, we ran a set of models with parameters randomly selected from the intervals in Table \ref{parameters}. We sampled parameters uniformly in logarithmic space, except for the temperature, which we sampled linearly. We allowed a wide range of values for the chemical age ($t$), the sulphur elemental abundance {in volatiles} (\text{[S/H]}), and the cosmic-ray ionisation rate ($\zeta_\mathrm{H_2}$), as these are parameters we wanted to constrain. The chemical age can take values between early chemistry (100 yr) to steady-state chemistry (10~Myr). The sulphur elemental abundance varies from no depletion ($1.5\times10^{-5}$) to high depletion ($7.5\times10^{-8}$), as estimated from different values of sulphur depletion in star-forming regions (\citealt{esplugues2014}; \citealt{vastel2018sulphur}; \citealt{Navarro-Almaida2020,Navarro-Almaida2021}). 
The cosmic-ray ionisation rate ranges between $10^ {-17}$~$\mathrm{s^{-1}}$ and $10^ {-15}$~$\mathrm{s^{-1}}$, values found in dense, evolved cores \citep{Caselli2002} and diffuse molecular gas \citep{NeufeldWolfire2017}, respectively. We considered the hydrogen density ($n_\text{H}$) to be between $2\times 10^4$~$\mathrm{cm^{-3}}$ and $2\times 10^6$~$\mathrm{cm^{-3}}$, and the temperature ($T$) {between} 10 and 50~K, which are the expected ranges in the envelope of these objects. Moreover, we fixed the visual extinction ($A_\mathrm{V}$) at 18 mag and the UV field ($\chi_\text{UV}$) at one {unit of} Draine field \citep{Draine1978}, assuming that the protostellar material is well shielded from external UV photons. {In addition, {because we are considering} the large-scale cold envelope of each source, we also assumed that the internal UV irradiation from the central Class 0 object was negligible, since the material in the inner regions can act as a shield blocking this radiation.}

To estimate the values of time, $n_\text{H}$, $\zeta_\mathrm{H_2}$, \text{[S/H]}, and temperature, we defined the square of the disagreement parameter ($D_\text{diff}$) to describe the goodness of each model:
\begin{equation}
    D_\text{diff}(t) = \dfrac{1}{n_\text{obs}}\sum_i \left[\log_{10} (X^i_\text{mod})(t)-\log_{10} (X^i_\text{obs})(t)\right]^2, 
\end{equation}
where $n_\text{obs}$ is the number of detected species in each source, $X^i_\text{mod}(t)$ is the abundance predicted by the model for the $i$ species at time $t$, and $X^i_\text{obs}(t)$ is the observed abundance of species $i$. The disagreement parameter is commonly used in astrochemistry when comparing observations with chemical models (e.g. \citealt{Wakelam2006}; \citealt{vastel2018sulphur}; \citealt{fuente2023}; \citealt{Taillard2025}).

As a first step, we aimed to determine the chemical age that best reproduces the observational data for HH\,212 and IRAS\,4A, as constraining this parameter first allows for more precise estimates of the other parameters. We ran the models and allowed the parameter $t$ to vary over a set of values within the previously described range. Figure~\ref{fig:modelos_tiempo} shows the $D_{\text{diff}}$ at each $t$ for a set of $10^3$ models in both objects. 
In both cases, we find the lowest $D_{\text{diff}}$ for a chemical age of $t$$\sim$$10^5$ yr. Nonetheless, for IRAS\,4A, this result is less well constrained, and values between $10^4$ and $10^5$~yr {might} also reproduce the observations with similar accuracy.

\begin{figure*}[t]
    \sidecaption
    \includegraphics[width=0.3\linewidth,valign=c]{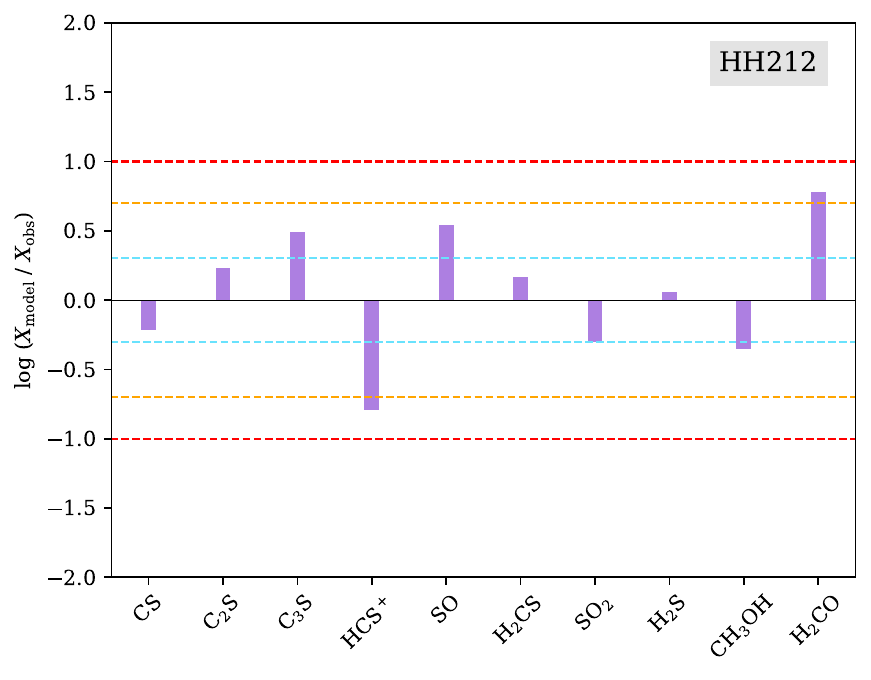}
    \includegraphics[width=0.3\linewidth,valign=c]{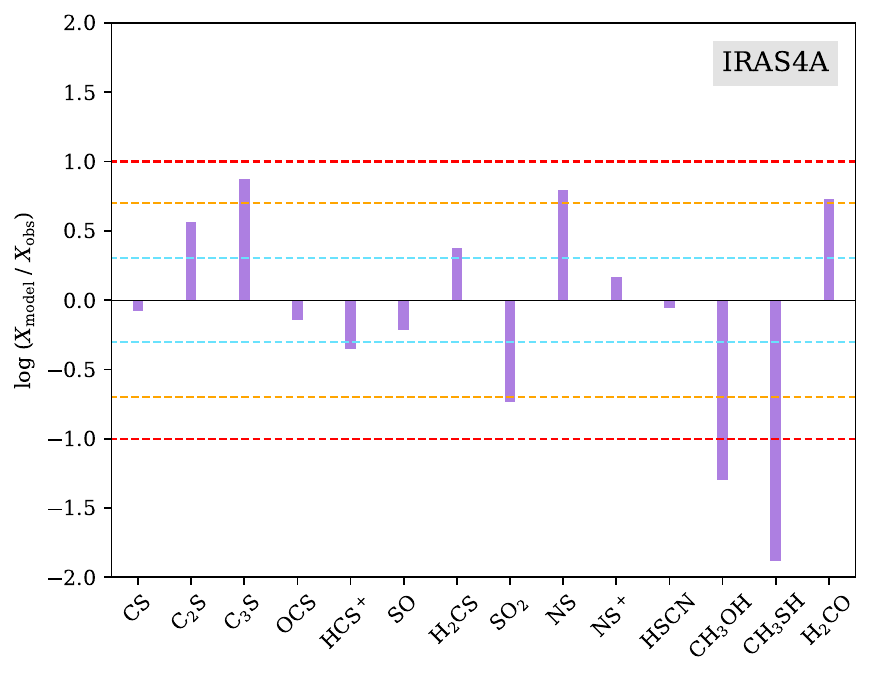}
    \caption{Abundance ratios between the best-fitting model (see Table \ref{modelling_results}) and the observed values in HH\,212 and IRAS\,4A. Dashed lines indicate discrepancies between the model and observations of a factor of two (blue), five (orange) and ten (red).}
    \label{fig:modelos_abundancias}
\end{figure*}

\begin{table}[t]
    \centering
    \caption{Parameters of the best-fitting models at $t=10^5$ yr.}
    \begin{tabular}{lll}
    \hline\hline         
\noalign{\smallskip}
&HH\,212 &IRAS\,4A\\

\noalign{\smallskip}
\hline  
\noalign{\smallskip}
       \multicolumn{3}{c}{$D_{\text{diff,\,min}}$}\\
       \noalign{\smallskip}
       \hline 
       \noalign{\smallskip}
        $n_\text{H}$ ($\mathrm{cm^{-3}}$)\ \ \ \ \ \ \ \  & $2.6\times 10^4$ & $1.7\times 10^5$ \\
        $\zeta_\mathrm{H_2}$ ($\mathrm{s^{-1}}$)&$2.5\times 10^{-16}$ & $1.2\times 10^{-17}$ \\
        \text{[S/H]}& $1.8\times 10^{-7}$&$8.8\times 10^{-8}$ \\
        $T$ (K)&28.0&26.2\\
        \noalign{\smallskip}
       \hline 
       \noalign{\smallskip}
       \multicolumn{3}{c}{$D_{\text{diff}}\in [D_{\text{diff,\,min}},\ 1.05\times D_{\text{diff,\,min}}]$}\\
       \noalign{\smallskip}
       \hline 
       \noalign{\smallskip}
        $n_\text{H}$ ($\mathrm{cm^{-3}}$)&$(2.6-3.8)\times 10^4$ &$(1.7-2.1)\times 10^5$\\
        $\zeta_\mathrm{H_2}$ ($\mathrm{s^{-1}}$)&$(2.5-3.1)\times 10^{-16}$ & $(1.0-3.0)\times 10^{-17}$\\
        \text{[S/H]} & $(1.7-2.8)\times 10^{-7}$&$(0.0088-1.1)\times 10^{-5}$\\
        $T$ (K)& $27.7-28.4$& $18.6-26.2$\\
        \noalign{\smallskip}
       \hline 
       
    \end{tabular}
    \label{modelling_results}
\end{table}

We then ran another set of models to constrain the other parameters while fixing the chemical age in both objects to $10^5$~yr. Table \ref{modelling_results} shows all the best-fitting parameters for HH\,212 and IRAS\,4A, as well as the parameters for which $D_\text{diff}$ differs from the minimum $D_\text{diff}$ by less than 5\%. 
The densities derived from the fitting are the same within the uncertainties as the values derived from the excitation studies of the observed molecules, indicating that we have achieved a consistent solution. When testing the results for a possible $t=10^4$~yr in IRAS\,4A, we find that the densities obtained are of the order of $10^6$~$\mathrm{cm^{-3}}$, much higher than the densities obtained in Sect.~\ref{h2densities}. This supports a chemical age $10^5$ yr in IRAS\,4A as the more accurate approximation. Moreover, the S$^{16}$O/S$^{18}$O ratio observed in IRAS\,4A also predicts a chemical age of $10^5$ yr when compared to the chemical model in \cite{O18fractionation}, as explained in Sect.~\ref{secisotopic}.
Returning to the parameters derived while fixing $t=10^5$ yr (Table~\ref{modelling_results}), we see that the estimated temperatures are similar in HH\,212 and IRAS\,4A, but the cosmic-ray ionisation rate is a factor of ten higher towards HH\,212, which drives the differences in chemical composition. {In fact, the range of gas temperatures derived in IRAS\,4A (18.6-26.2~K) is close to the dust temperature at the time of  mantle formation in NGC\,1333 IRAS\,4, as determined by \citet[$\sim$17~K]{DeSimone2022}, which is also consistent with the average temperature of the southern NGC\,1333 diffuse cloud. In contrast, the dust temperatures in the L1630 cloud hosting HH\,212 are 15-20~K \citep{Savva2003}, similar to those of NGC\,1333, and only slightly lower than our derived gas temperature in HH\,212.} {Regarding} the sulphur elemental abundance, it is higher by a factor of approximately two towards HH\,212, which could be related to the higher cosmic-ray ionisation rate.

Fig.~\ref{fig:modelos_abundancias} compares our best-fit model with the abundances derived from the observations. In HH\,212, the abundances are fitted within a factor of ten. The fitting is worse towards IRAS\,4A, where the discrepancies {in }CH$_3$OH and CH$_3$SH are larger than a factor of ten. This poor fit is likely due to the complexity of this target, with a large hot corino and several bipolar outflows. This complex physical structure makes it very difficult to discern the abundances that stem from the cold envelope from those in the hot corino and the walls of the bipolar outflows. {Moreover, the abundances of SO$_2$ and CH$_3$OH, which are known to be major components of the ice mantles on dust grains, are under-reproduced in IRAS\,4A. These compounds could have been released into the gas-phase due to evaporation or grain sputtering by the jets of the protostar, a scenario that is not accounted for by the chemical model.}

\begin{figure*}[t]
    \centering
    \includegraphics[width=
    \hsize]{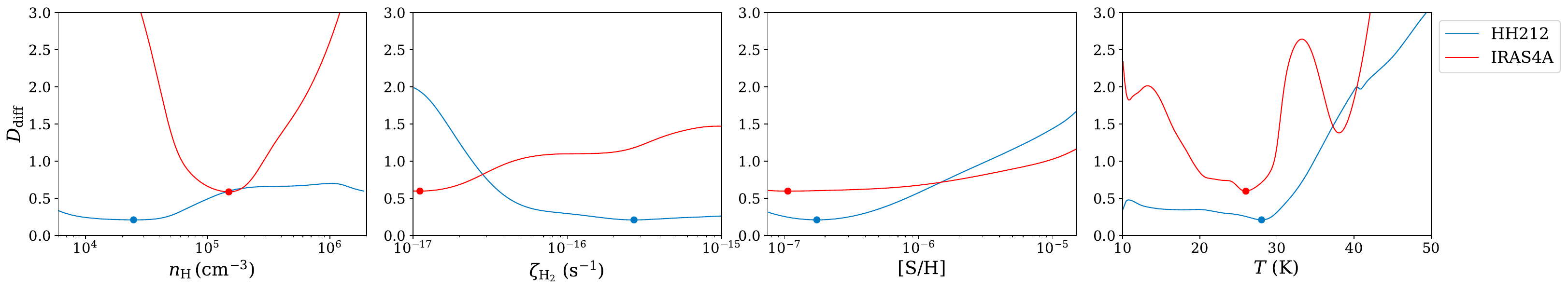}
    \caption{Variation of $D_{\text{diff}}$ with each parameter for HH\,212 and IRAS\,4A, with all other parameters fixed. Points indicate the minimum $D_{\text{diff}}$ value for each parameter.}
    \label{fig:modelos_errores}
\end{figure*}

To test the robustness of the fitting, we varied one parameter while keeping all others constant and equal to the best-fitting values, and calculated the variations in $D_\text{diff}$. Following this procedure, we investigated the reliability of our estimates of the temperature, density, cosmic-ray ionisation rate, and sulphur elemental abundances
(see Fig.~\ref{fig:modelos_errores}). {The temperature and density are better constrained in the fitting than the values of [S/H] and $\zeta_\mathrm{H_2}$.} Regarding $\zeta_\mathrm{H_2}$, while the value of  $\zeta_\mathrm{H_2}$ appears to be very low ($\sim$10$^{-17}$~s$^{-1}$ towards IRAS\,4A), values of $\zeta_\mathrm{H_2}$ $>$ 5 $\times$ 10$^{-17}$ s$^{-1}$ could account for the observations towards HH\,212. For IRAS\,4A, the variation of $D_\text{diff}$ with [S/H] is shallow, with values ranging from $\sim$0.5 to  $\sim$1.0  when [S/H] increases by a factor of $\sim$100.
Fig.~\ref{fig:SH} shows model-derived abundances for each of the observed molecules in IRAS\,4A at different sulphur elemental abundances, while fixing the other parameters to the values listed in Table \ref{modelling_results} for IRAS\,4A. As expected, the abundances of all sulphur compounds increase with [S/H], but the correlation is not linear. Indeed, a variation of two orders of magnitude in [S/H] produces changes of less than a factor of ten in the predicted abundances of the observed sulphur compounds, making a robust determination of [S/H] challenging. Moreover, the fit to the abundances is poor for some molecules, especially for CH$_3$SH, whose abundance is severely underestimated by the chemical model predictions. Although most of the observed abundances are best fitted with a high sulphur depletion 
([S/H] = 1.0 $\times$ 10$^{-7}$), SO$_2$ and CH$_3$SH require higher sulphur elemental abundances. In fact, the high abundances of SO$_2$ and CH$_3$SH cannot be fitted even by assuming undepleted sulphur. This may be due to the fact that part of their emission originates in warm components of the envelope, either the warm walls of the bipolar outflow or the hot corino. These two molecules are expected to be abundant in warm regions \citep{Esplugues2013}. However, our chemical models have difficulty reproducing the abundances of all sulphur compounds (see e.g. \citealt{Moral-Almansa2026}). The limited ability of our models to reproduce all sulphur-bearing species implies that the results depend on the specific selection of species in the fitting. Since we are essentially using all observable sulphur compounds, we consider this fit to be the best possible with the current knowledge of sulphur chemistry.

\section{Discussion}

\subsection{Isotopic ratios\label{secisotopic}}

\begin{figure}[t]
    \centering
    \includegraphics[width=\linewidth]{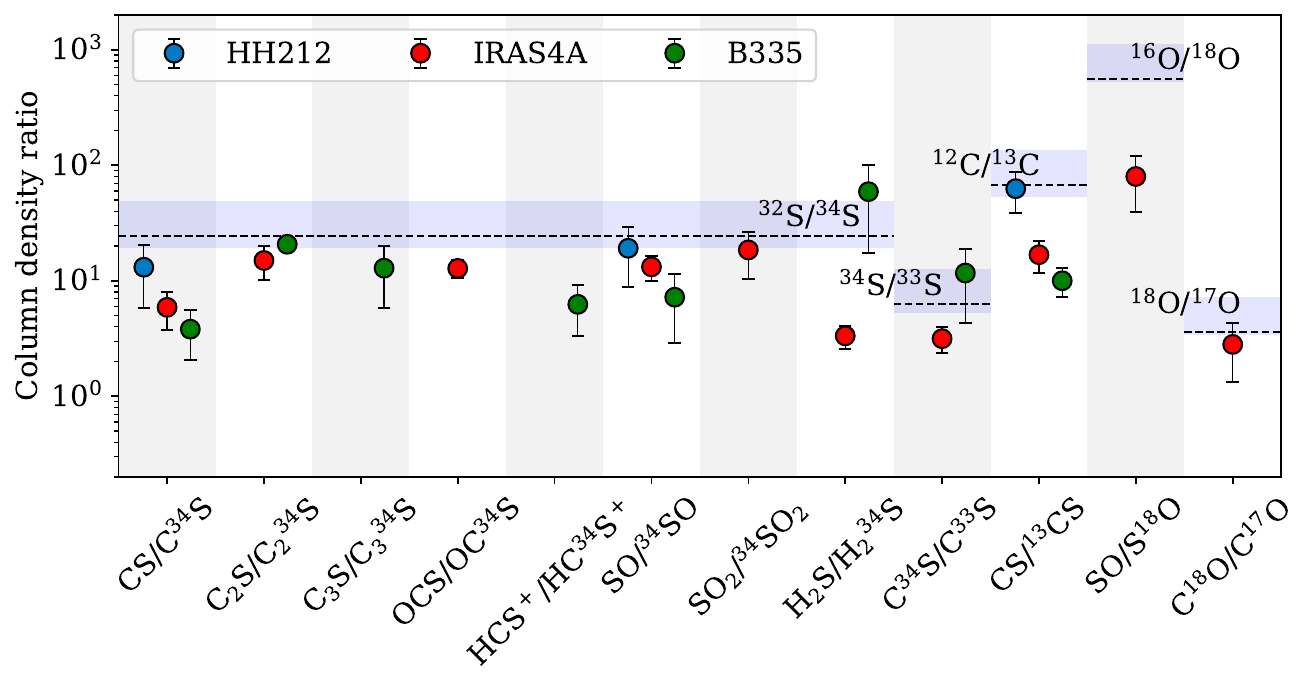}
    \caption{Isotopologue ratios in HH\,212, IRAS\,4A, and B\,335 \citep{esplugues2023evolution}. The dashed line shows the ISM elemental ratios of $^{32}$S/$^{34}$S ($\sim$24.4; \citealt{ratioS}), $^{34}$S/$^{33}$S ($\sim$6.3; \citealt{ratioS}), $^{12}$C/$^{13}$C ($\sim$68; \citealt{ratioC}), $^{16}$O/$^{18}$O ($\sim$557; \citealt{ratioO18}), and $^{18}$O/$^{17}$O ($\sim$3.6; \citealt{ratioO18}), with the uncertainties indicated by blue shading.}
    \label{fig:isotopic}
\end{figure}

From the column densities associated with the narrow component obtained in Table \ref{ncol} for objects HH\,212 and IRAS\,4A, as well as those {calculated} in \cite{esplugues2023evolution} for B\,335, we derived isotopologue ratios (Fig. \ref{fig:isotopic}) and compared them with their corresponding elemental ratios in the ISM: $^{32}$S/$^{34}$S, $^{34}$S/$^{33}$S \citep{ratioS}, $^{12}$C/$^{13}$C \citep{ratioC}, $^{16}$O/$^{18}$O, and $^{18}$O/$^{17}$O \citep{ratioO18}.
We find that some $^{32}$S/$^{34}$S isotopologue ratios are lower than expected. This is the case for CS, OCS, SO, and H$_2$S in IRAS\,4A, and for CS, HCS$^+$, and SO in B\,335, suggesting that the $^{32}$S compound may be optically thick. For CS, this is further supported by the fact that the $^{12}$CS/$^{13}$CS ratio is also lower than expected in IRAS\,4A and B\,335, and that this species is usually slightly saturated in star-forming regions (e.g. \citealt{LinkeGoldsmith1980}). We find the $^{32}$SO/$^{34}$SO isotopic ratio to vary little with time in gas-grain chemical models of dense molecular clouds \citep{O18fractionation}, and it has a similar value to the $^{32}$S/$^{34}$S ratio in the local ISM, in diffuse molecular clouds \citep{LucasLiszt1998}, and in the Solar System \citep{Lodders2003}. This ratio shows a positive gradient with increasing galactocentric radius \citep{Yu2020}, but this is not expected to affect our results, since our targets are located in the solar neighbourhood. For these reasons, the $^{34}$S isotopologue is considered a reliable proxy for deriving abundances when the main $^{32}$S species is optically thick.
Conversely, the C$^{34}$S/C$^{33}$S  ratio in IRAS\,4A is also below the ISM isotopic ratio, as well as the $^{34}$S/$^{33}$S ratio found in comets \citep{Calmonte2017} and the accepted solar value \citep{Ding2001}, suggesting that $^{34}$S could also be somewhat optically thick.
In addition, the S$^{16}$O/S$^{18}$O ratio is much lower than the ISM isotopic ratio derived by \cite{ratioO18}, {even when deriving the S$^{16}$O from the $^{34}$S$^{16}$O value to correct for possible opacity effects.}
As shown by \cite{O18fractionation}, the S$^{16}$O/S$^{18}$O ratio varies significantly with time in chemical models  and reaches values similar to IRAS\,4A at $\sim$$1\times 10^5$ yr. Our S$^{16}$O/S$^{18}$O ratio is also similar to those observed in cold cores in the same study. Consequently, this isotopologue could be used as an evolutionary tracer, but it is not reliable for deriving abundances of the main SO isotopologue.

\subsection{Relative O/S ratio}

\begin{figure}[t]
    \centering
    \includegraphics[width=\linewidth]{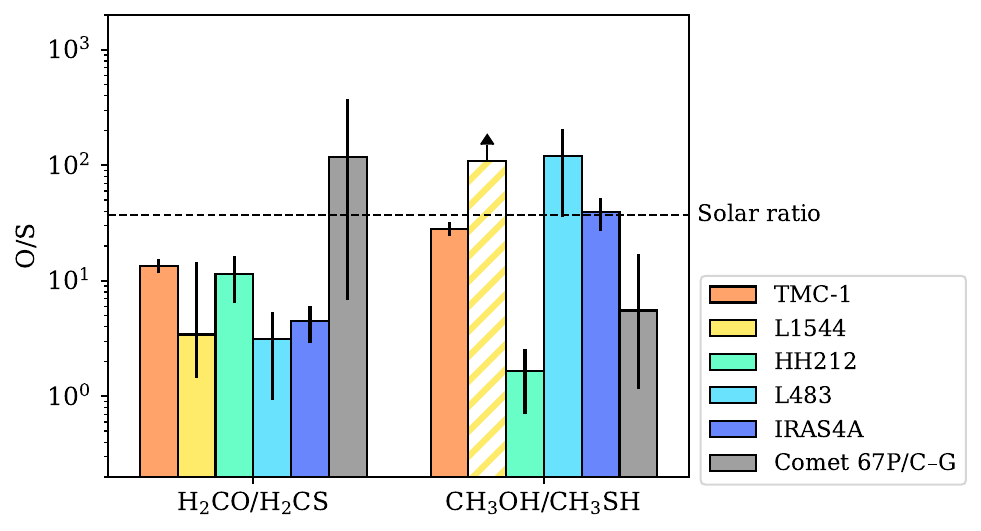}
    \caption{Relative O/S ratio for the H$_2$CO/H$_2$CS and CH$_3$OH/CH$_3$SH pairs and upper limits (striped bars) for TMC-1 \citep{TMC1_3}, L1544 (\citealt{L1544_2}; \citealt{L1544}; \citealt{vastel2018sulphur}), HH\,212 (this work), L483 (\citealt{agundez2019sensitive}), IRAS\,4A (this work), and comet 67P/Churyumov-Gerasimenko \citep{cometa}. We also show the solar ratio O/S$\sim$37 (\citealt{OSratiosol}; \citealt{OSratiosol2}; \citealt{OSratiosol3}). }
    \label{fig:o/s}
\end{figure}

To understand sulphur depletion, we studied the observed sulphur abundance relative to oxygen across different stages of star formation. In Fig. \ref{fig:o/s} we include the H$_2$CO/H$_2$CS and CH$_3$OH/CH$_3$SH ratios of the starless core TMC-1 \citep{TMC1_3}, {the prestellar core L1544 (\citealt{L1544_2}; \citealt{L1544}; \citealt{vastel2018sulphur}), the Class 0 protostars} HH\,212 (this work), L483 (\citealt{agundez2019sensitive}), IRAS\,4A (this work), and comet 67P/Churyumov-Gerasimenko \citep{cometa}. From Fig. \ref{fig:o/s} we find that the CH$_3$OH/CH$_3$SH abundance ratio is higher than the H$_2$CO/H$_2$CS ratio in the low-mass star-forming sources TMC-1, L1544, L483, and IRAS\,4A. This is in agreement with \cite{TMC1_3}, who find that the H$_x$CO/H$_x$CS abundance increases with hydrogenation ($x$) in TMC-1, likely because hydrogenation is less efficient for S-bearing than O-bearing molecules. 
{Nevertheless}, in HH\,212 and comet 67P/C-G we observe the opposite trend. In particular, the CH$_3$OH/CH$_3$SH ratio in HH\,212 is
as low as $\sim$1.6, making it an outlier compared to the other sources. Although CH$_3$SH is only a tentative detection in the envelope of HH\,212, we find that high-angular-resolution interferometric observations also detect CH$_3$SH emission in the atmosphere of the compact disc \citep{lee2017c}. The CH$_3$OH/CH$_3$SH abundance ratio in the disc atmosphere is $\sim$3.4, similar to that found in our single-dish observations. This confirms HH\,212 as having a particularly low CH$_3$OH/CH$_3$SH ratio.  {As comet 67P/C-G {also} shows a low CH$_3$OH/CH$_3$SH ratio, this may indicate similarities between HH\,212 and the early Solar System. Indeed, some studies suggest that the Solar System {could} have formed near high-mass stars \citep{Adams2010,Pfalzner2015}, similar to the Orion cloud where HH\,212 is located. However, further observations are needed to explore possible variations of this ratio in such objects, as well as observations of other low-mass stars forming in Orion.}

Moreover, for all sources in Fig. \ref{fig:o/s}, the O-bearing species (H$_2$CO and CH$_3$OH) are more abundant than their S-bearing counterparts (H$_2$CS and CH$_3$SH), as expected given that the oxygen elemental abundance is 30 times greater than that of sulphur. However, this is not always the case for this class of molecules; for instance, S-bearing molecules containing carbon chains are much more abundant than their O-bearing equivalents \citep{TMC1_3}.
This has also been found for HS$_2$ and its oxygen counterpart HSO (\citealt{Fuente2017}; \citealt{esplugues2025}) in different types of regions, as well as for the H$_2$C$_3$O and H$_2$C$_3$S pair \citep{Esplugues2026}. This suggests a markedly different chemistry between oxygen and sulphur analogue species, although both elements belong to the chalcogens.

{Overall, taking into account the uncertainties and the outlying CH$_3$OH/CH$_3$SH ratio in HH\,212, we find that the O/S ratio is fairly similar across the sources in Fig.~\ref{fig:o/s}. As the O/S ratio is intrinsically related to the availability of sulphur, this suggests that sulphur depletion remains constant in the early stages of star formation and in the cold envelopes of Class 0 protostars.}

\subsection{{Abundance ratios of sulphur-bearing species in the cold envelope}}

Sulphur species have been widely proposed and used as evolutionary indicators during the pre-stellar and protostellar stages \citep{BerginLanger1997, Kontinen2000, B1b, vidal2017, Vidal2018, vastel2018sulphur, NScoeff, esplugues2023evolution, Quitian-Lara2024, Ghosh2024}. Figure~\ref{fig:comparacion} compares representative abundance ratios as estimated in the protostellar envelopes of the ECHOS programme, with results obtained from other spectral surveys observed towards pre-stellar cores and low-mass protostars at comparable spatial resolution, {deriving them from their $^{34}$S isotopologues when available to eliminate opacity effects.} We do find an overall trend with the SO/CS and SO$_2$/C$_2$S ratios increasing from the pre-stellar to the Class 0 phase, {especially for SO$_2$/C$_2$S,} while the HCS$^+$/CS ratio {could be slightly decreasing.} However, there are some ratios that cannot be explained only in the basis of protostellar evolution. For instance, explaining the difference between the chemistry of TMC-1 and L1544 remains challenging only on the basis of the evolutionary stage \citep{Agundez2013}. Also, B1-b is one of the youngest Class 0 protostars detected thus far \citep{Gerin2015, Fuente2017} and the measured abundance ratios are similar to those observed in the more evolved object IRAS\,4A.  This suggests that the environment could also play a role in setting these ratios, as B1-b and  IRAS\,4A are both located in active star forming regions in Perseus. Significant chemical differences are also found between the evolved Class 0 stars HH\,212 and IRAS\,4A, especially in the SO/CS ratio. 

\subsection{Understanding chemical abundances using machine learning interpretability}

\begin{figure}[t]
    \centering
    \includegraphics[width=0.91\linewidth]{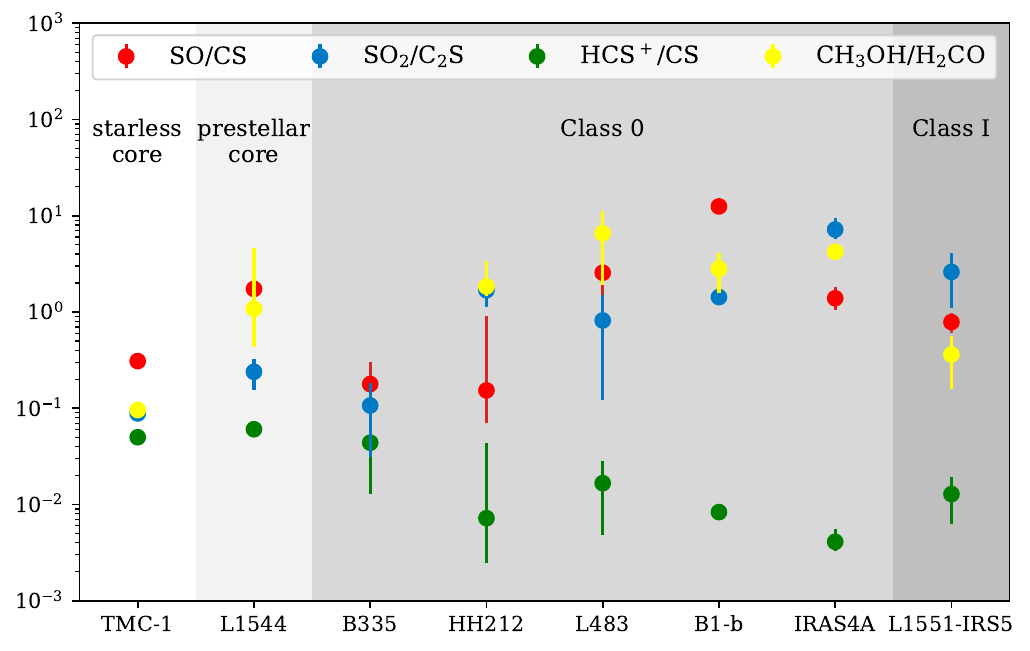}
    \caption{Ratios SO/CS, SO$_2$/C$_2$S, HCS$^+$/CS, and CH$_3$OH/H$_2$CO for a sample of forming stars at different stages: TMC-1 (\citealt{TMC1_3}), L1544 (\citealt{L1544_2}; \citealt{L1544}; \citealt{vastel2018sulphur}), B\,335 (\citealt{esplugues2023evolution}), HH\,212 (this work), L483 (\citealt{agundez2019sensitive}), B1-b (\citealt{B1b_2}; \citealt{B1b_3}; \citealt{B1b}), IRAS\,4A (this work), and L1551-IRS5 (\citealt{marchand2024chemical}).}
    \label{fig:comparacion}
\end{figure}

    {{In the following,} we investigated the origin of the chemical differences observed in the previous section.}
    Factors like density, temperature, molecular budget, and external radiation are modified as star formation develops, inducing changes in the observed chemical content. Figure~\ref{fig:comparacion} shows how the SO/CS, SO$_{2}$/C$_{2}$S, and HCS$^{+}$/CS {ratios} behave at different evolutionary stages, and Sect. \ref{sect:modeling} presents a model of the {the observational }results using a chemical emulator \citep{AsensioRamos2024} of the gas-grain chemical code Nautilus \citep{Ruaud2015}. In this context, machine-learning interpretability using Shapley values and SHapley Additive exPlanations \citep[SHAP;][]{Lundberg2017} may help us identify the emulator input parameters that most strongly influence the chemical composition we observe and how they relate to the evolutionary stage of the observed target.

\begin{figure}[t]

    \centering
    \includegraphics[width=0.9\hsize]{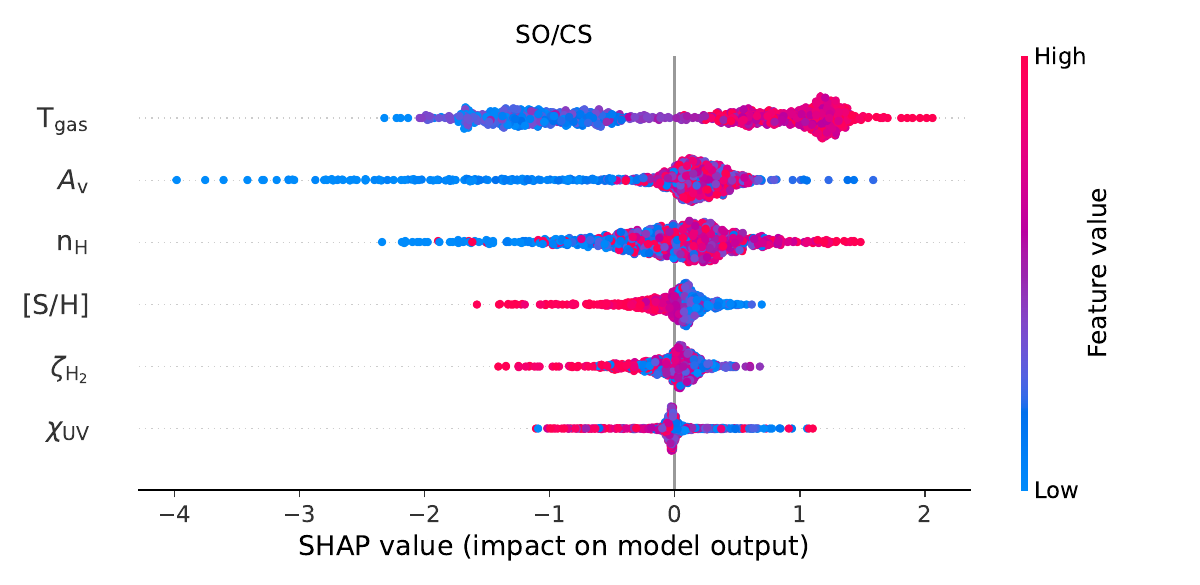}\\
    \includegraphics[width=0.9\hsize]{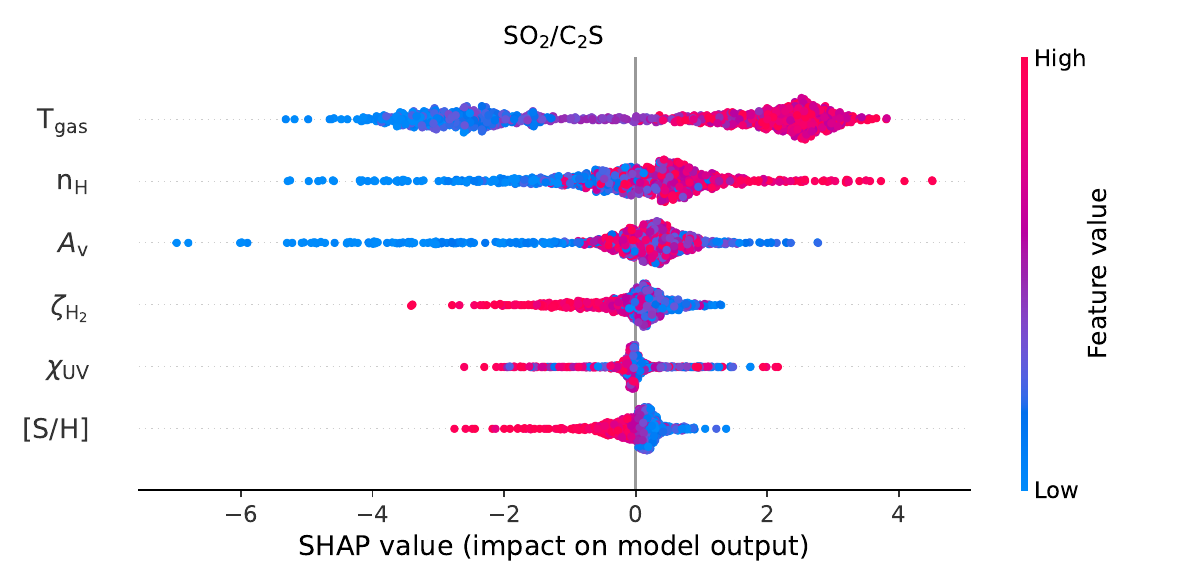}\\
    \includegraphics[width=0.9\hsize]{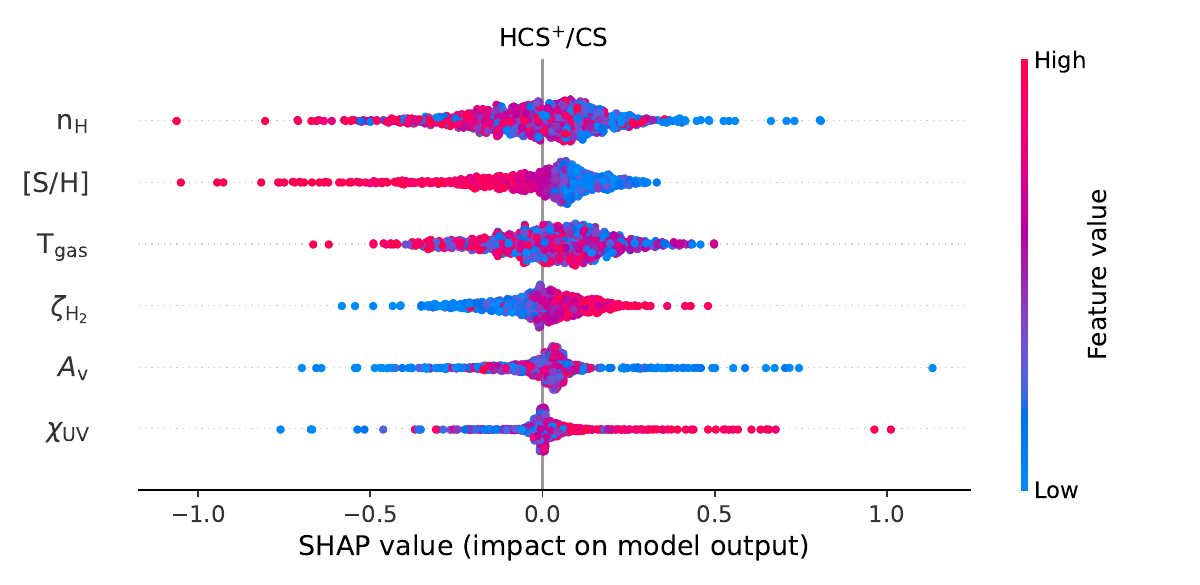}
    \caption{Beeswarm plot showing  SHAP values for the parameters of the chemical emulator obtained from 1000 predictions of the SO/CS, SO$_{2}$/C$_{2}$S, and HCS$^{+}$/CS ratios.}
    \label{fig:shap}

\end{figure}

    Shapley values \citep{Shapley1953} are a method from game theory that quantify the importance of a given parameter (feature) by computing the expected marginal contribution of that feature to a certain prediction. It is increasingly {being} used in astrochemistry as chemical emulators become available \citep[see, e.g.][]{Heyl2023, AsensioRamos2024}. We applied this technique to investigate the most important parameters in setting the observed SO/CS, SO$_{2}$/C$_{2}$S, and HCS$^{+}$/CS molecular ratios. We sampled the parameter space of the chemical emulator with 1000 points and computed the SHAP values of the emulator features for each prediction. Fig.~\ref{fig:shap} shows a beeswarm plot of SHAP values for the features of the 1000 predictions of the SO/CS ratio. In this case, gas temperature is the most important parameter, with higher temperatures leading to higher SO/CS ratios as a consequence of their positive SHAP values. We find the same trend with density. This is consistent with the observed trend in Fig. \ref{fig:comparacion}: as star formation proceeds, the protostar and its surroundings experience higher temperatures and densities, leading to the increasing SO/CS we observe. We find the same trends for the SO$_{2}$/C$_{2}$S ratio (Fig.~\ref{fig:shap}), which behaves similarly to SO/CS in the sample of observed objects. Finally, we show the SHAP values of the features in the prediction of the HCS$^{+}$/CS ratio in Fig. \ref{fig:shap}. In this case, the negative SHAP values of higher densities indicate that they reduce the HCS$^{+}$/CS ratio. This again agrees with the observed HCS$^{+}$/CS ratio, as it decreases as the protostellar system evolves. However, the initial sulphur budget is the second most important parameter setting the HCS$^{+}$/CS ratio, with lower sulphur budgets enhancing the HCS$^{+}$/CS ratio. One possible explanation is the low electron density expected with a decreasing sulphur budget{, as sulphur atoms are easily ionised in the gas and become important electron donors \citep{esplugues2023evolution,fuente2023,AsensioRamos2024}}, which would reduce the electronic recombination of HCS$^{+}$. The low HCS$^{+}$/CS ratio observed towards Barnard 1b, a dense core known for its enhanced sulphur chemistry \citep{B1b}, supports this trend. In addition, a high HCS$^{+}$/CS ratio is also associated with an increased cosmic-ray ionisation rate due to the impact on ionised species. This analysis therefore agrees with the use of these molecular ratios as molecular clocks through the star formation process, especially for the SO/CS and SO$_2$/C$_2$S ratios. In the case of HCS$^{+}$/CS, this ratio can also be affected by other environmental factors. 

\subsection{Comparing cold and warm sulphur chemistry}

    Protostellar envelopes provide an excellent opportunity to follow the evolution of sulphur chemistry during the star formation process with different components covering a wide range of temperatures (from $\sim$10 to $>$100 K) and densities ($n_\mathrm{H_2}$ from $\sim$10$^4$ to $>$10$^6$ cm$^{-3}$). Although these components are usually not resolved with single-dish telescopes, line profiles and multi-transition studies of abundant species provide kinetic information that disentangles the chemical composition of the cold envelope {from
that of the hot core and bipolar outflow.} Comparison with interferometric studies {allows} us to determine variations in {the} gas chemical composition with spatial scale.  {In this
section,} we compare the abundances derived in the cold protostellar envelope with those observed using high-angular resolution interferometric observations towards IRAS\,4A and HH\,212.

    The hot corino and bipolar outflows associated with IRAS\,4A have been widely studied using PdBI, {NOrthern Extended Millimeter Array} (NOEMA) and ALMA interferometers \citep{Santangelo2015, Taquet2015, lopez-sepulcre2017, Kushwahaa2023, Artur2023}. \citet{Kushwahaa2023} carried out a complete study of the H$_2$S and OCS emission in the warm region of IRAS\,4A using the ALMA compact array, and estimate $N$(H$_2$S$)=3.4\times10^{16}$~cm$^{-2}$ and  $N$(OCS$)=1.8\times10^{16}$~cm$^{-2}$ assuming a source size of 2$''$. These estimates agree within a factor of two with those derived in this work for the warm $N$(OC$^{34}$S) and $N$(H$_2$$^{34}$S), assuming the same source size. From NOEMA observations with an angular resolution of $2.4''\times1.8''$, \citet{Taquet2020} estimate $N$(SO$_2) = 1.4\times10^{15}$~cm$^{-2}$ in the hot corino. To derive the abundances of these sulphur compounds in the warm innermost 2$''$ region, we assumed a source average value of $N$(H$_2)= 2.3\times10^{24}$~cm$^{-2}$ \citep{Taquet2015}. The abundances of H$_2$S, OCS, and SO$_2$ in the warm region are $1.5\times10^{-8}$, $7.8\times10^{-9}$, and $6.1\times10^{-10}$, respectively. If most of the sulphur is locked in these species, these estimates are consistent with the value of $\text{[S/H]} = 8.8\times10^{-8}$ in the cold envelope derived with our chemical model (Table~\ref{modelling_results}). This would mean that sulphur is locked in some type of \text{(semi-)refractory} material (e.g. sulphur chains) or remains in atomic form along the envelope of this protobinary. High-sensitivity and high-angular-resolution spectral surveys are needed to  obtain a more accurate view of the sulphur budget at disc ($<$100~au) scales.

    The Class 0 object HH\,212 is one of the best studied in the Orion molecular cloud and is therefore a good candidate to investigate the evolution of sulphur chemistry during star formation in a giant molecular cloud where massive stars form.  \citet{Podio2015} reported high-angular-resolution observations of SO and SO$_2$ using ALMA. Most of the SO and SO$_2$ emission came from the energetic bipolar outflow, with abundances between 10$^{-7}$ and 10$^{-6}$ in the jet. These authors also provided an estimate of the abundance of SO in the compact disc ($R$$\sim$0.2$''$; \citealp{Codella2014}) of $X$(SO$)\sim 10^{-8}$ to 10$^{-7}$. \cite{lee2017c} report CH$_3$SH emission in this disc with an average column density of $N$(CH$_3$SH$)=1.0\times10^{17}$ cm$^{-2}$ in a beam of $\sim$0.04$''$ targeting the southern part of the disc atmosphere. The abundances derived in HH\,212 are consistent with a higher value of $\text{[S/H]}\sim 1.8 \times10^{-7}$ in the envelope of this protostar. This is not unexpected, since several studies point to a low sulphur depletion in the Orion molecular cloud \citep{fuente2023}. Interestingly, we estimate a high cosmic-ray molecular ionisation rate in this source, which may be related to the energetic bipolar outflow driven by it.

    Our data are therefore consistent with significant sulphur depletion {($\sim$100)} in the cold and warm regions of these protostars. This differs from the low sulphur depletion values observed in the photodissociation regions (PDRs) formed in the vicinity of massive stars $(\sim${2-10;} \citealt{Goicoechea2006, fuente2024, fuente2025}). {Moderate values of sulphur depletion are also reported in the hot cores associated with massive protostars where dust temperatures reach $\sim$400~K \citep{Crockett2014}. In contrast, similar sulphur depletion as in our study is found in the extinction peaks of starless cores \citep{fuente2023}. } {Recently, \cite{Miranzo-Pastor2025} presented observations of H$_2$S and OCS in a wide sample of hot corinos in Perseus using the NOEMA interferometer.  The abundances of H$_2$S and OCS range from a few $10^{-9}$ to a few $10^{-7}$ in Class~0 protostars. Higher abundances ($>$$10^{-6}$) are only detected in young Class~I objects. Yet, the hot corinos were not fully resolved at the angular resolution of $\sim$400~au, which implies that sulphur depletion remains high during the Class~0 stage of low-mass star formation at these scales.} 
    One possibility is that sulphur is locked into sulphur compounds that can survive the physical conditions prevailing in {Class~0} hot corinos, which are not as extreme as those of PDRs and hot cores. However, the chemical composition of the ice is very sensitive to the physical conditions in the parent molecular cloud, in particular to the dust temperature, promoting a different composition of the sulphur budget. Because the binding energies differ among molecules, this diversity in the ice composition would also imply a diversity in the gas composition during the hot-core phase.

\section{Summary and conclusions}

    We used unbiased spectral surveys carried out with the Yebes-40m and IRAM-30m telescopes as part of the ECHOS programme to characterise sulphur chemistry in the cold envelopes of the protostars NGC\,1333\,IRAS\,4A and HH\,212. Our goal was to identify chemical diagnostics that determine the physical conditions in these protostellar envelopes and help establish the evolution of matter during the formation of a low-mass star. Since sulphur is a reactive atom, the abundances of sulphur compounds are highly sensitive to both the local physical conditions and the dynamical history of the core and are therefore potential evolutionary probes in star-forming regions.

    \begin{itemize}
    
        \item We estimated column densities of CS, C$^{34}$S, C$^{33}$S, $^{13}$CS, C$_2$S, C$_2$$^{34}$S, C$_3$S, C$_3$$^{34}$S, OCS, OC$^{34}$S, OC$^{33}$S, HCS$^+$, HC$^{34}$S$^+$, SO, $^{34}$SO, S$^{18}$O, H$_2$CS, SO$_2$, $^{34}$SO$_2$, NS, NS$^+$, H$_2$S, H$_2$$^{34}$S, HSCN, HNCS, CH$_3$SH, C$^{18}$O, C$^{17}$O, CH$_3$OH, and H$_2$CO using LTE and non-LTE calculations. Comparison of the abundances of these species with those towards B \,335 shows a differentiated chemistry. While sulphur-bearing species containing carbon chains (C$_2$S, C$_3$S) are more abundant in B\,335 due to its early chemistry, sulphur oxides and nitrogen-bearing species are more abundant in IRAS\,4A. 

        \item We used our data to explore the isotopic ratios in the targeted protostellar envelopes. The C$^{34}$S/C$^{33}$S ratio is below the ISM isotopic ratio for IRAS\,4A, which is contrary to that found in comets \citep{Calmonte2017}. In this same target, the S$^{16}$O/S$^{18}$O ratio is $\sim$100, about five times lower than the ISM $^{16}$O/$^{18}$O isotopic ratio. This value is similar to those found towards cold cores by \cite{O18fractionation}. Chemically, isotopic fractionation reactions yield S$^{16}$O/S$^{18}$O$\sim$100 at typical ages of $\sim$0.1 Myr, {as found in} \cite{O18fractionation}.

        \item {For both Class 0 objects HH\,212 and IRAS\,4A, the H$_2$CO/H$_2$CS ratio in their envelopes is ten and 27 times lower than in comet 67P/C–G, respectively. However, while the IRAS\,4A CH$_3$OH/CH$_3$SH estimate is seven times higher than the comet value, this ratio in HH\,212 is around three times lower. Although CH$_3$SH in HH\,212 is a tentative detection, if confirmed, it would imply a particularly low CH$_3$OH/CH$_3$SH ratio in this target, differing from other low-mass star-forming regions and showing closer similarity to the early Solar System.}

        \item Sulphur-bearing species have been widely used as tracers of protostellar evolution. Our data strengthens this interpretation, identifying the SO/CS, SO$_2$/CS$_2$, and HCS$^+$/CS ratios as useful chemical diagnostics. The SO/CS and SO$_2$/CS$_2$ ratios increase, while the HCS$^+$/CS ratio decreases from the prestellar phase to the {cold envelope of the} Class 0 phase. Shallower changes in these ratios are observed within the group of Class 0 protostars. Computing the SHAP values of a neural emulator of the chemical code Nautilus, we find that the SO/CS and SO$_2$/CS$_2$ ratios increase {in these envelopes} due to the rising gas temperature and density, while the HCS$^+$/CS ratio {appears to decrease} with density. However, these changes are difficult to explain only in evolutionary terms, suggesting the influence of additional parameters such as environmental effects or differences in the collapse history on the sulphur chemistry. 

        \item  We performed multi-species fitting using the neural emulator of Nautilus to investigate the physical and chemical conditions driving the evolution of the chemical composition of the cold envelope in IRAS\,4A and HH\,212. Both targets are well explained by gas temperatures of $\sim$25 K, but show different densities ($2.6\times10^4\mathrm{\ cm^{-3}}$ in HH\,212 and $1.7\times10^5\mathrm{\ cm^{-3}}$ in IRAS\,4A) and cosmic-ray ionisation {fluxes} ($2.5\times10^{-16}\mathrm{\ s^{-1}}$ for HH\,212 and $1.2\times10^{-17}\mathrm{\ s^{-1}}$ for IRAS\,4A).

        \item 
        We estimated the sulphur elemental abundance in the {cold} envelope of IRAS\,4A and HH\,212. Our calculations indicate that {sulphur depletion remains high in these envelopes, at a factor of $\sim$100. These values are consistent with those found in starless cores \citep{fuente2023}, confirming the similarity of their chemistry and suggesting that shocks associated with {the} bipolar outflows do not significantly affect the sulphur content of the protostellar envelopes.}

    \end{itemize}

    We conclude that the Class 0 object IRAS\,4A, located in the Perseus cloud, has a higher density and lower cosmic-ray ionisation rate in its envelope than the Class 0 protostar HH\,212, {found} in the Orion region. However, both envelopes are similar in gas temperature and sulphur depletion. These differences could explain the observed sulphur chemistry, although it remains unclear whether they are purely evolutionary in origin or also influenced by environmental factors.

\section*{Data availability}

{The supporting material file with supplementary tables and figures, along with the data in ascii format, is available in Zenodo via \href{https://doi.org/10.5281/zenodo.20138324}{https://doi.org/10.5281/zenodo.20138324}.}

\begin{acknowledgements}
This work is supported by ERC grant SUL4LIFE, GA No. 101096293. Funded by the European Union. Views and opinions expressed are however those of the author(s) only and do not necessarily reflect those of the European Union or the European Research Council Executive Agency. Neither the European Union nor the granting authority can be held responsible for them. 
The project has been carried out with observations
from the 30m radio telescope of the Institut de radioastronomie millimétrique
(IRAM) and from the 40m radio telescope of the National Geographic Institute of Spain (IGN) at Yebes Observatory. {G.E., A.F., M.R.B., and P.R.M.
acknowledge support from the Spanish grant PID2022-137980NB-I00, funded
by MCIN/AEI/10.13039/501100011033/FEDER UE.
A.A.R. and C.W.P. acknowledge funding from the Agencia Estatal de Investigación del Ministerio de Ciencia, Innovación y Universidades (MCIU/AEI) under grant
``Polarimetric Inference of Magnetic Fields'' and the European Regional Development Fund (ERDF) with reference PID2022-136563NB-I00/10.13039/501100011033.  The
authors thank the anonymous referee for their valuable suggestions.}
\end{acknowledgements}

\bibliographystyle{aa} 
\bibliography{0_referencias.bib}

\begin{appendix}

\onecolumn

\section{Additional tables and figures}
\nopagebreak

\begin{table*}[h!]
\caption{Yebes-40m and IRAM-30m telescope efficiency data along the covered frequency range.}               
\centering          
\label{table:tablaeficiencias}
\begin{tabular}{c c c c}     
\hline\hline                    
\noalign{\smallskip}
Telescope  & Frequency  &  $\eta$$_{\mathrm{MB}}$ & HPBW \\
           & (GHz)      &                         & ($\arcsec$)\\
\noalign{\smallskip}
\hline              
\noalign{\smallskip}
Yebes 40 m &  32.4 & 0.61 & 54.4  \\
           & 34.6 & 0.58 & 51.0  \\
           & 36.9 & 0.57 & 47.8  \\
           & 39.2 & 0.55 & 45.0  \\
           & 41.5 & 0.53 & 42.5  \\
           & 43.8 & 0.52 & 40.3  \\
           & 46.1 & 0.49 & 38.3   \\
           & 48.4 & 0.47 & 36.4   \\
\noalign{\smallskip}
\hline 
\noalign{\smallskip}
IRAM 30 m  & 86  & 0.81 & 28.6  \\
           & 115 & 0.78 & 21.4  \\
           & 145 & 0.73 & 17.0  \\
           & 210 & 0.63 & 11.7  \\
           & 230 & 0.59 & 10.7  \\
           & 280 & 0.49 &  8.8  \\
\noalign{\smallskip}
\hline
\end{tabular}
\end{table*}

\begin{figure*}[h!]
    \sidecaption
    \includegraphics[width=0.3\linewidth,valign=c]{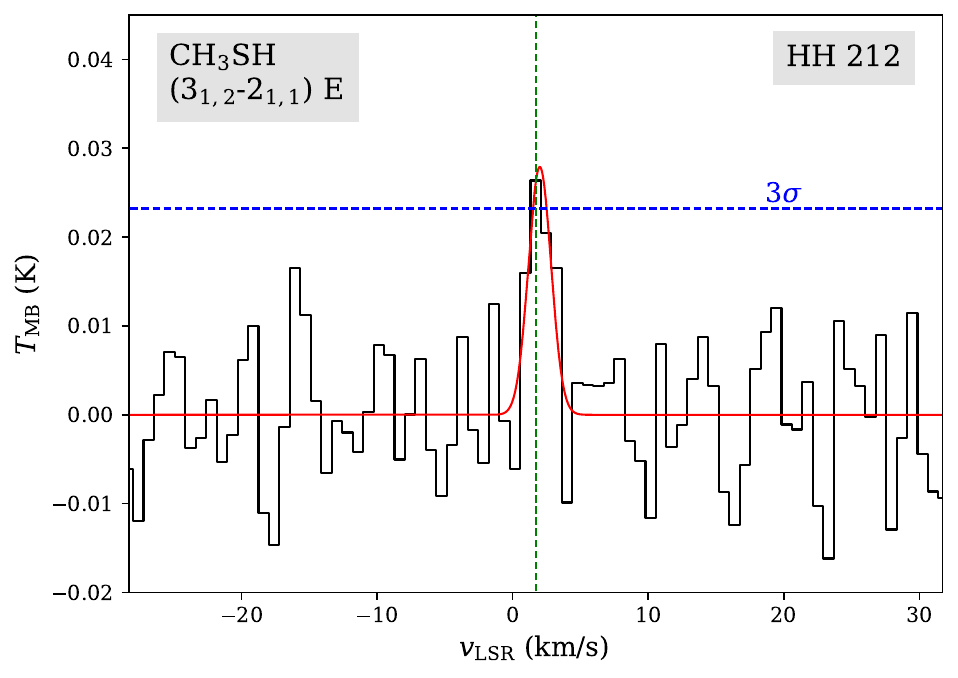}
    \caption{Tentative detection of CH$_3$SH in HH\,212. The solid red line shows the Gaussian fit, the dashed blue line represents the 3$\sigma$ noise level and the green one indicates the systemic velocity of the source, which is 1.7 km s$^{-1}$.}
    \label{CH3SH_tentativo}
\end{figure*}

\begin{figure*}[h!]
    \centering
    \includegraphics[width=\hsize]{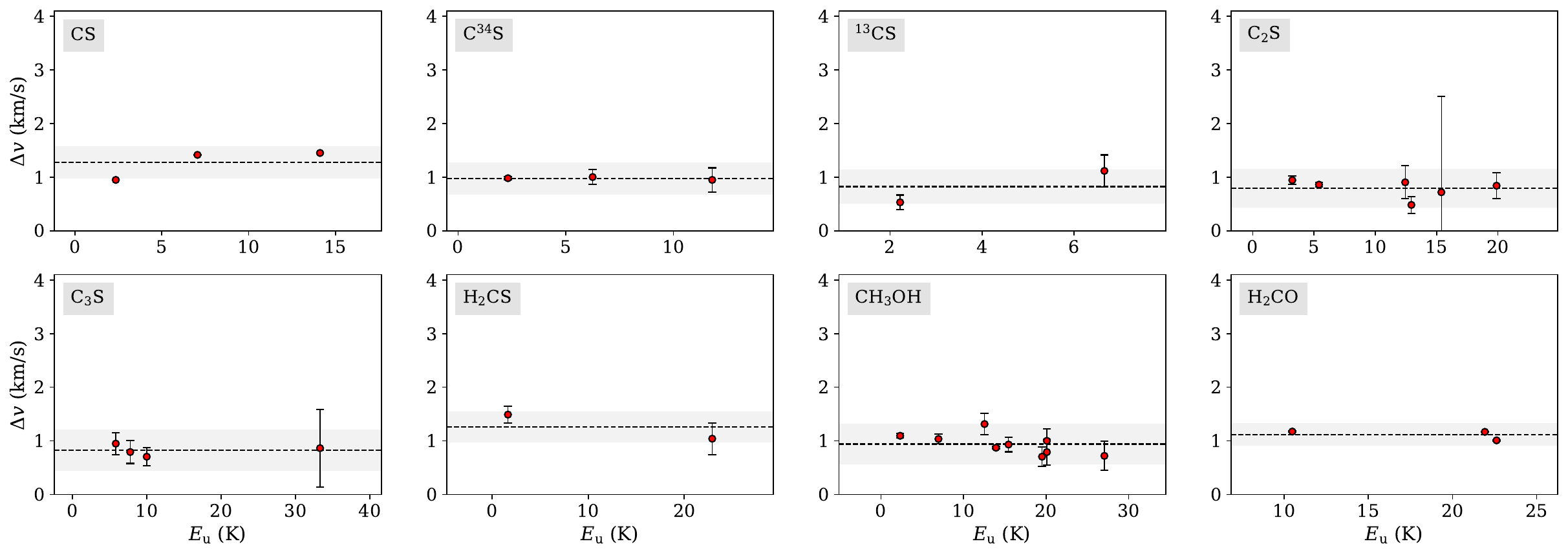}
    \caption{Linewidth ($\Delta v$) as a function of upper energy level $(E_\text{u})$ for species in HH\,212 that are not contaminated by the outflow. Dashed horizontal lines represent the mean linewidth in each graph, while the shaded area indicates the lowest channel resolution in the detected transitions.}
    \label{EuHH212}
\end{figure*}

\begin{figure*}[h!]
    \centering
    \includegraphics[width=\hsize]{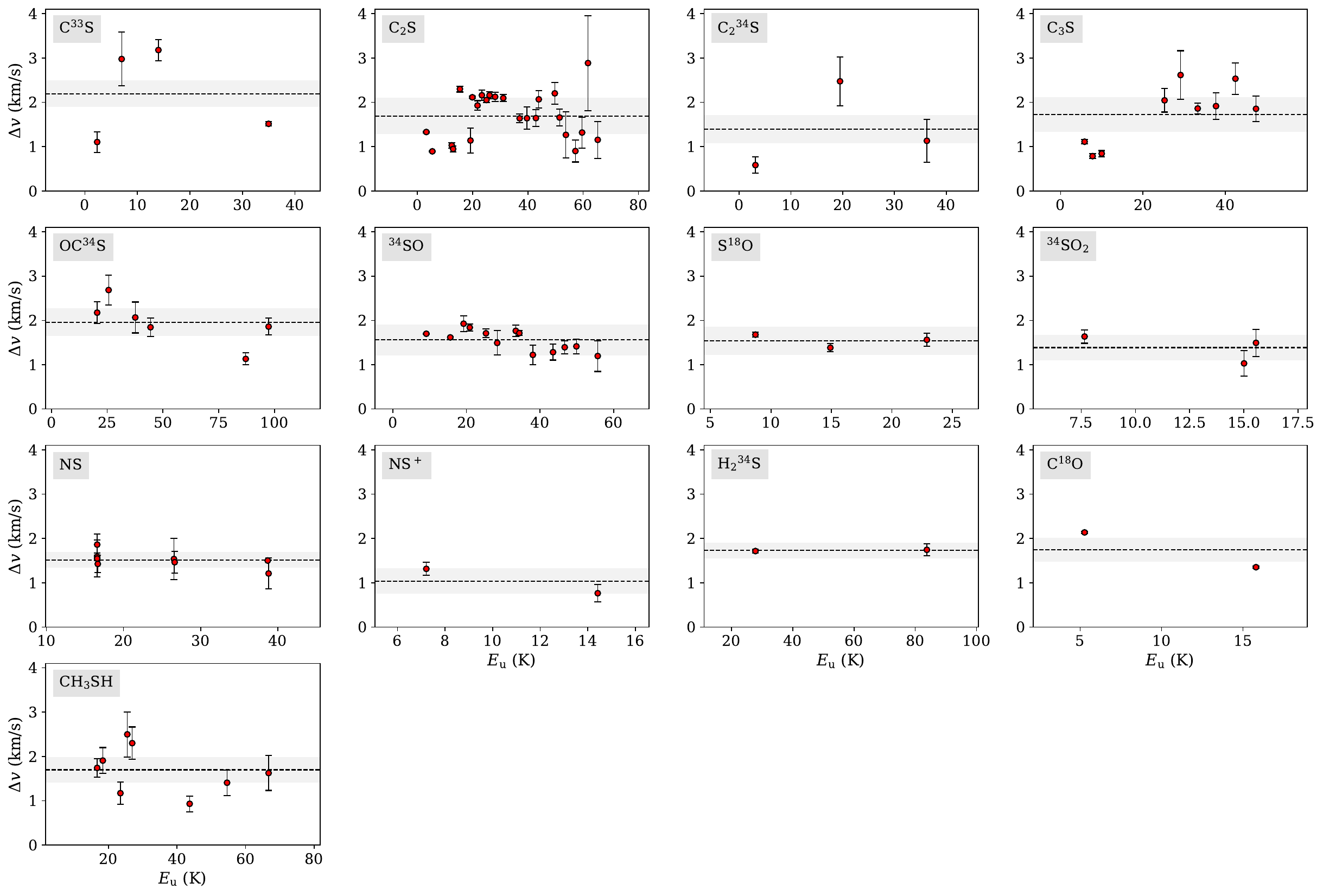}
    \caption{Linewidth ($\Delta v$) as a function of upper energy level $(E_\text{u})$ for species in IRAS\,4A that are not contaminated by the outflow. Dashed horizontal lines represent the mean linewidth in each graph, while the shaded area indicates the lowest channel resolution in the detected transitions.}
    \label{EuIRAS4A}
\end{figure*}

\begin{figure*}[h!]
    \centering
    \includegraphics[width=0.33\hsize]{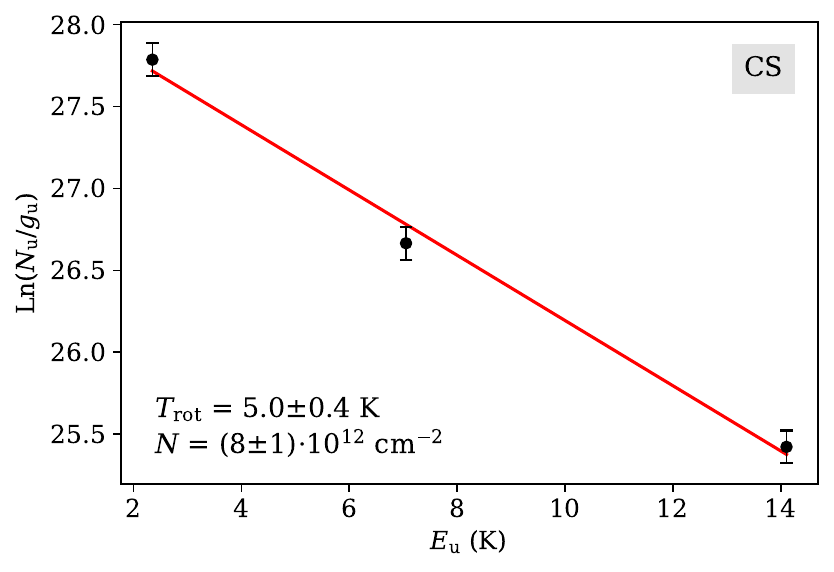}
    \includegraphics[width=0.33\hsize]{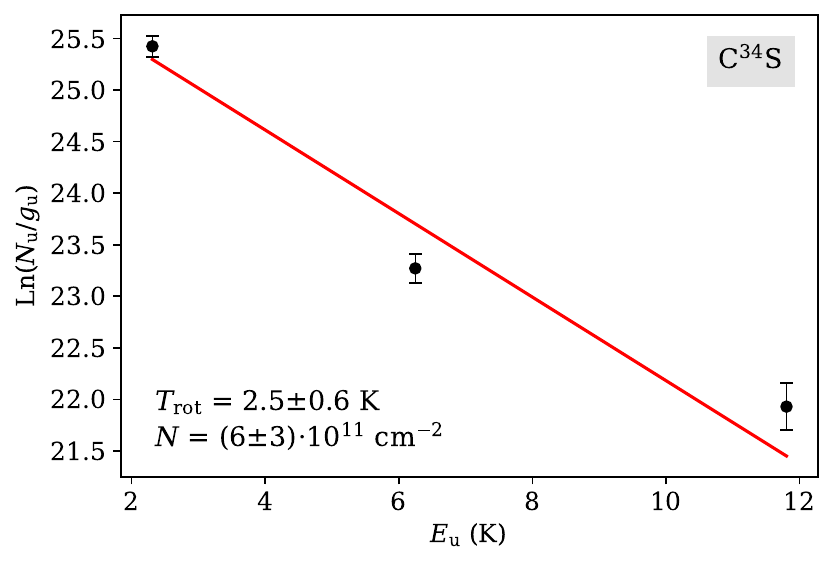}
    \includegraphics[width=0.33\hsize]{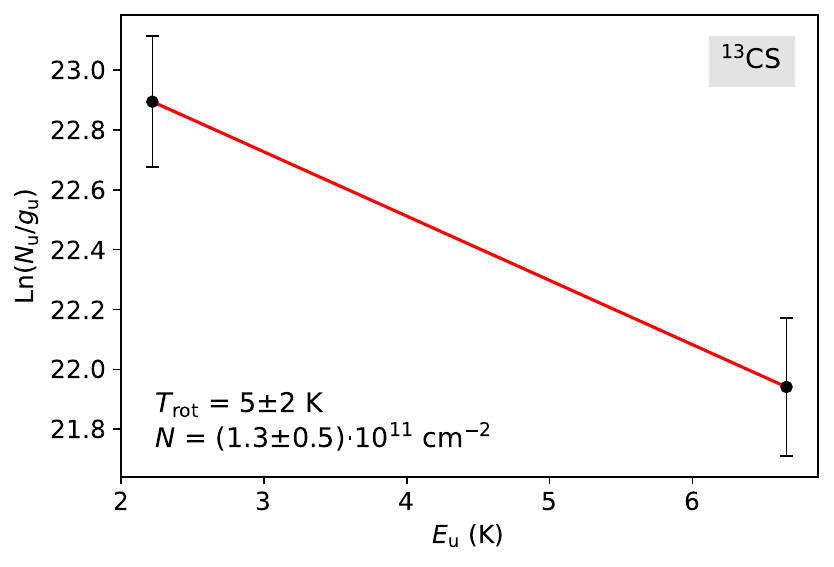}
    \\
    \includegraphics[width=0.33\hsize]{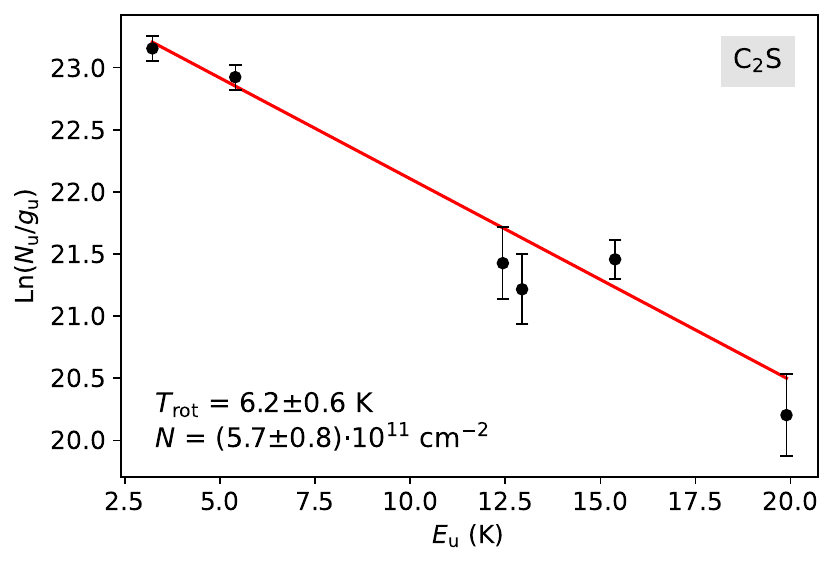}  
        \includegraphics[width=0.33\hsize]{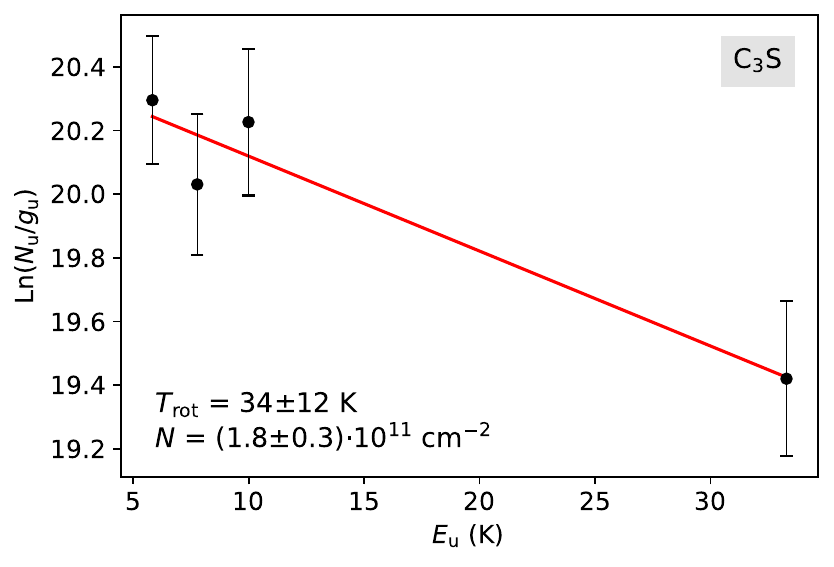}
        \includegraphics[width=0.33\hsize]{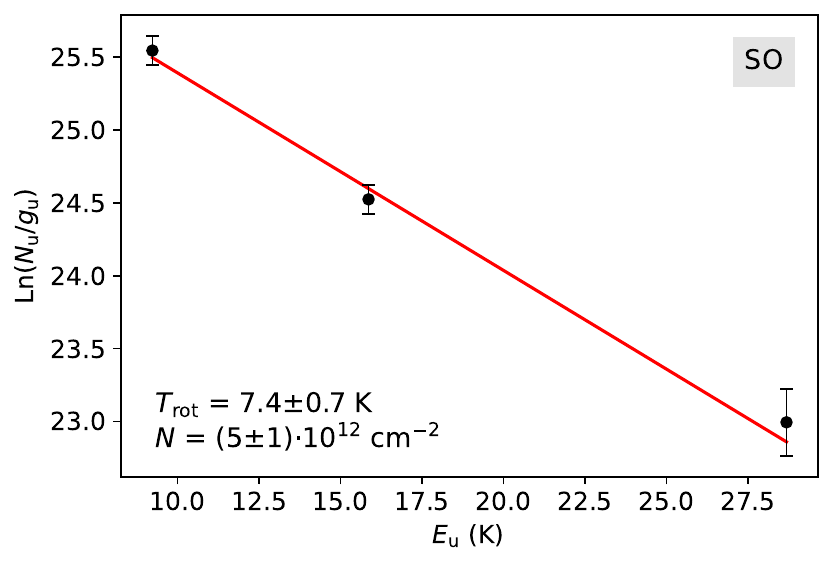}\\
        
\caption{Rotational diagrams for the narrow component of species in HH\,212.}
\label{diagramasHH212}
\end{figure*}

\begin{figure*}[h!]
\addtocounter{figure}{-1}
    \centering
    \includegraphics[width=0.33\hsize]{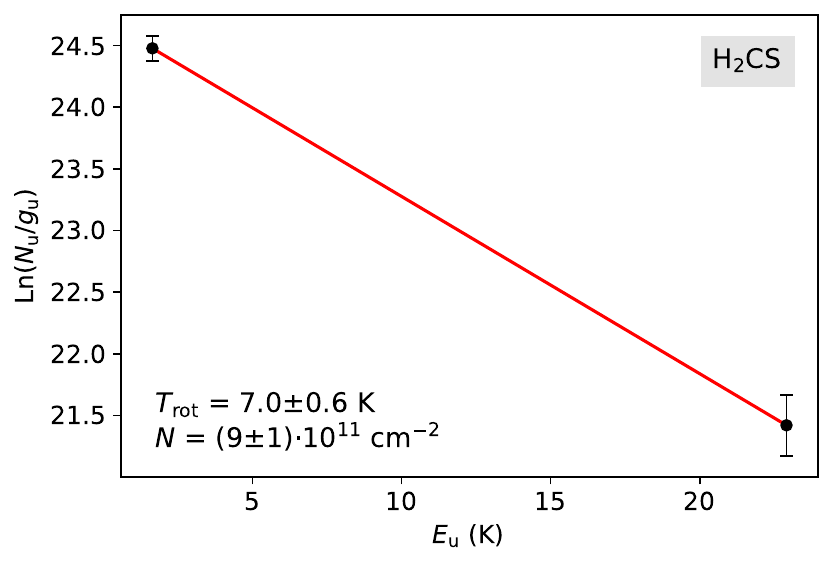}
        \includegraphics[width=0.323\hsize]{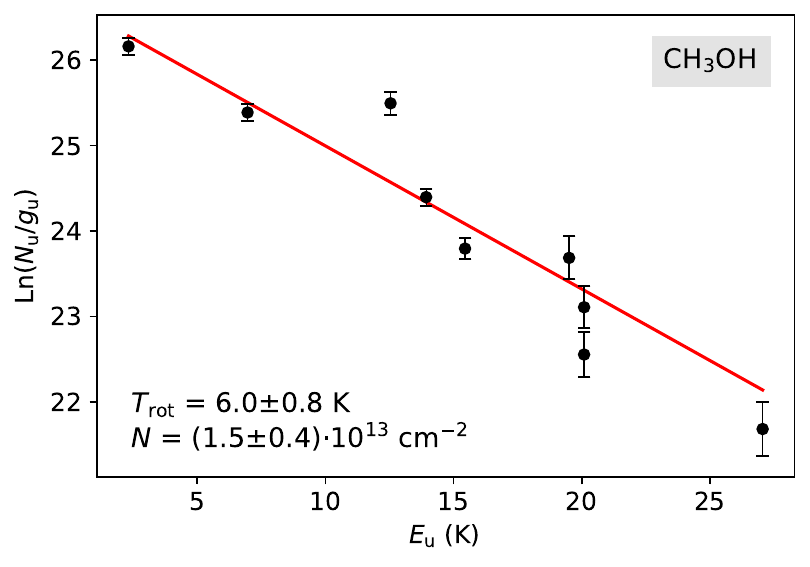}
        \includegraphics[width=0.33\hsize]{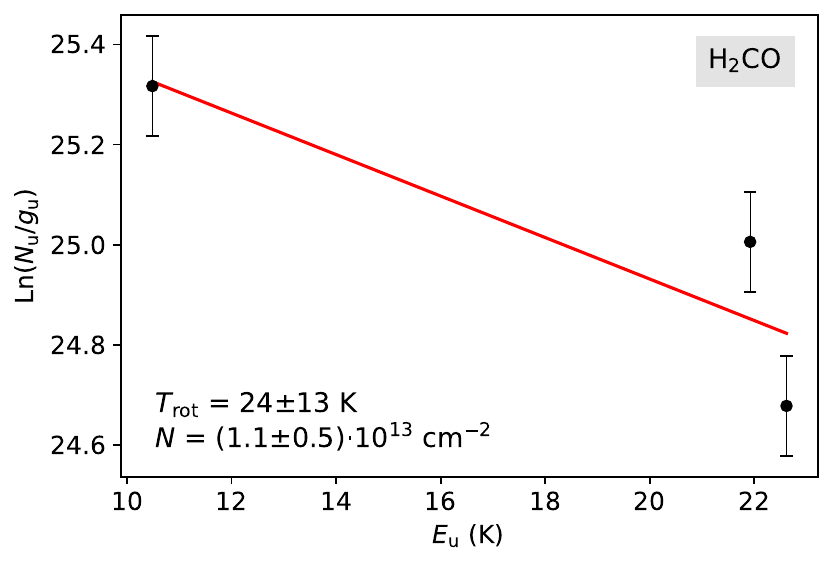}\\
\caption{continued.}
\label{diagramasHH212_2}
\end{figure*}

\begin{figure*}[h]
    \centering
    
    \includegraphics[width=0.33\hsize]{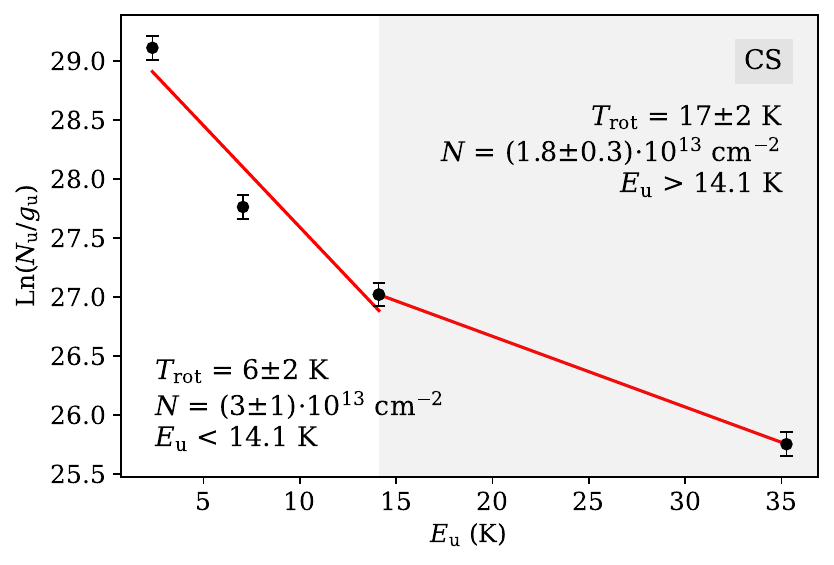}
    \includegraphics[width=0.33\hsize]{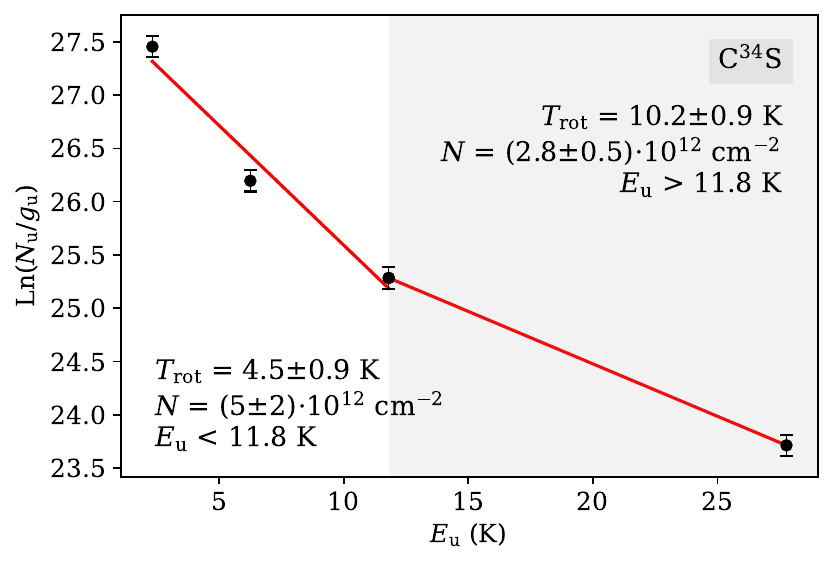}
    \includegraphics[width=0.33\hsize]{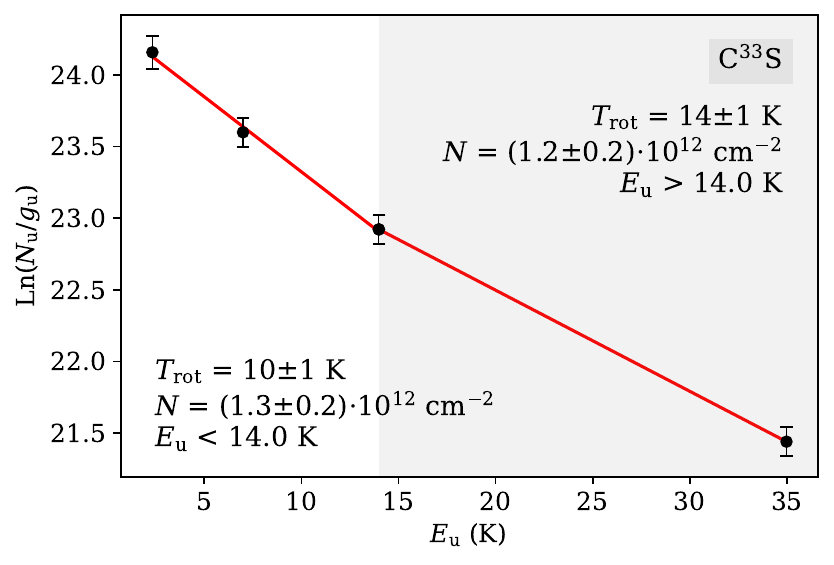}
    \\
    \includegraphics[width=0.33\hsize]{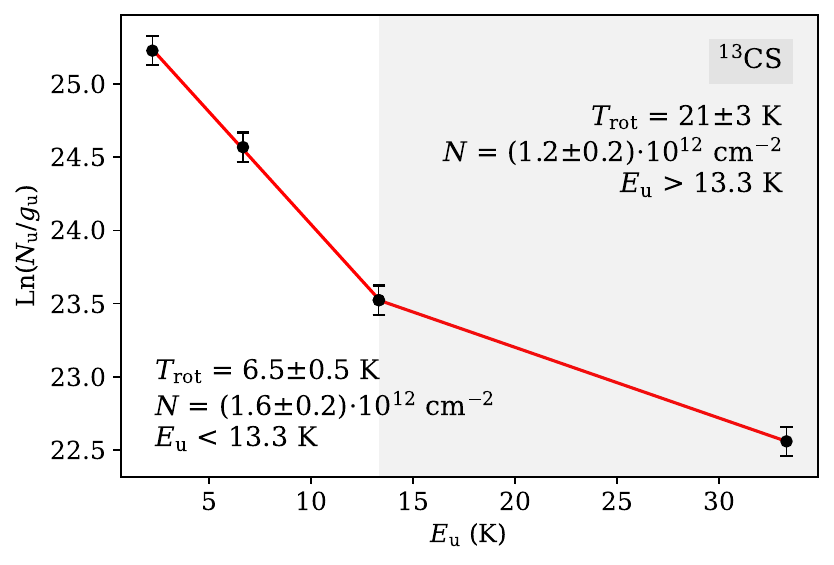}
    \includegraphics[width=0.323\hsize]{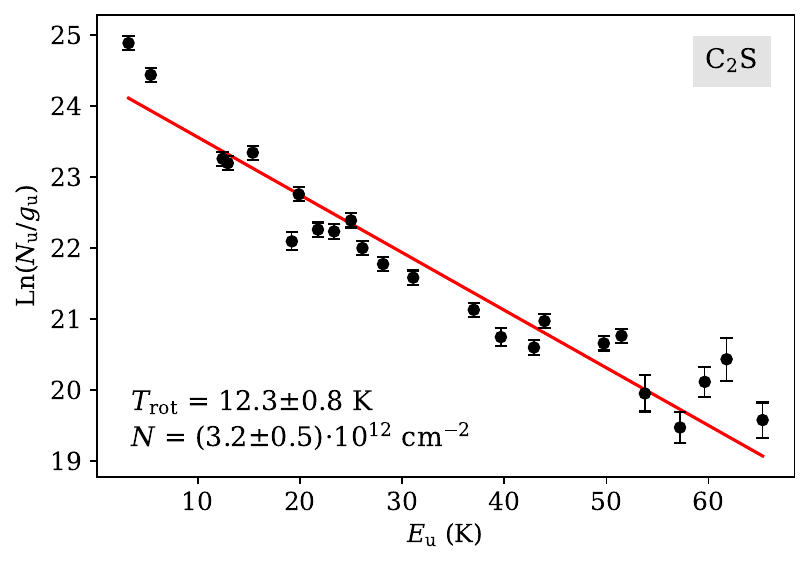}
    \includegraphics[width=0.33\hsize]{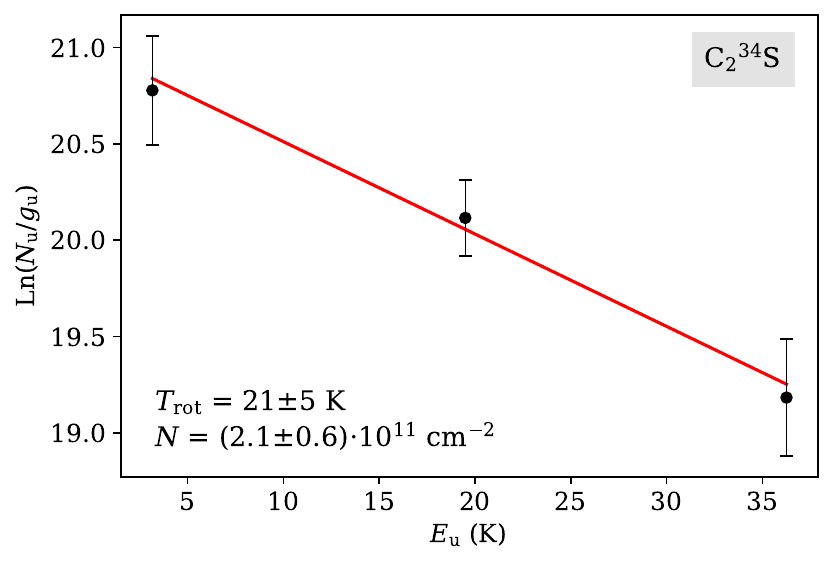}\\
    \includegraphics[width=0.33\hsize]{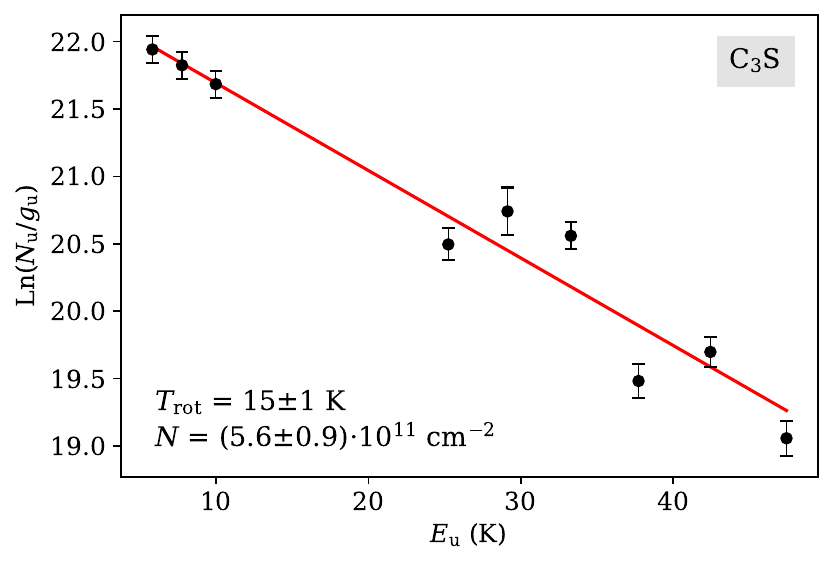}
    \includegraphics[width=0.33\hsize]{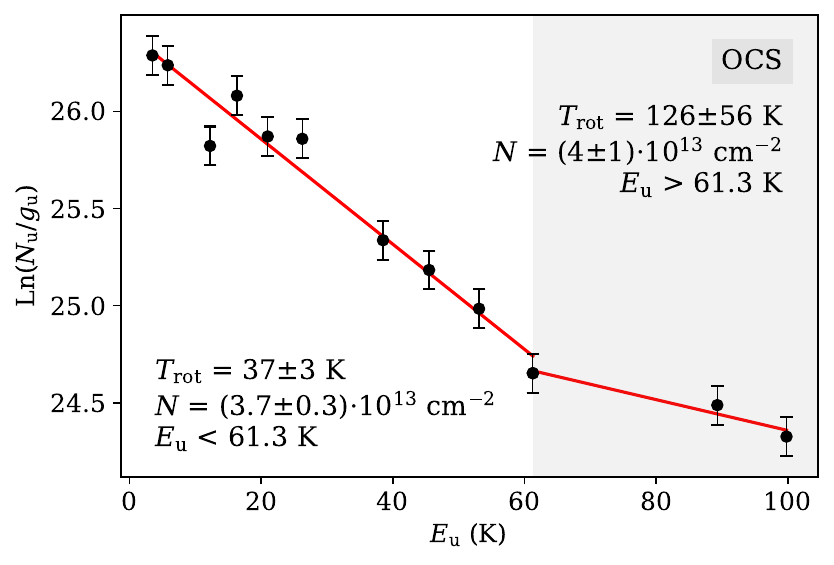}
    \includegraphics[width=0.33\hsize]{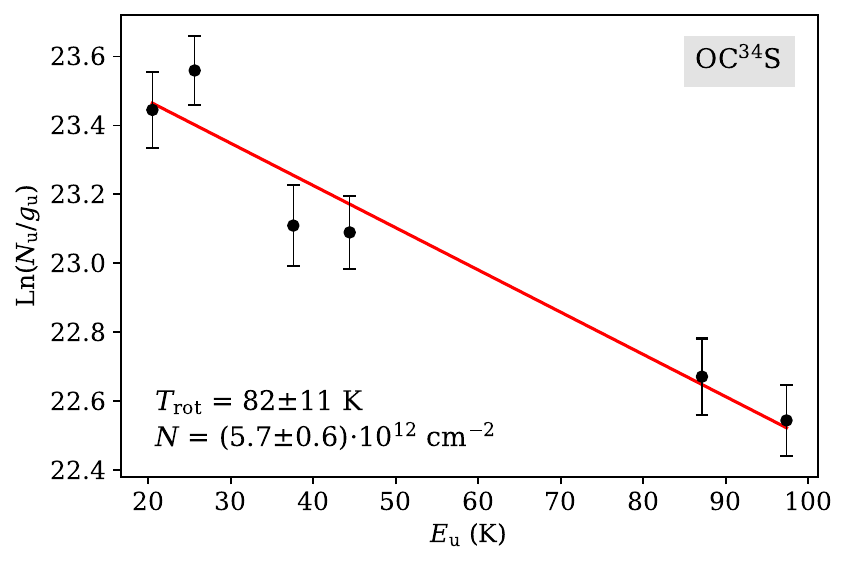}
    \\
    \includegraphics[width=0.33\hsize]{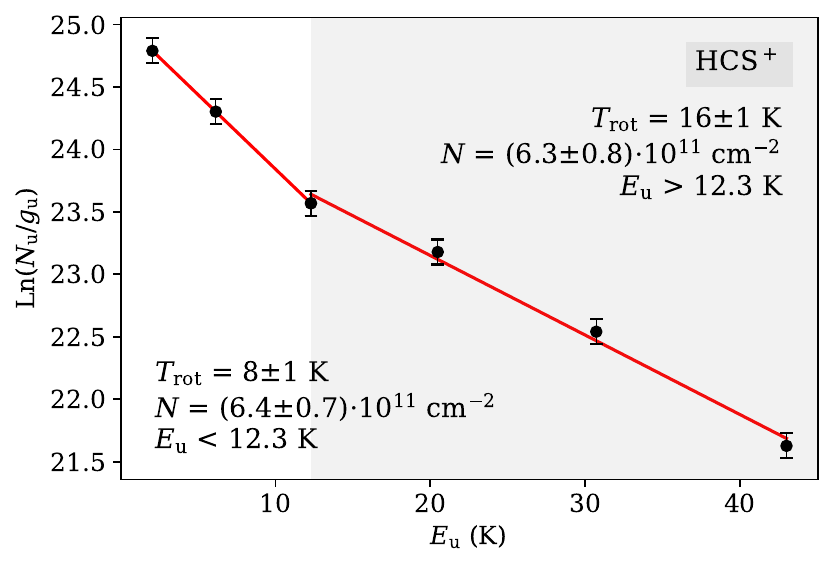}
    \includegraphics[width=0.33\hsize]{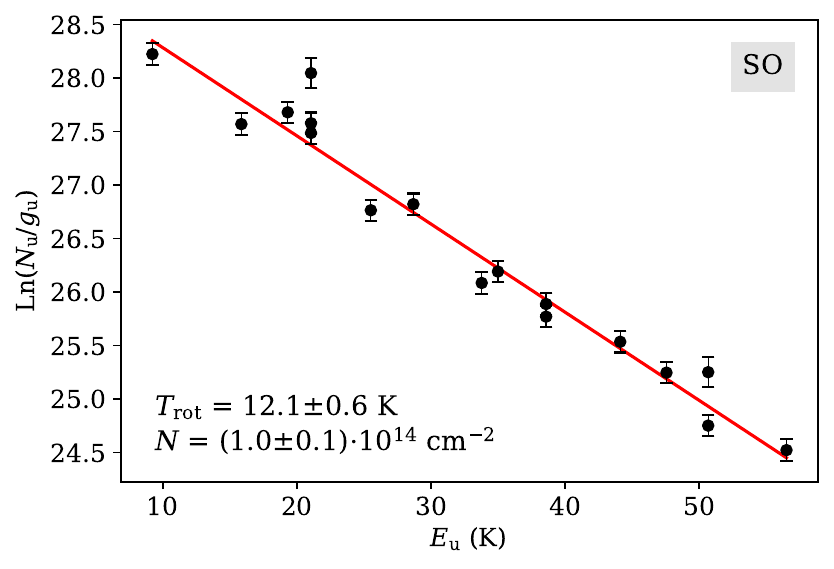}
    \includegraphics[width=0.323\hsize]{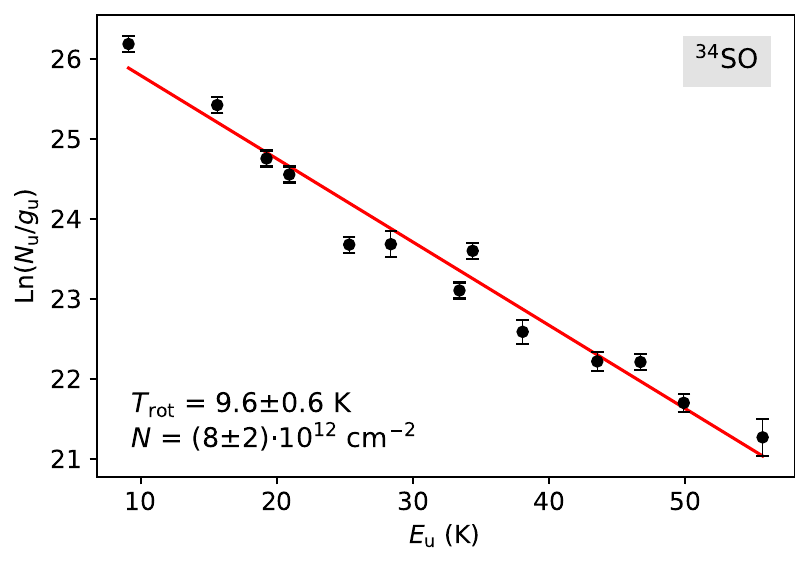}
    \\

\caption{Rotational diagrams for the narrow component of species in IRAS\,4A.}
\label{diagramasIRAS4A1}
\end{figure*}

\begin{figure*}[h]
\addtocounter{figure}{-1}
\centering
    \includegraphics[width=0.34\hsize]{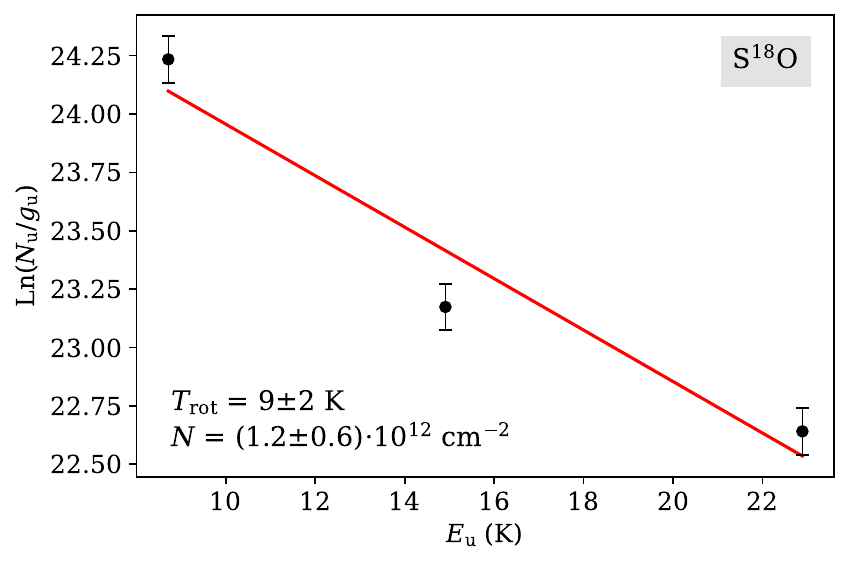}
    \includegraphics[width=0.323\hsize]{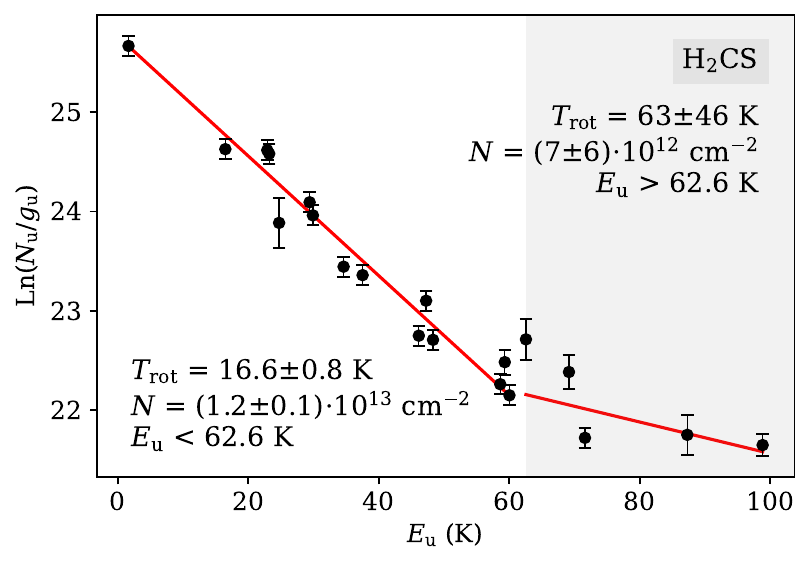}
    \includegraphics[width=0.323\hsize]{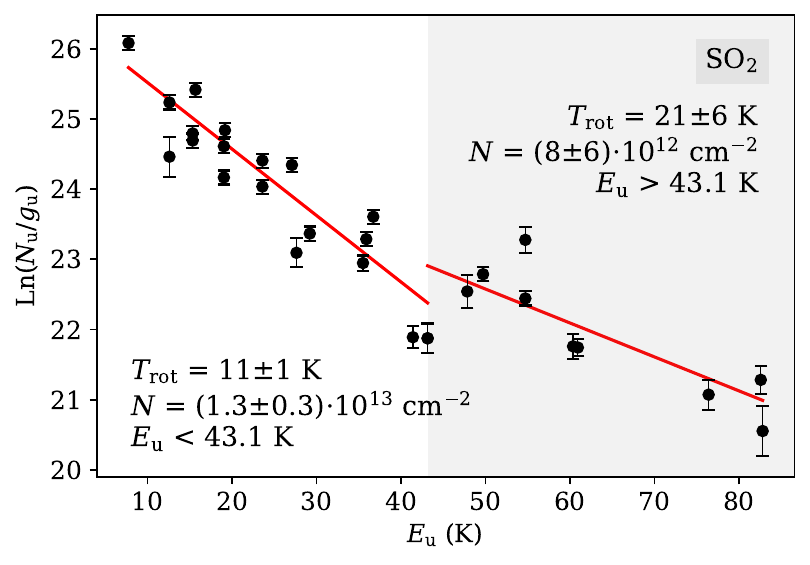}
    \\
    
    \includegraphics[width=0.33\hsize]{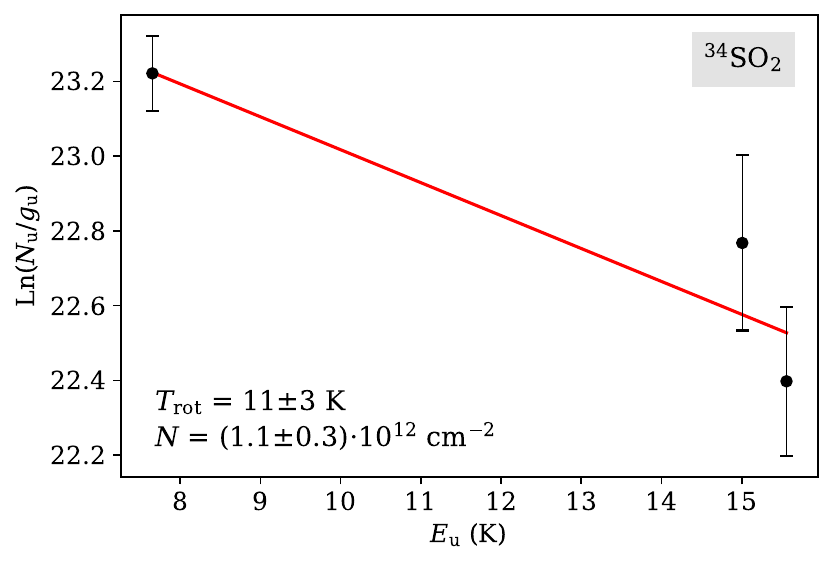}
    \includegraphics[width=0.33\hsize]{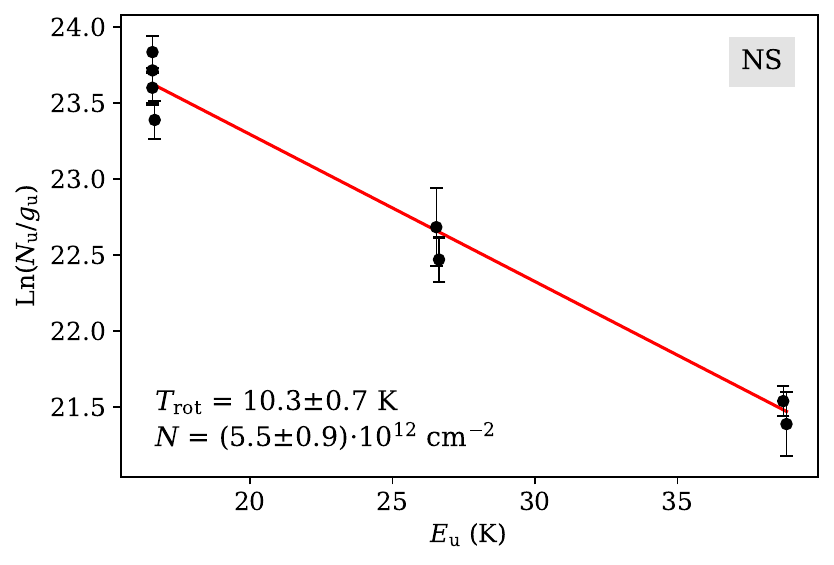}
    \includegraphics[width=0.33\hsize]{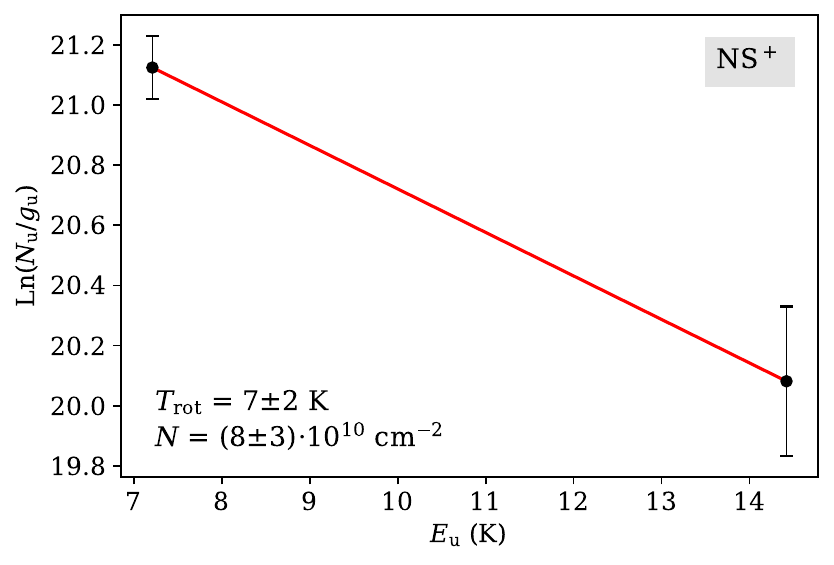}\\
    \includegraphics[width=0.326\hsize]{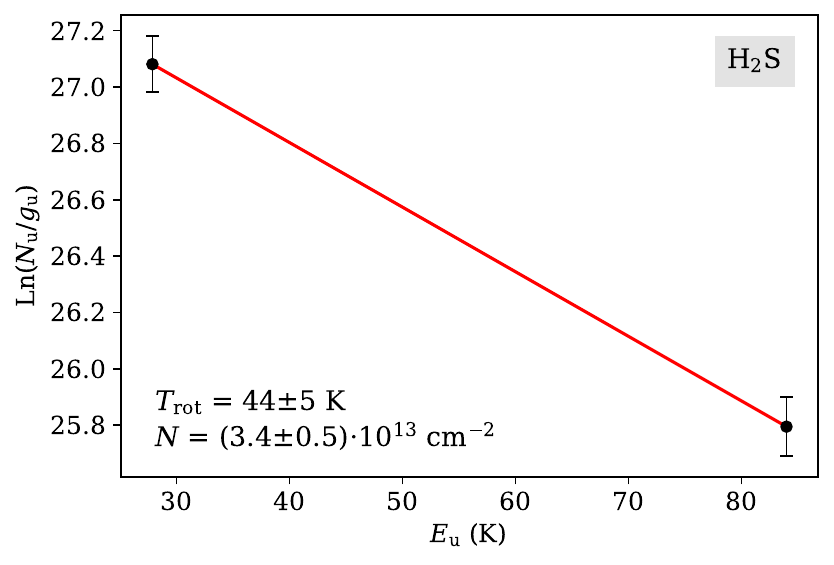}
    \includegraphics[width=0.326\hsize]{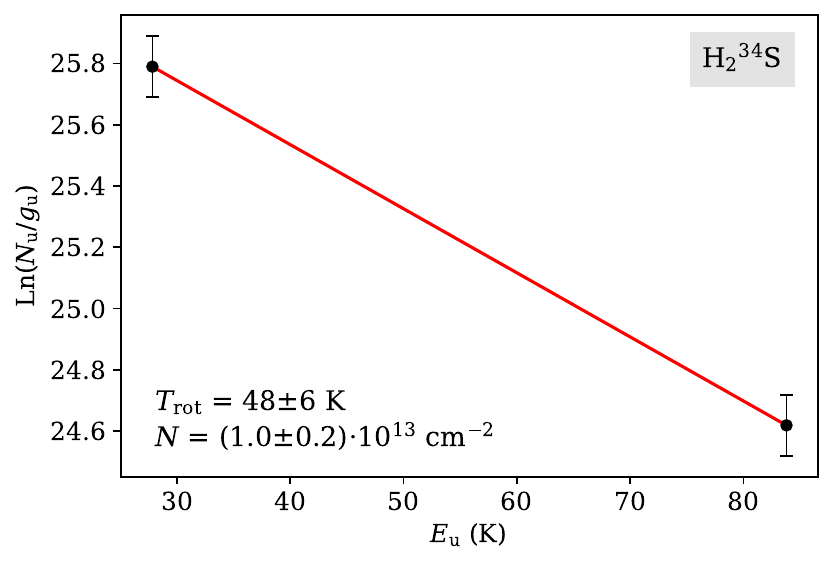}
    \includegraphics[width=0.332\hsize]{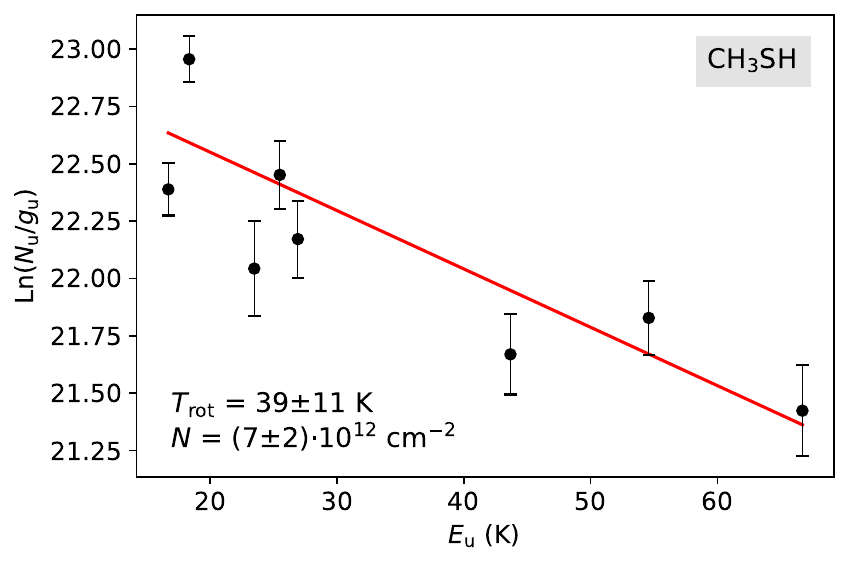}\\
    \includegraphics[width=0.33\hsize]{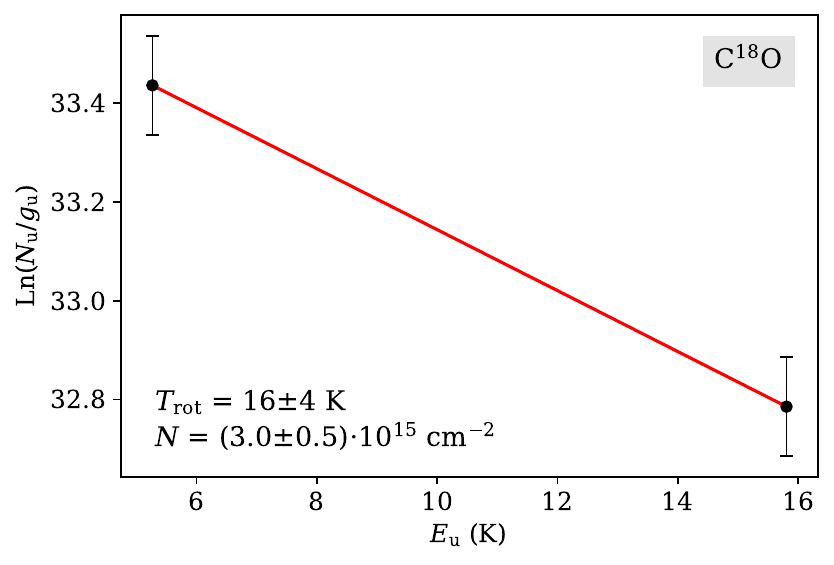}
    \includegraphics[width=0.323\hsize]{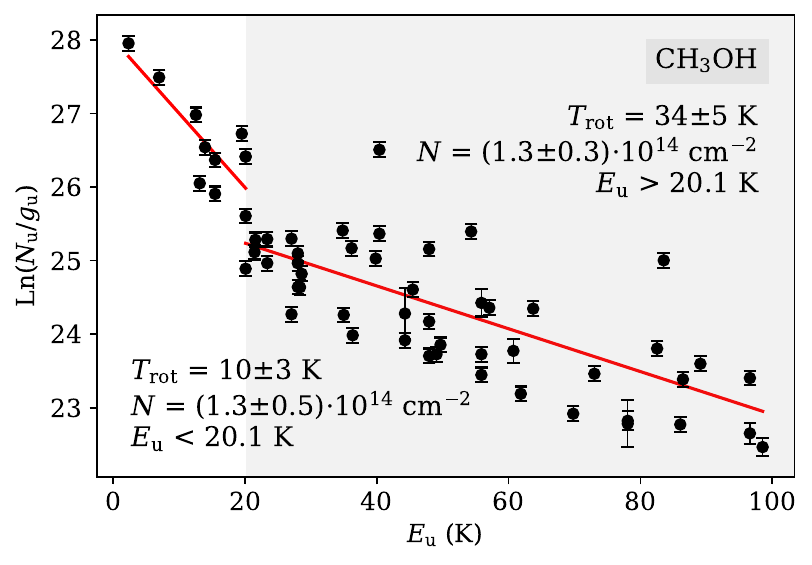}
    \includegraphics[width=0.323\hsize]{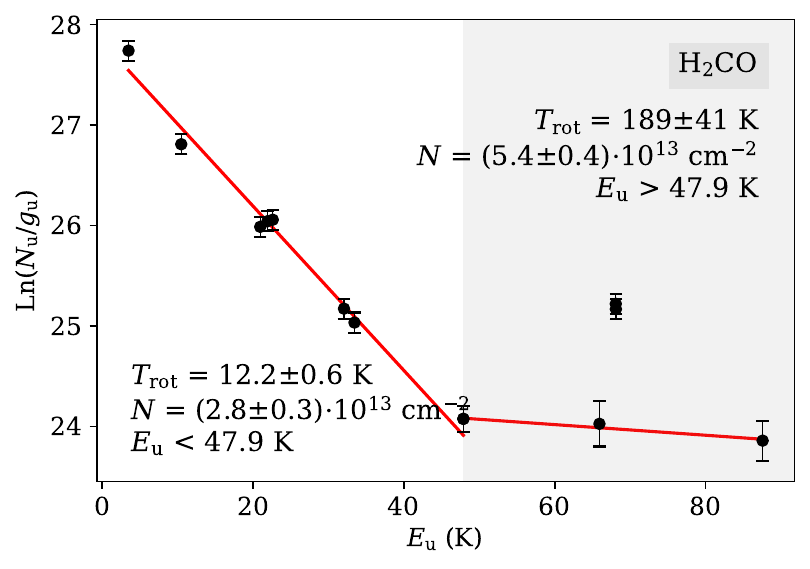}

\caption{continued.}
\label{diagramasIRAS4A2}
\end{figure*}

\begin{figure*}[h]
    \centering
    \includegraphics[width=0.33\hsize]{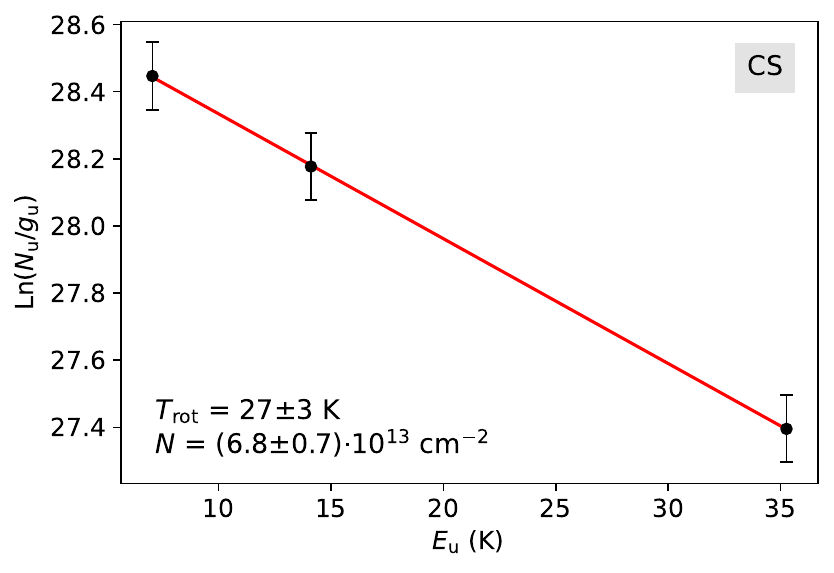}
    \includegraphics[width=0.33\hsize]{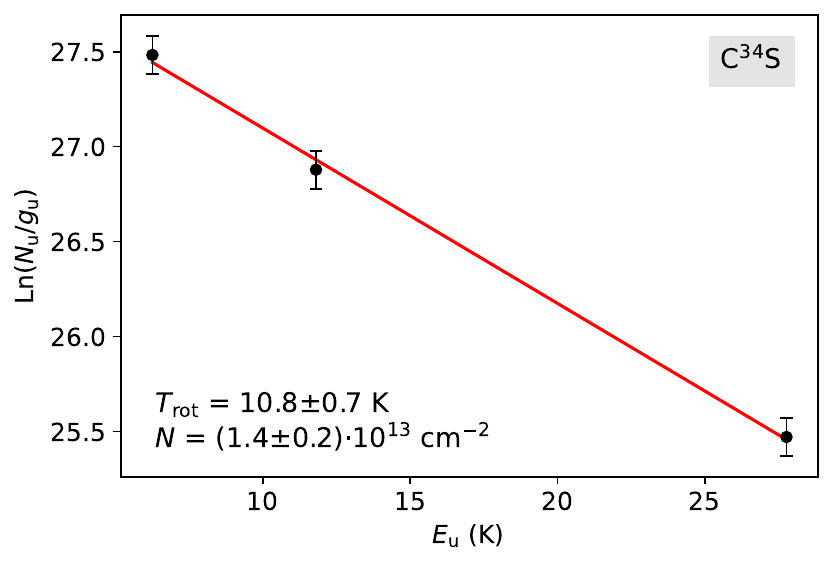}
    \includegraphics[width=0.33\hsize]{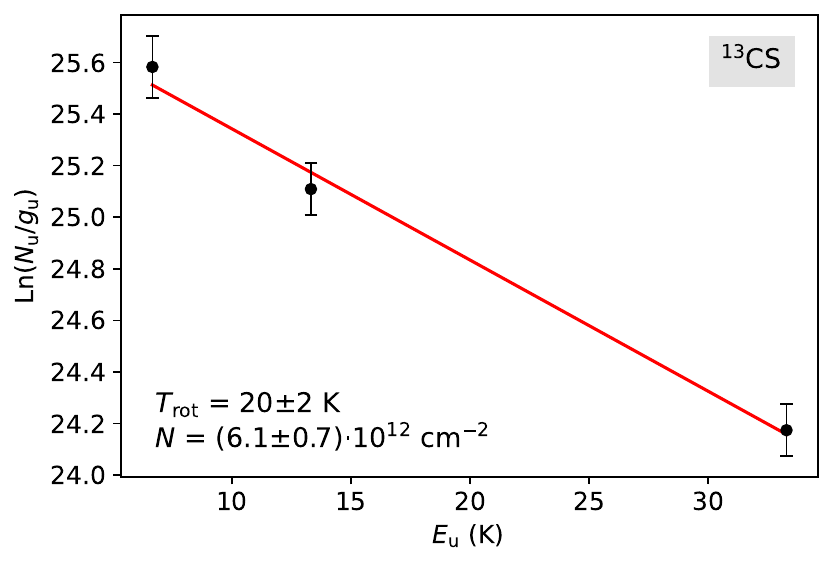}
    
\caption{Rotational diagrams for the wide component of species in IRAS\,4A.}
\label{diagramasIRAS4Awide}
\end{figure*}

\begin{figure*}[h]
\addtocounter{figure}{-1}
    \centering
    
    \includegraphics[width=0.33\hsize]{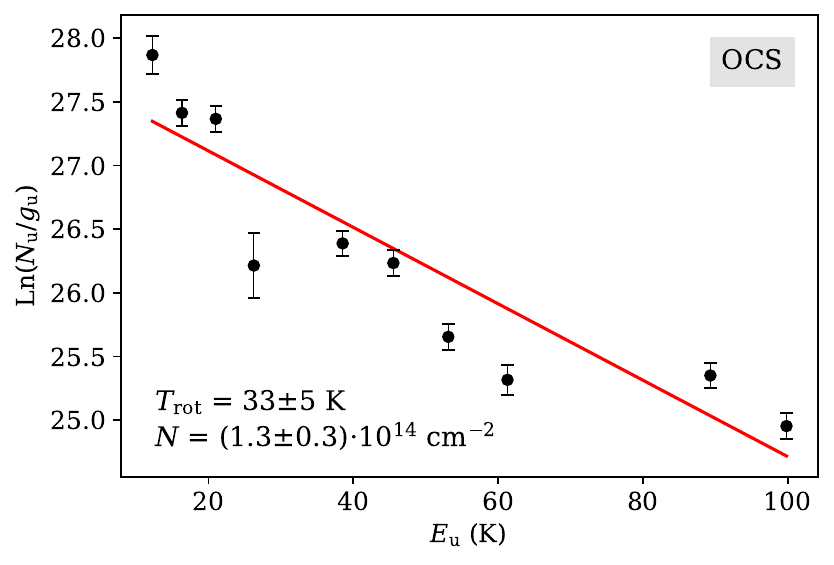}  
        \includegraphics[width=0.33\hsize]{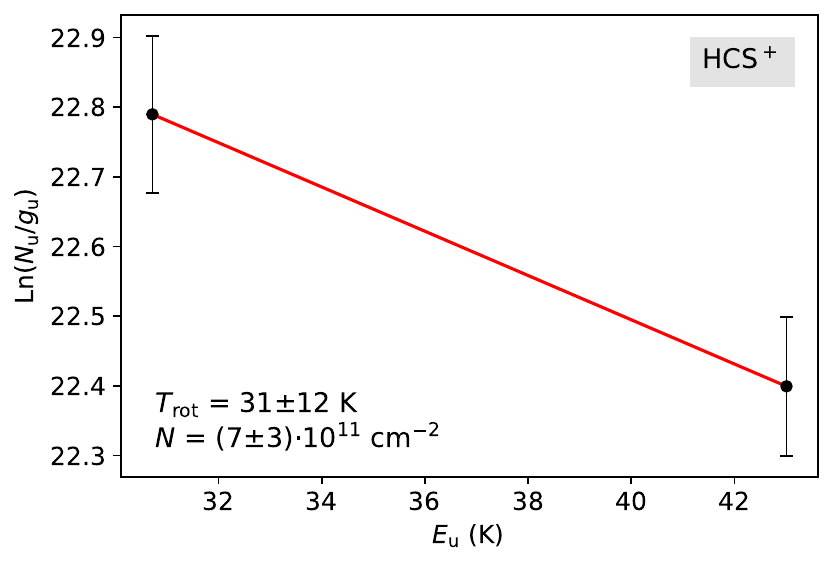}
        \includegraphics[width=0.33\hsize]{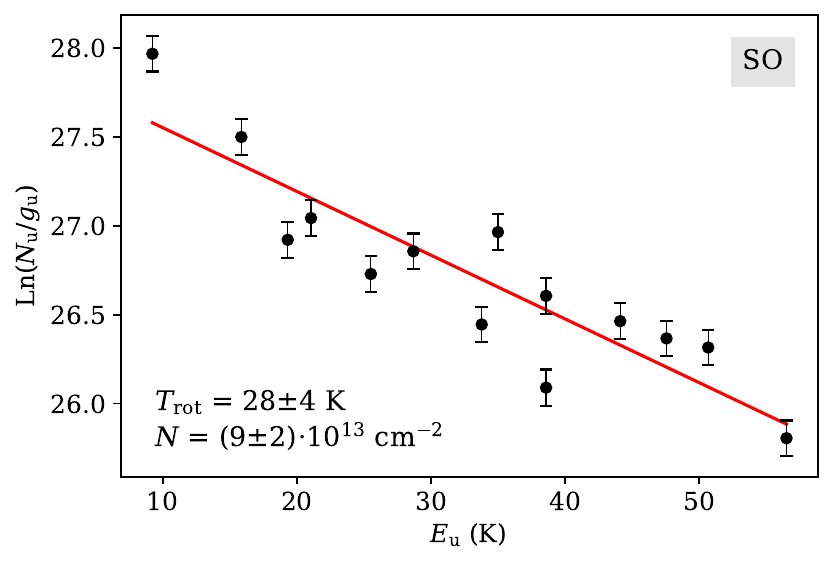}\\
        \includegraphics[width=0.33\hsize]{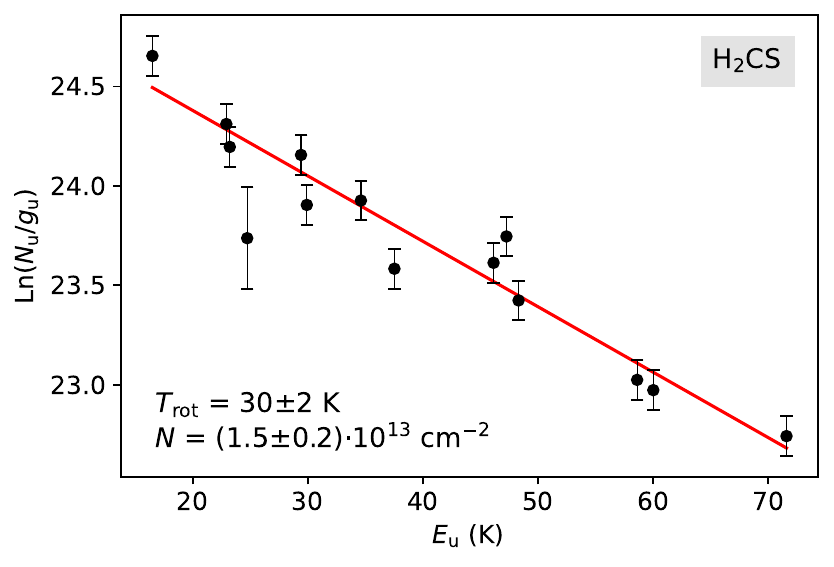}
        \includegraphics[width=0.323\hsize]{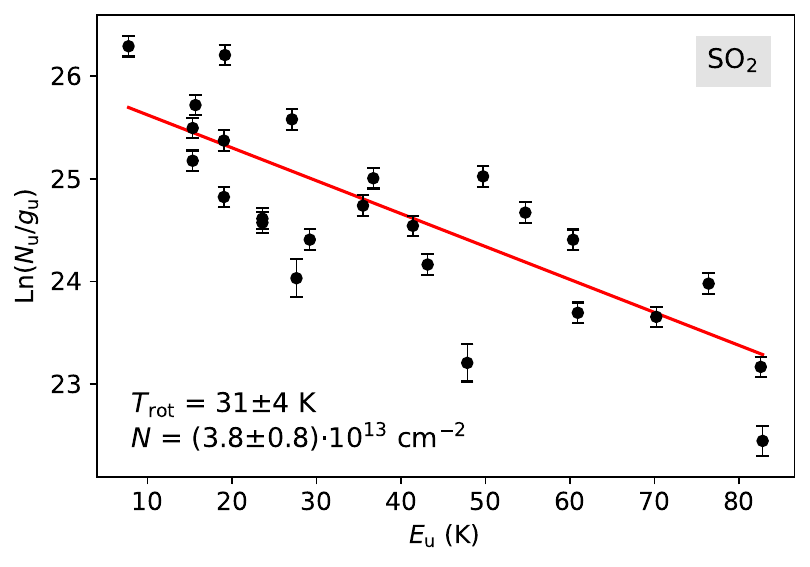}
        \includegraphics[width=0.33\hsize]{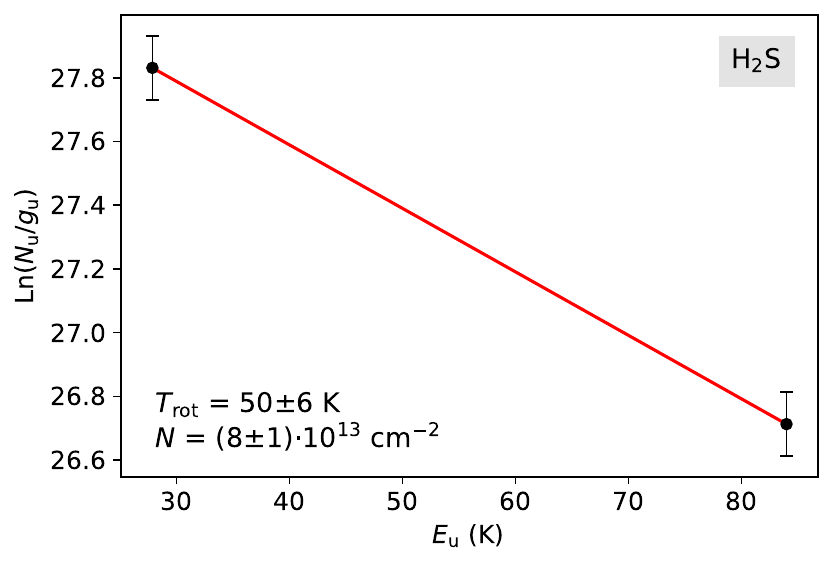}\\
        \includegraphics[width=0.323\hsize]{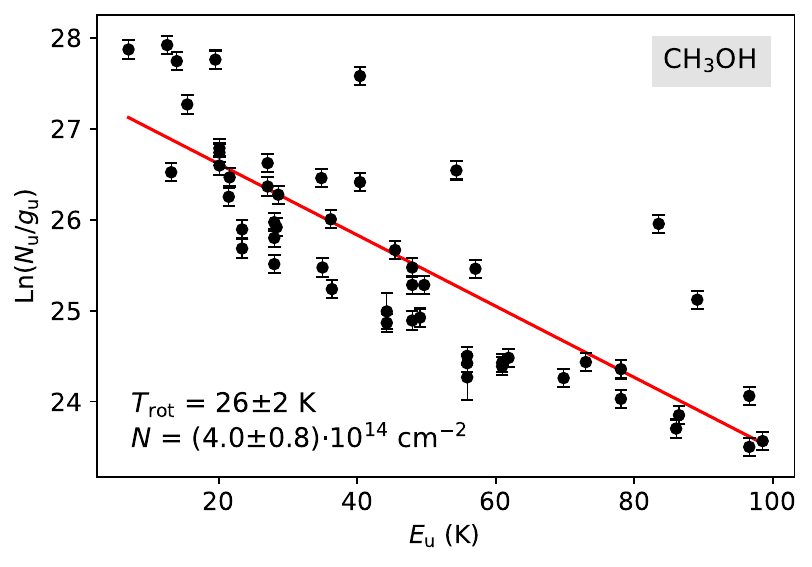}
        \includegraphics[width=0.323\hsize]{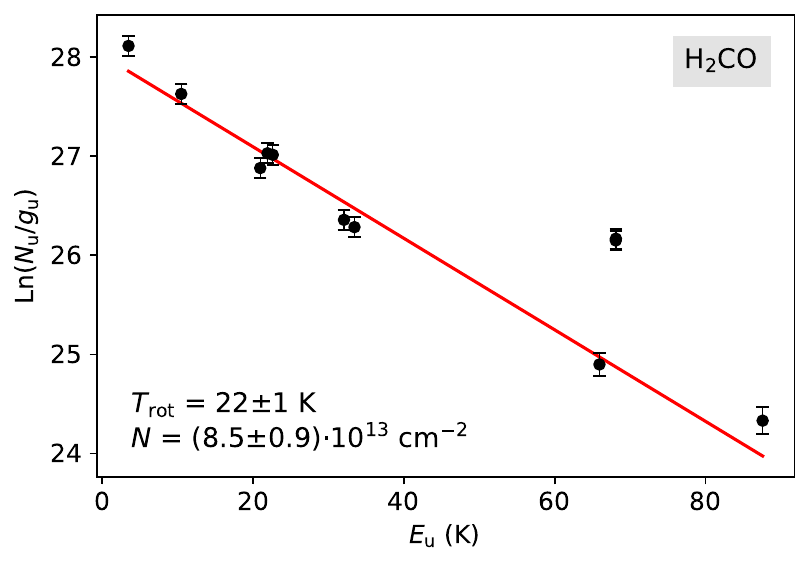}
\caption{continued.}
\label{diagramasIRAS4Awide2}
\end{figure*}

\begin{figure}
    \begin{minipage}[t]{0.5\textwidth}
    \centering
        \includegraphics[width=\hsize,valign=c]{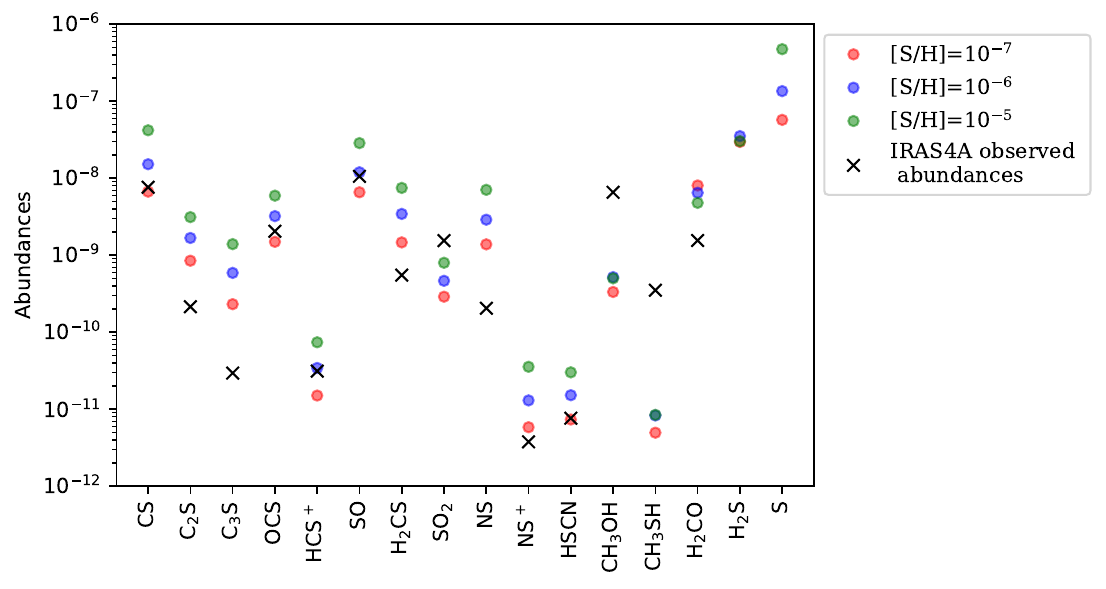}
    \caption{Abundances derived with the chemical model for values $\text{[S/H]}=10^{-7},10^{-6}$ and $10^{-5}$, while fixing the other parameters to the values listed in Table \ref{modelling_results} for IRAS\,4A.}
    \label{fig:SH}
    \end{minipage}
    \hfill
    
\end{figure}

\begin{table*}[h!]
\centering
\caption{LTE column densities $(N_\text{rot})$, rotational temperatures $(T_\text{rot})$, and non-LTE column densities $(N_\chi)$ derived from the narrow component of searched molecules in HH\,212 and IRAS\,4A.}
\begin{tabular}{l|lll|lll}
\hline\hline         \noalign{\smallskip}
     \multicolumn{1}{c}{}   & \multicolumn{3}{c}{HH\,212} & \multicolumn{3}{c}{IRAS\,4A} \\
     \noalign{\smallskip}
\hline
\noalign{\smallskip}
\multicolumn{1}{l|}{Species} & \multicolumn{1}{c}{$T_\text{rot}\ \mathrm{(K)}$}   & $N_\text{rot}\ (\times\,10^{12}\ \mathrm{cm^{-2})}$ & $N_\chi\ \mathrm{(\times\,10^{12}\ cm^{-2})}$  & \multicolumn{1}{c}{$T_\text{rot}\ \mathrm{(K)}$}   & $N_\text{rot}\ \mathrm{(\times\,10^{12}\ cm^{-2})}$ & $N_\chi\ \mathrm{(\times\,10^{12}\ cm^{-2})}$     \\
\noalign{\smallskip}
\hline
\noalign{\smallskip}
CS                      &  $ 5.0 \pm 0.4 $ & $ 8 \pm 1 $ &   $ 9.5 _{ -0.9 }^{+ 0.4 }$ &  $6 \pm 2  $ & $ 30 \pm 10 $ &  $ 37_{ -9}^{ +16}$\\[2.6pt]
               &       &   &    &  $17 \pm 2  $ & $ 18 \pm 3 $ &   \\[2.6pt]
$\mathrm{C{^{34}S}}$   &  $2.5 \pm 0.6  $ & $ 0.6 \pm 0.3 $ &  $ 1.7_{- 0.8 }^{+ 8.3 }$&  $4.5 \pm 0.9  $ & $ 5 \pm 2 $&  $ 6.0_{ -0.9}^{ +1.6}$\\[2.6pt]
    &    &   &   &  $10.2 \pm 0.9  $ & $ 2.8 \pm 0.5 $&   \\[2.6pt]
$\mathrm{C^{33}S}$     &  $ 5.0$$^{(a)}$ & $ <0.64$$^{(b)}$ & $< 0.11 $$^{(b)}$ &  $ 10 \pm 1 $ & $1.3\pm0.2$ &  $ 1.7_{- 0.4 }^{+ 0.2 }$ \\[2.6pt]
      &   &   &  &  $ 14 \pm 1 $ & $ 1.2 \pm 0.2 $&    \\[2.6pt]
$\mathrm{^{13}CS}$   &  $ 5 \pm 2 $ &  $0.13 \pm 0.05 $&  $0.133_{-0.009}^{+0.017}$ &  $ 6.5 \pm 0.5 $ & $1.6\pm0.2$ &  $ 1.68 _{- 0.02 }^{+ 0.01 }$\\[2.6pt]
    &    &  &    &   $21 \pm 3  $ & $ 1.2 \pm 0.2 $&   \\[2.6pt]
$\mathrm{C_2S}$    &  $6.2 \pm 0.6  $ & $ 0.57 \pm 0.08 $&  $ 0.63_{ -0.05 }^{ +0.10 }$ &  $12.3 \pm 0.8  $ & $ 3.2 \pm 0.5 $&   $ 3.9_{- 0.8 }^{+ 0.4}$\\[2.6pt]
$\mathrm{C_2{^{34}S}}$  &  $ 6.2 ^{(a)}$ & $ < 0.047 ^{(b)}$ & $< 0.091 ^{(b)}$&  $21\pm 5$ & $0.21\pm0.06$&  $ 0.17_{- 0.02 }^{+ 0.04 }$\\[2.6pt]
$\mathrm{C_3S}$   &  $34\pm 12$& $0.18\pm0.03$&  - &  $15 \pm 1 $ & $0.56\pm0.09$&  -\\[2.6pt]
$\mathrm{C_3{^{34}S}}$  &  $ 34 ^{(a)}$ & $ < 0.092 ^{(b)}$ & -   &  $ 15 ^{(a)}$ & $ < 0.052 ^{(b)}$ & - \\[2.6pt]
$\mathrm{OCS}$ & $ 10 ^{(a)}$ & $ < 0.84 ^{(b)}$ & $< 1.6 ^{(b)}$ &  $37 \pm 3  $ & $ 37 \pm 3 $&  $ 39_{- 8 }^{+ 4}$ \\[2.6pt]
  &   &   &   &   $126\pm 56$ & $40\pm10$&   \\[2.6pt]
$\mathrm{OC^{34}S}$ &  $ 10 ^{(a)}$ & $ < 0.60 ^{(b)}$ & $< 1.2 ^{(b)}$ &  $ 82 \pm 11 $ & $ 5.7 \pm 0.6 $&  -\\[2.6pt]
$\mathrm{OC^{33}S}$ &  $ 10 ^{(a)}$ & $ < 1.8 ^{(b)}$ & $< 1.7 ^{(b)}$&  $ 82 ^{(a)}$ & $ < 3.0 ^{(b)}$ &-\\[2.6pt]
$\mathrm{HCS^+}$ &$ 10 ^{(a)}$ & $ 0.3 \pm 0.2 $  &   $ 0.3 _{- 0.1}^{+ 0.1 }$&  $ 8 \pm 1 $ & $0.64\pm0.07$&  $ 0.60_{- 0.07 }^{+ 0.15}$\\[2.6pt]
  &  &    &   &   $ 16 \pm 1 $ & $ 0.63 \pm 0.08 $&   \\[2.6pt]
$\mathrm{HC^{34}S^+}$&  $ 10 ^{(a)}$ & $ < 0.12 ^{(b)}$ & $< 0.13 ^{(b)}$&  $ 8 ^{(a)}$ & $ < 0.093 ^{(b)}$ & $< 0.17 ^{(b)}$\\[2.6pt]
$\mathrm{SO}$ &  $7.4 \pm 0.7  $ & $ 5 \pm 1 $ &  $ 5.2_{ -0.9}^{+ 2.8}$&  $ 12.1 \pm 0.6 $ & $100\pm10$&  $ 91 _{- 5 }^{+ 4}$\\[2.6pt]
$\mathrm{^{34}SO}$ &  $ 7.4 ^{(a)}$ & $ 0.3 \pm 0.1 $ &$ 0.26_{- 0.05}^{+ 0.14 }$ &  $9.6\pm 0.6$ & $8\pm2$&  $ 8_{- 2 }^{+ 1 }$\\[2.6pt]
$\mathrm{S^{18}O}$   &  $ 7.4 ^{(a)}$ & $ < 0.21 ^{(b)}$ & $< 0.22 ^{(b)}$&  $ 9 \pm 2 $ & $ 1.2 \pm 0.6 $&  $ 1.3_{- 0.3 }^{+ 0.4 }$\\[2.6pt]
$\mathrm{H_2CS}$ &  $7.0 \pm 0.6  $ & $ 0.9 \pm 0.1 $  &  $ 1.5_{- 0.4 }^{+ 0.7 }$& $16.6\pm 0.8$ & $12\pm1$&  $ 11_{- 1}^{+ 3 }$\\[2.6pt]
  &    &    &   &  $63\pm 46$ & $7\pm6$&   \\[2.6pt]
$\mathrm{SO_2}$  &  $ 10 ^{(a)}$ & $ 0.9 \pm 0.5 $ &  $ 1.1_{- 0.3 }^{+ 0.8 }$ &  $11\pm 1$ & $13\pm3$&  $ 15_{- 2 }^{+ 2 }$\\[2.6pt]
   &    &   &    &  $21\pm 6$ & $8\pm7$&   \\[2.6pt]
$\mathrm{^{34}SO_2}$  &  $ 10 ^{(a)}$ & $ < 0.38 ^{(b)}$ &$< 0.39 ^{(b)}$&  $11\pm 3$ & $1.1\pm0.3$&  $1.2_{-0.2}^{+0.3}$\\[2.6pt]
$\mathrm{NS}$ &$ 10 ^{(a)}$ & $ < 1.5 ^{(b)}$ &$< 2.6 ^{(b)}$&  $10.3\pm 0.7$ & $5.5\pm0.9$&  $ 3.9_{- 0.4 }^{+ 0.6 }$\\[2.6pt]
$\mathrm{NS^+}$ &  $ 10 ^{(a)}$ & $ < 0.051 ^{(b)}$ &$< 0.047 ^{(b)}$ &  $7\pm 2$ & $0.08\pm0.03$&  $ 0.072_{- 0.002 }^{+ 0.012 }$\\[2.6pt]
$\mathrm{H_2S}$ & $ 10 ^{(a)}$ & $ 2 \pm 1 $ &  $ 7_{- 3}^{+ 11}$  &  $44\pm 5$ & $34\pm5$ & -  \\[2.6pt]
$\mathrm{H_2{^{34}S}}$ &  $ 10 ^{(a)}$ & $ < 0.56 ^{(b)}$ & $< 2.0 ^{(b)}$& $48\pm 6$ & $10\pm2$& -\\[2.6pt]
$\mathrm{HSCN}$  &  $ 10 ^{(a)}$ & $ < 0.060 ^{(b)}$ &-&  $ 37 ^{(a)}$ & $ 0.15 \pm 0.07 $&-\\[2.6pt]
$\mathrm{HNCS}$  &  $ 10 ^{(a)}$ & $ < 0.16 ^{(b)}$ &-&  $ 37 ^{(a)}$ & $ < 0.71 ^{(b)}$&-\\[2.6pt]
  CH$_3$SH & $ 10 ^{(a)}$ & $9\pm4$ & -  & $39\pm11$ &$7\pm2$ &-
  \\\noalign{\smallskip}
  \hline\noalign{\smallskip}

$\mathrm{C^{18}O}$ &  - & - & -&  $16\pm 4$ & $3000\pm500$& $ 3200_{- 300 }^{+ 700}$ \\[2.6pt]
$\mathrm{C^{17}O}$ &  $ 10 ^{(a)}$ & $ 80 \pm 40 $&  $ 110 _{- 30 }^{+ 30 }$ &  $ 16 ^{(a)}$ & $ 1100 \pm 500 $&  $ 1100_{- 300 }^{+ 300}$\\[2.6pt]
$\mathrm{CH_3OH}$ &  $6.0 \pm 0.8  $ & $ 15 \pm 4 $ &  $ 15_{- 2 }^{+ 9}$&  $10 \pm 3  $ & $ 130 \pm 50$&  $ 125_{- 9 }^{+ 8 }$\\[2.6pt]
  &    &   &  &  $34 \pm 5  $ & $ 130 \pm 30 $&   \\[2.6pt]

$\mathrm{H_2CO}$ &  $24 \pm 13  $& $ 11 \pm 5 $&  $ 8_{- 1}^{+ 4}$&   $12.2\pm 0.6$&  $28\pm3$ &  $ 30 _{- 4 }^{+ 6}$\\
  &    &  &   &   $189\pm 41$ &  $54\pm4$&  \\\noalign{\smallskip}

\hline

\end{tabular}

\tablefoot{$^{(a)}$Rotational temperature has been fixed. $^{(b)}$Column density upper limit for the undetected species.}
\label{ncol}
\end{table*}

\begin{table}[]
    \centering
    \begin{minipage}[t]{0.49\linewidth}
    \centering
    \caption{LTE column densities $(N_\text{rot})$ and rotational temperatures $(T_\text{rot})$ derived from the wide component of detected molecules in HH\,212 and IRAS\,4A, and mass of the molecules associated with the wide component relative to the narrow component for each species ($m_\text{wide}/m_\text{narr}$).}
    \label{ncolwide}
    \begin{tabular}{llll}
\hline\hline         
\noalign{\smallskip}
\multicolumn{1}{l}{Species} & \multicolumn{1}{c}{$T_\text{rot}\ \mathrm{(K)}$}   & $N_\text{rot}\ \mathrm{(\times\,10^{12}\ cm^{-2})}$ &
$m_\text{wide}/m_\text{narr}$$^{(a)}$ 
\\
\noalign{\smallskip}
\hline
\noalign{\smallskip}
\multicolumn{4}{c}{HH\,212}\\
\noalign{\smallskip}
\hline
\noalign{\smallskip}
SO        & {$7.4^{\ (b)}$}&$1.3\pm0.7$ &0.27 \\

\noalign{\smallskip}
\hline
\noalign{\smallskip}
\multicolumn{4}{c}{IRAS\,4A}\\
\noalign{\smallskip}
\hline
\noalign{\smallskip}
CS        &$27\pm 3$    & $68\pm7$ & 1.5\\
C$^{34}$S$^{(c)}$ &$10.8\pm 0.7$& $14\pm2$ & 0.15\\
$^{13}$CS$^{(c)}$ &$20\pm 2$      &$6.1\pm0.7$ & 0.63\\
OCS$^{(c)}$       &$33\pm 5$ & $130\pm30$ &0.091\\
HCS$^+$   &$31\pm 12$    & $0.7\pm0.3$ & 0.51\\
SO        &$28\pm 4$& $90\pm20$ & 0.86\\
H$_2$CS   &$30\pm 2$& $15\pm2$& 0.81\\
SO$_2$    &$31\pm 4$&$38\pm8$& 1.8\\
H$_2$S    &$50\pm 6$ &$80\pm10$& 2.4\\
CH$_3$OH  &$26\pm 2$& $400\pm80$& 1.5\\
H$_2$CO &$22\pm 1$&$85\pm9$& 1.0\\
\noalign{\smallskip}
\hline

\end{tabular}
\tablefoot{{$^{(a)}$$(N_\text{rot}^\text{wide}\cdot {\theta_\text{s}^\text{wide}}^2) / (N_\text{rot}^\text{narr}\cdot  {\theta_\text{s}^\text{narr}}^2)$, where $\theta_\text{s}^\text{wide}=\theta_\text{s}^\text{narr}$ in the case of extended emission. $^{(b)}$Rotational temperature has been fixed.}{ $^{(c)}$Assuming $\theta_\text{s}^\text{wide}=11''$, and $\theta_\text{s}^\text{narr}=30''$, the HPBW of the IRAM-30m telescope at 3~mm.}}
    
    \end{minipage}
\hfill
    \begin{minipage}[t]{0.49\linewidth}\centering
    \caption{References for the collisional rate coefficients used in RADEX.}
    \label{table:colisiones}
    \begin{tabular}{ll}    
\hline\hline                    
\noalign{\smallskip}
Species  & Reference  \\
\noalign{\smallskip}
\hline              
\noalign{\smallskip}
CS & \cite{CScoeff}\\
C$_2$S & \cite{C2Scoeff}\\
OCS & \cite{OCScoeff}\\
HCS$^+$ & \cite{HCS+coeff}\\
SO & \cite{SOcoeff}\\
o-H$_2$CS & Scaled from o-H$_2$CO\\
p-H$_2$CS & Scaled from p-H$_2$CO\\

SO$_2$ & \cite{SO2coeff}\\
NS & \cite{NScoeff}\\
NS$^+$ & \cite{NS+coeff}\\
o-H$_2$S & \cite{H2Scoeff}\\
p-H$_2$S & \cite{H2Scoeff}\\
C$^{18}$O& \cite{C17OC18Ocoeff}\\
C$^{17}$O& \cite{Yang2010}\\
A-CH$_3$OH & \cite{Dagdigian24}\\
E-CH$_3$OH & \cite{Dagdigian24}\\
o-H$_2$CO & \cite{H2COcoeff}\\
p-H$_2$CO & \cite{H2COcoeff}\\
\noalign{\smallskip}
\hline

\end{tabular}

\end{minipage}
\end{table}

\begin{table*}[h]
\centering
\caption{Non-LTE column densities and H$_2$ densities of species in HH\,212 for different kinetic temperatures.}
\begin{tabular}{lllllll}
\hline\hline
\noalign{\smallskip}
                             & \multicolumn{6}{c}{HH\,212}                                                                                                                                                                                                                                                                                                \\
\noalign{\smallskip}\hline\noalign{\smallskip}
                 & \multicolumn{3}{c}{$n_\mathrm{H_2}\ \mathrm{(\times\,10^5\ cm^{-3})}$}                                                                                 & \multicolumn{3}{c}{$N_\chi\ (\times\,10^{12}\ \mathrm{cm^{-2})}$}\\
\noalign{\smallskip}\hline\noalign{\smallskip}
\multicolumn{1}{l}{${T_\text{k}\ \mathrm{(K)}}$}    & \multicolumn{1}{c}{10} & \multicolumn{1}{c}{20} & \multicolumn{1}{c}{30} & \multicolumn{1}{c}{10} & \multicolumn{1}{c}{20} & \multicolumn{1}{c}{30} \\ 
\noalign{\smallskip} \hline \noalign{\smallskip}

CS  &  $ 1.84 _{ -0.57 }^{ +0.64 }$  &  $ 0.791 _{ -0.089 }^{+ 0.161 }$  &  $ 0.60 _{- 0.17 }^{ +0.19 }$  &  $ 11_{ -1 }^{+ 2 }$  &  $ 9.5_{ -0.9 }^{+ 0.4 }$  &  $ 8.7 _{ -0.9 }^{ +1.9 }$\\ \noalign{\smallskip}
    C$^{34}$S  &  $ 0.19 _{- 0.17 }^{+ 0.21 }$  &  $ 0.072 _{ -0.062 }^{ +0.108 }$  &  $0.054_{-0.044}^{+0.075}$  &  $ 1.3_{- 0.5 }^{ +8.2 }$  &  $ 1.7_{- 0.8 }^{+ 8.3 }$  &  $ 1.7 _{- 0.8 }^{ +7.2 }$\\ \noalign{\smallskip}
$^{13}$CS  &  $1.60_{-0.58}^{+0.86}$  &  $0.791_{-0.23}^{+0.34}$ &  $0.60_{-0.14}^{+0.12}$  &  $0.15_{-0.01}^{+0.02}$  &  $0.133_{-0.009}^{+0.017}$ &  $0.133_{-0.005}^{+0.008}$\\ \noalign{\smallskip}
C$_2$S  &  $ 0.52 _{ -0.17 }^{ +0.13 }$  &  $ 0.202 _{- 0.053 }^{+ 0.032 }$  &  $ 0.139 _{- 0.047 }^{+ 0.041 }$  &  $ 0.72 _{- 0.05 }^{ +0.12}$  &  $ 0.63 _{ -0.05 }^{ +0.10 }$  &  $ 0.6 _{ -0.1 }^{ +0.1 }$\\ \noalign{\smallskip}
HCS$^+$$^{(a)}$  &  -  &  -  &  -  &  $ 0.3_{- 0.1}^{+ 0.1 }$  &  $ 0.3 _{- 0.1}^{+ 0.1 }$  &  $ 0.3_{- 0.1}^{+ 0.2}$\\ \noalign{\smallskip}
SO  &  $ 1.60 _{ -0.87 }^{ +4.29}$  &  $ 0.63 _{- 0.38 }^{+ 0.42 }$  &  $ 0.43 _{- 0.26 }^{ +0.27 }$  &  $ 6_{ -1}^{ +1}$  &  $ 5.2_{ -0.9}^{+ 2.8}$  &  $ 5 _{- 1}^{ +3}$\\ \noalign{\smallskip}
$^{34}$SO$^{(a)}$  & -  &  -  &  -  & $ 0.28_{- 0.06 }^{+ 0.10 }$ &  $ 0.26_{- 0.05}^{+ 0.14 }$  &  $ 0.26 _{- 0.05 }^{+ 0.16 }$\\ \noalign{\smallskip}
H$_2$CS  &  $ 0.079 _{- 0.029 }^{+ 0.051 }$  &  $ 0.045 _{- 0.021 }^{+ 0.022 }$ &  $ 0.026 _{- 0.016 }^{+ 0.029 }$  &  $ 1.5 _{- 0.4 }^{+ 0.5 }$  &  $ 1.5 _{- 0.4 }^{+ 0.7 }$  &  $ 1.9_{- 0.9 }^{+ 2.1 }$\\ \noalign{\smallskip}
SO$_2$$^{(a)}$  &  -  &  -  &  -  &  $ 1.4 _{- 0.6}^{+ 1.0 }$  &  $ 1.1_{- 0.3 }^{+ 0.8 }$  &  $ 1.0_{- 0.3 }^{+ 0.7 }$\\ \noalign{\smallskip}
o-H$_2$S$^{(a)}$  &  -  &  -  &  -  &  $ 4_{- 3}^{+ 7 }$  &  $ 5_{- 3}^{+ 10 }$  &  $ 6 _{- 3 }^{+ 11 }$\\ \noalign{\smallskip}
C$^{17}$O$^{(a)}$  &  -  &  -  &  -  &  $ 90 _{- 20}^{+ 20}$  &  $ 110 _{- 30 }^{+ 30 }$  &  $ 130_{- 40}^{+ 40}$\\ \noalign{\smallskip}
A-CH$_3$OH  &  $0.356_{-0.096}^{+0.465}$  &  $0.139_{-0.044}^{+0.056}$ &  $0.095_{-0.021}^{+0.038}$  &  $10_{-1}^{+1}$  &  $9_{-1}^{+1}$  &  $8.7_{-1.3}^{+0.9}$\\ \noalign{\smallskip}
E-CH$_3$OH  & $0.43_{-0.24}^{+0.41}$  &  $0.20_{-0.16}^{+0.41}$  &  $ 0.115 _{- 0.085 }^{+ 0.250 }$  &  $7_{-2}^{+3}$  &  $6_{-2}^{+9}$  &  $ 7_{- 2}^{+ 9}$\\ \noalign{\smallskip}

\hline\noalign{\smallskip}
\multicolumn{1}{l}{${T_\text{k}\ \mathrm{(K)}}$}    & \multicolumn{1}{c}{40} & \multicolumn{1}{c}{} & \multicolumn{1}{c}{} & \multicolumn{1}{c}{40} & \multicolumn{1}{c}{} & \multicolumn{1}{c}{} \\ 
\noalign{\smallskip} \hline \noalign{\smallskip}
H$_2$CO  &  $ 1.9 _{- 1.2 }^{+ 5.1 }$  &     &     &  $ 8 _{- 1}^{+ 4}$  &     &   \\ \noalign{\smallskip}

\hline
\end{tabular}
\tablefoot{$^{(a)}$Species with a single detected transition where $n_\mathrm{H_2}$ was fixed as the SO value at each $T_\text{k}$.}
\label{RADEXHH212}
\end{table*}

\begin{table*}[h]
\centering
\caption{Non-LTE column densities and H$_2$ densities of species in IRAS\,4A for different kinetic temperatures.}
\begin{tabular}{lllllll}

\hline\hline
\noalign{\smallskip}
                             & \multicolumn{6}{c}{IRAS\,4A}                                                                                                                                                                                                                                                                                                \\
\noalign{\smallskip}\hline\noalign{\smallskip}
                 & \multicolumn{3}{c}{$n_\mathrm{H_2}\ \mathrm{(\times\,10^5\ cm^{-3})}$}                                                                                 & \multicolumn{3}{c}{$N_\chi\ (\times\,10^{12}\ \mathrm{cm^{-2})}$}\\
\noalign{\smallskip}\hline\noalign{\smallskip}
\multicolumn{1}{l}{${T_\text{k}\ \mathrm{(K)}}$}    & \multicolumn{1}{c}{10} & \multicolumn{1}{c}{20} & \multicolumn{1}{c}{30} & \multicolumn{1}{c}{10} & \multicolumn{1}{c}{20} & \multicolumn{1}{c}{30} \\ 
\noalign{\smallskip} \hline \noalign{\smallskip}

CS  &  $ 3.9 _{ -2.4}^{ +83.2}$  &  $ 1.27 _{- 0.59}^{ +1.39}$  &  $ 0.83 _{- 0.34}^{+ 0.74}$ &  $ 50_{ -10}^{ +30}$  &  $ 37_{ -9}^{ +16}$  &  $ 40 _{ -10 }^{+ 10}$\\ \noalign{\smallskip}
C$^{34}$S  &  $ 2.12 _{- 1.1}^{+ 1.9}$  &  $ 0.91 _{- 0.40 }^{ +0.51}$  &  $ 0.60 _{- 0.22}^{+ 0.38}$  &  $ 7 _{- 1 }^{+ 2}$  &  $ 6.0 _{ -0.9}^{ +1.6}$  &  $ 6 _{ -1}^{ +1}$\\ \noalign{\smallskip}
C$^{33}$S  &  -$^{(a)}$  &  $ 1.93 _{- 0.63 }^{+ 1.59}$  &  $ 1.33 _{- 0.45 }^{+ 0.95 }$  &  -$^{(a)}$  &  $ 1.7 _{- 0.4 }^{+ 0.2 }$  &  $ 1.5 _{- 0.2 }^{+ 0.3 }$\\ \noalign{\smallskip}
$^{13}$CS  &   $3.24 _{- 0.77 }^{+ 1.91}$  &  $ 1.389 _{- 0.080 }^{+ 0.083 }$  &  $ 0.91 _{- 0.16 }^{+ 0.28}$  &  $ 1.93 _{- 0.26 }^{+ 0.09}$  &  $ 1.68 _{- 0.02 }^{+ 0.01 }$  &  $ 1.68 _{- 0.22 }^{+ 0.08 }$\\ \noalign{\smallskip}
C$_2$S  &  -$^{(a)}$  &  $ 0.63 _{- 0.15 }^{+ 0.68}$  &  $ 0.43 _{- 0.19 }^{+ 0.46}$  &  -$^{(a)}$  &  $ 3.9_{- 0.8}^{+ 0.4}$  &  $ 3.4_{- 0.8 }^{+ 1.2}$\\ \noalign{\smallskip}
C$_2$$^{34}$S  &  -$^{(a)}$  &  -$^{(a)}$  &  $ 1.93 _{- 0.85 }^{+ 3.16}$  &  -$^{(a)}$  &  -$^{(a)}$  &  $ 0.17_{- 0.02 }^{+ 0.04}$\\ \noalign{\smallskip}
HCS$^+$  &  $ 1.9 _{- 1.0 }^{+ 29.0 }$  &  $ 0.52 _{- 0.20 }^{+ 0.65}$  &  $ 0.36 _{- 0.10 }^{+ 0.29}$  &  $ 0.69_{- 0.05 }^{+ 0.12 }$  &  $ 0.60 _{- 0.07 }^{+ 0.15}$  &  $ 0.60 _{- 0.07 }^{+ 0.11}$\\ \noalign{\smallskip}
SO  &  -$^{(a)}$  &  $ 3.39 _{- 0.66 }^{+ 0.70}$  &  $ 1.93 _{- 0.68 }^{+ 0.47}$  &  -$^{(a)}$  &  $ 91 _{- 5 }^{+ 4}$  &  $ 79 _{- 7 }^{+ 17 }$\\ \noalign{\smallskip}
$^{34}$SO  &  -$^{(a)}$  &  $ 0.91 _{- 0.25 }^{+ 0.52 }$  &  $ 0.52 _{- 0.13 }^{+ 0.29}$  &  -$^{(a)}$  &  $ 8 _{- 2 }^{+ 1}$  &  $ 8 _{- 2 }^{+ 1 }$\\ \noalign{\smallskip}
S$^{18}$O  &  -$^{(a)}$  &  $ 0.75 _{- 0.46 }^{+ 1.55 }$  &  $ 0.43 _{- 0.19 }^{+ 0.63 }$  &  -$^{(a)}$  &  $ 1.3 _{- 0.3 }^{+ 0.4 }$  &  $ 1.3 _{- 0.3 }^{+ 0.4}$\\ \noalign{\smallskip}
H$_2$CS  & -$^{(a)}$   &  $ 10.5_{- 7.4 }^{+ 15.1 }$  &  $ 2.81 _{- 0.98 }^{+ 1.31 }$  &  -$^{(a)}$   &  $ 11_{- 1}^{+ 3 }$   &  $ 11_{- 1 }^{+ 2 }$\\ \noalign{\smallskip}

SO$_2$  &  -$^{(a)}$   &  $ 3.39 _{- 0.67 }^{+ 2.01 }$  &  $ 2.81 _{- 0.94 }^{+ 0.59 }$  &  -$^{(a)}$   &  $ 15 _{- 2 }^{+ 2 }$  &  $ 12.7_{- 0.7 }^{+ 2.3 }$\\ \noalign{\smallskip}
$^{34}$SO$_2$  &  -$^{(a)}$   &  $ 39 _{- 31 }^{+ 57}$   &  $ 39 _{- 33 }^{+ 59}$   &  -$^{(a)}$   &  $1.2_{-0.2}^{+0.3}$  &  $ 1.4 _{- 0.3 }^{+ 0.3}$\\ \noalign{\smallskip}
NS  &  -$^{(a)}$  &  $ 15.3 _{- 5.8 }^{+ 7.0}$  &  $ 15.3 _{- 7.2 }^{+ 4.1 }$  &  -$^{(a)}$  &  $ 3.9 _{- 0.4 }^{+ 0.6 }$  &  $ 3.4 _{- 0.3 }^{+ 0.7 }$\\ \noalign{\smallskip}
NS$^+$  &  $ 1.10_{- 0.20 }^{+ 0.24 }$  &  $ 0.429 _{- 0.140 }^{+ 0.057 }$  &  $ 0.295 _{- 0.022 }^{+ 0.021 }$  &  $ 0.083_{- 0.002 }^{+ 0.001 }$   &  $ 0.072_{- 0.002 }^{+ 0.012 }$  &  $ 0.072 _{- 0.002 }^{+ 0.001 }$\\ \noalign{\smallskip}
C$^{18}$O  &  -$^{(a)}$  &  $ 0.30_{- 0.19}^{+ 1.25 }$  &  $ 0.115 _{- 0.065 }^{+ 0.823}$$^{(b)}$  &  -$^{(a)}$  &  $ 3200_{- 300 }^{+ 700}$  &  $ 3200 _{- 600 }^{+ 1100 }$\\ \noalign{\smallskip}
C$^{17}$O$^{(b)}$  &  -$^{(a)}$  &  -  &  -  &  -$^{(a)}$  &  $ 1100_{- 300 }^{+ 300}$  &  $ 1200_{- 300 }^{+ 300 }$\\ \noalign{\smallskip}

A-CH$_3$OH  &  -$^{(a)}$  &  $ 0.202 _{- 0.028 }^{+ 0.068 }$  &  $ 0.139 _{- 0.025 }^{+ 0.050 }$  &  -$^{(a)}$ & $ 66 _{- 3 }^{+ 5 }$ &  $ 66_{- 7 }^{+ 4 }$\\ \noalign{\smallskip}
E-CH$_3$OH  & -$^{(a)}$  &  $ 0.91 _{- 0.45 }^{+ 1.36 }$  &  $ 0.91 _{- 0.43 }^{+ 1.04 }$  &  -$^{(a)}$  &  $ 60 _{- 8 }^{+ 7 }$  &  $ 66_{- 10 }^{+ 8 }$\\ \noalign{\smallskip}
H$_2$CO  &  -$^{(a)}$  &  $ 10.5 _{- 4.8 }^{+ 13.8 }$  &  $ 6.0 _{- 2.2 }^{+ 2.7 }$  &  -$^{(a)}$  &  $ 30 _{- 4 }^{+ 6}$  &  $ 30 _{- 4 }^{+ 5 }$\\ \noalign{\smallskip}

\hline\noalign{\smallskip}
\multicolumn{1}{l}{${T_\text{k}\ \mathrm{(K)}}$}    & \multicolumn{1}{c}{40} & \multicolumn{1}{c}{} & \multicolumn{1}{c}{} & \multicolumn{1}{c}{40} & \multicolumn{1}{c}{} & \multicolumn{1}{c}{} \\ 
\noalign{\smallskip} \hline \noalign{\smallskip}

OCS  &  $ 2.8 _{- 2.6 }^{+ 4.2 }$  &     &     &  $ 39_{- 8}^{+ 4}$  &     &   \\ \noalign{\smallskip}

\hline
\end{tabular}
\tablefoot{$^{(a)}$Rotational temperature higher than kinetic temperature. $^{(b)}$Species with a single detected transition where $n_\mathrm{H_2}$ was fixed as the SO value at each $T_\text{k}$.}
\label{RADEXIRAS4A}
\end{table*}
\begin{table}[h]
    \begin{minipage}[t]{0.49\linewidth}
    \centering
    \caption{Chemical model parameters.}
    \label{parameters}
    \begin{tabular}{ll}
\hline\hline         
\noalign{\smallskip}

$t$ & $10^2-10^7$ yr\\
$n_\text{H}$ & $(0.02-2)\times10^6$ $\mathrm{cm^{-3}}$\\
$A_\mathrm{V}$ & 18 mag \\
$\zeta_\mathrm{H_2}$ & $(0.01-1)\times10^{-15}$ $\mathrm{s^{-1}}$\\
\text{[S/H]} & $(0.0075-1.5)\times 10^{-5}$ \\
$\chi_\text{UV}$ (Draine) & 1\\
$T$ & $10-50$ K \\

\noalign{\smallskip}
\hline
\noalign{\smallskip}
\end{tabular}
    
    \end{minipage}
\hfill

\end{table}

\end{appendix}

\end{document}